\documentclass[12pt]{article}
\usepackage{color}
\usepackage{epsf}
\usepackage{graphicx}
\usepackage{a4}
\usepackage{amsmath}
\usepackage{amssymb}
\usepackage[title,titletoc,toc]{appendix}
\usepackage{tikz}
\usetikzlibrary{shapes,snakes}
\usetikzlibrary{shapes.misc}

\tikzset{cross/.style={cross out, draw=black, minimum size=2*(#1-\pgflinewidth), inner sep=0pt, outer sep=0pt},
cross/.default={1pt}}

\def\hybrid{\topmargin 0pt
        \oddsidemargin 0pt
        \headheight 0pt \headsep 0pt
        \textwidth 6.25in       
        \textheight 9.5in       
        \marginparwidth .875in
        \parskip 5pt plus 1pt   \jot = 1.5ex}

\catcode`\@=11
\def\marginnote#1{}

\newcount\hour
\newcount\minute
\newtoks\amorpm
\hour=\time\divide\hour by60
\minute=\time{\multiply\hour by60 \global\advance\minute by-\hour}
\edef\standardtime{{\ifnum\hour<12 \global\amorpm={am}%
        \else\global\amorpm={pm}\advance\hour by-12 \fi
        \ifnum\hour=0 \hour=12 \fi
        \number\hour:\ifnum\minute<10 0\fi\number\minute\the\amorpm}}
\edef\militarytime{\number\hour:\ifnum\minute<10 0\fi\number\minute}

\def\draftlabel#1{{\@bsphack\if@filesw {\let\thepage\relax
   \xdef\@gtempa{\write\@auxout{\string
      \newlabel{#1}{{\@currentlabel}{\thepage}}}}}\@gtempa
   \if@nobreak \ifvmode\nobreak\fi\fi\fi\@esphack}
        \gdef\@eqnlabel{#1}}
\def\@eqnlabel{}
\def\@vacuum{}
\def\draftmarginnote#1{\marginpar{\raggedright\scriptsize\tt#1}}

\def\draft{\oddsidemargin -.5truein
        \def\@oddfoot{\sl preliminary draft \hfil
        \rm\thepage\hfil\sl\today\quad\militarytime}
        \let\@evenfoot\@oddfoot \overfullrule 3pt
        \let\label=\draftlabel
        \let\marginnote=\draftmarginnote
   \def\@eqnnum{(\theequation)\rlap{\kern\marginparsep\tt\@eqnlabel}%
\global\let\@eqnlabel\@vacuum}  }

\def\numberbysection{\@addtoreset{equation}{section}
        \def\theequation{\thesection.\arabic{equation}}}

\def\titlepage{\@restonecolfalse\if@twocolumn\@restonecoltrue\onecolumn
     \else \newpage \fi \thispagestyle{empty}\c@page\z@
        \def\thefootnote{\fnsymbol{footnote}}
	\setcounter{page}{0} }
\def\endtitlepage{\if@restonecol\twocolumn \else  \fi
        \def\thefootnote{\arabic{footnote}}
        \setcounter{footnote}{0}}  
\definecolor{c1}{rgb}{1, 0, 0}
\definecolor{c2}{rgb}{0, 1, 0}
\definecolor{c3}{rgb}{0, 0, 1}
\definecolor{c4}{rgb}{1, 0, 1}
\definecolor{c5}{rgb}{0, 1, 1}

\catcode`@=12
\relax

\def\beq{\begin{equation}}
\def\eeq{\end{equation}}
\def\bea{\begin{eqnarray}}
\def\eea{\end{eqnarray}}
\def\EQ{\begin{equation}}
\def\EN{\end{equation}}

\relax
\numberbysection
\hybrid

\begin{document}
\begin{titlepage}
\begin{center}
{\large\bf Local logarithmic correlators  as limits of Coulomb gas integrals}\\[.3in] 

{\bf  Raoul \ Santachiara$^{1}$ and Jacopo Viti$^2$}\\
  $^1$ {\it LPTMS \&CNRS\footnote[4]{Unit\'e mixte de 
             recherche du CNRS UMR 8626}, 
             Universit\'e Paris-Sud\\
             B\^atiment 100\\
          F-91405 Orsay, France, \\
    e-mail: {\tt raoul.santachiara@lptms.u-psud.fr}. }\\
   $^2${\it LPT \'Ecole Normale Sup\'erieure \&CNRS\\
   24 Rue Lhomond, 75005 Paris, France\\
   email: \tt{jacopo.viti@lpt.ens.fr}} 
\end{center}
\centerline{(Dated: \today)}
\vskip .2in
\centerline{\bf ABSTRACT}
\begin{quotation}
We will describe how logarithmic singularities arise as limits of Coulomb Gas integrals.
Our approach will combine  analytic properties 
of the time-like Liouville structure constants,
together with the recursive formula of the Virasoro conformal blocks. Although the Coulomb Gas formalism forces a diagonal
coupling between the chiral and antichiral sectors of the Conformal Field Theory (CFT),  we present
new results for the multi-screening
integrals which  are potentially interesting for applications to critical statistical systems described
by Logarithmic CFTs.
In particular our findings extend and complement previous results, derived with Coulomb Gas methods, at $c=0$ and $c=-2$.

\vskip 0.5cm 
\noindent
\end{quotation}
\end{titlepage}
\section{Introduction}
The Conformal Field Theory (CFT) approach aims at constructing
two-dimensional quantum field theories whose correlation
functions satisfy the infinite set of
constraints imposed by conformal invariance \cite{DiFra,ItSaZu}.
A CFT is characterized by a set of states and operators which are 
determined by the
representations of the Virasoro algebra, or
of other chiral algebras if additional symmetries are present. A consistent CFT has to
satisfy general requirements of
a quantum field theory, in particular  
the conformal bootstrap equations expressing the associativity of
the operator algebra \cite{BPZ}.         

The CFT approach provides access to the study of critical phases of statistical systems
in two dimensions.
Certain properties of critical statistical models, such as the spin or the energy
correlation functions in the Ising and in the Potts model,
are given by a particular family of CFTs, the minimal models. These
are exactly solvable CFTs which are constructed
from a finite number of  degenerate Virasoro representations, the correlation functions of which satisfy
certain differential equations \cite{BPZ}.

The two-dimensional critical phases can also be characterized by
geometrical objects which take the form of
conformally invariant random fractals \cite{Durev}. Examples of conformally invariant random fractals
are the Potts spin domain walls or the
Fortuin and Kasteleyn \cite{FK1, FK2} clusters. The question is then if one can use CFT methods as a tool
to capture the geometric properties of critical phases, or more in general, to describe the
conformal random geometry.
In the last twenty years, there has been an intense effort in this direction, and a
series of remarkable results followed: the fractal dimension of many random paths as well as
different geometric exponents controlling, for instance, the reunion probability of an
ensemble of self-avoiding walks, have been obtained \cite{Durev}.
These results have been possible also thanks to the development of a palette of advanced techniques,
such as Coulomb Gas (CG) methods \cite{Ni} or Stochastic Loewner Evolutions\cite{BaBe}.
However, despite many successes, the CFT approach to the
study of conformally invariant fractals is in many respects, unsatisfactory.
On the one hand, many of the results found so far are
for quantities related to two-point correlation functions, while the fine structure of CFT
fully manifests itself only at the level of three- and four-points correlation functions.
The only exceptions are the observables that satisfy some known differential equation,
such as the probability measure associated to an SLE interface
\cite{BaBe}. On the other hand,
besides notable exceptions\cite{GaKa,GaKao,Ka,GaRuWo}, the only
models which are known to satisfy the conformal bootstrap
and which describe statistical models
are the minimal models. The minimal models are too simple to capture 
the geometrical properties of critical phases whose description requires instead   
representations of the Virasoro algebra with more complicated features than those of the minimal models.
In particular, differently from
the case of the minimal models, the theory may contain indecomposable representations with a non-diagonalizable dilatation
operator: this leads
to the appearance of logarithms in the correlation functions \cite{Gu}.

These last years have seen extensive investigations on logarithmic CFTs (LCFT)\cite{Ca_rev_log,CrRi,Gu_rev, GJRSV}.
A better comprehension of LCFT representation modules
and their fusion rules has been possible from the study of non semi-simple representations of the
Virasoro algebra \cite{FuHwSeTi,FeGaSeTio,FeGaSeTi,PeRa} and from lattice approaches
\cite{PeRaZu,ReSa,GaJaVaSa}. However,
the construction of a consistent CFT, obtained by combining the chiral and anti-chiral spaces of representation
with indecomposable Virasoro modules, is an hard problem and is much less under control.
Recent progress in this direction
come from the study of the $c=0$ bulk theories \cite{VaGaJaSa,Ri} and by the use of the category theory
\cite{FuScSt}.

Other insights into the CFTs describing conformal invariant fractals came from a completely new perspective. 
It was conjectured in \cite{DeVi} that the probability that a percolation cluster,
or more in general an FK cluster,
connects three given points is given at criticality by the structure constants
computed in \cite{Za,KoPe}, see eq. (\ref{Zam_structure}). This conjecture found a strong numerical
verification for the percolation clusters \cite{Ziff} and the FK clusters \cite{DPSV}.
In the following we will refer to these structure constants
as the time-like Liouville structure constants since they are obtained within the time-like Liouville
theory \cite{HaMaWi}, corresponding
at the classical level to the analytic continuation of the usual  
Liouville theory \cite{TeLiouville, Zam_lectures}.
Their relevance in the percolation properties of Potts models,
together with the fact that they represent the only consistent \cite{DPSV}
analytic continuation of the minimal model structure constants \cite{DoFa,DoFa2,DoFa3},
strongly suggests the time-like Liouville structure
constants  are the basic building blocks for constructing new non-minimal Virasoro CFTs, at least when additional
symmetries do not play any non-trivial role \cite{DPSV2}.

A natural question then is whether and how the logarithmic behavior of the correlation functions can be spotted directly from the time-like Liouville
structure constants. An analogous question has been answered  in \cite{Zamhem} for the Liouville (DOZZ) 
structure constants \cite{DoOt,ZaZa}. In this paper we consider this problem  analyzing the four-point correlation functions
which posses an integral representation, the so-called CG integral representations \cite{DoFa,DoFa2}.
The CG integrals provide correlation functions in which the chiral anti-chiral sectors of the CFT are glued symmetrically.
This is in general not the case in LCFTs. There are nevertheless notable examples in which logarithmic CG correlation functions and thus
diagonal LCFTs have been applied in the  computation of bulk critical geometrical properties \cite{Salog,HaYa,Ya,JDPSW, DeVi,DPSV}
and their validity checked numerically \cite{JDPSW, Ziff, DPSV}. Our analysis extends in particular these results
to the multi-screening case and can be potentially relevant for the study of critical bulk systems. 

This paper is organized as follows. In Section~2 we introduce the basic CFT
tools, needed for our analysis: the recursive formula for the conformal blocks  \cite{Zam_rec1, Zam_rec2} and the time-like
Liouville structure constants \cite{Za, KoPe}. In Section~3, we examine in details the one-screening CG
integral and show that the mechanism of cancellation of the leading singularity and the generation of logarithmic terms is related to a precise relation between the
residue of the conformal block and the structure constants, see eq. \eqref{simp_rel_log}. Such considerations are
generalized in Sec.~4 to the multi-screening case for irrational values of the central charge, see in particular eqs.
(\ref{canc_sing},~\ref{canc_sing0}), whereas in Sec.~5 the special cases $c=-2$ and $c=0$ are analyzed. After summarizing our findings
in the Conclusions in Sec.~6, two appendices complete the paper.

\section{Conformal blocks and structure constants in Virasoro CFT}
In the first part of this introductory section we will review  the definition of conformal block in CFT. In the second part we will discuss, within the
Coulomb gas framework, the structure constants for a Virasoro CFT.
\subsection{Algebric data: conformal blocks}
Conformal symmetry implies the existence of an holomorphic  $T(z)$ and an antiholomorphic $\bar{T}(\bar{z})$ stress-energy
tensor, whose coefficients $L_n$ and $\bar{L}_n$ in the Laurent expansion
\begin{equation}
\label{Laurent}
T(z)=\sum_{n=-\infty}^{\infty}\frac{L_n}{z^{n+2}},\quad \bar{T}(\bar{z})=\sum_{n=-\infty}^{\infty}\frac{\bar{L}_n}{\bar{z}^{n+2}}
\end{equation}
are the generators of the conformal algebra $\mathcal{A}$ acting on the field $\Phi(z,\bar{z})$. Coordinate transformations
which involve only analytic or antianalytic functions lead to independent
field variations and, as a consequence, $\mathcal{A}$ is the tensor product of the two commuting  Virasoro algebra of central
charge $c$ 
\begin{equation}
\left[L_n,L_m\right]=(n-m)L_{n+m}+\frac{c}{12} n(n^2-1)\delta_{n+m,0},~\left[\bar{L}_n,\bar{L}_m\right]=(n-m)\bar{L}_{n+m}+
\frac{c}{12} n(n^2-1)\delta_{n+m,0}.
\label{vir}
\end{equation}
The action on the Hilbert space of the primary field \cite{BPZ} $\Phi(z,\bar{z})$
of conformal dimension $\Delta+\bar{\Delta}$ is equivalent to that
of the tensor product $\phi_{\Delta}(z)\otimes\bar{\phi}_{\bar{\Delta}}(\bar{z})$, with $\phi_{\Delta}$
(resp. $\bar{\phi}_{\bar{\Delta}}$) primary fields  of conformal dimension $\Delta$ (resp. $\bar{\Delta}$).
In the following we will always assume $\Delta=\bar{\Delta}$, which in particular corresponds to the case of scalar operators. The 
Hilbert space of a CFT based on a semi-simple Virasoro representation is a direct sum of Verma modules $\mathcal{V}_{\Delta}\otimes\mathcal{\bar{V}}_{\bar{\Delta}}$ and
each of them, for example $\mathcal{V}_{\Delta}$, is built from the highest weight state
$|\Delta\rangle\equiv\phi_\Delta(0)|0\rangle$, upon applications of the
generators $L_{n}$ ($n<0$). A standard basis for the Verma module $\mathcal{V}_{\Delta}$ is the basis of the descendant states
$L_{-n_1}\dots L_{-n_k}|\Delta\rangle$, with $n_{i+1}\geq n_{i}>0$. A descendant at level $D$, with $D=\sum_{i=1}^k n_i$,  is indexed by  the partition ${N}=\{n_1,\dots,n_k\}$ of $D$ and indicated with the notation $|\{N\}\rangle$. An inner product can be defined in the Verma module $\mathcal{V}_{\Delta}$ through
\begin{equation}
 (L_n)^{\dagger}\equiv L_{-n},\quad \langle\Delta|\equiv\lim_{z\rightarrow\infty}z^{2\Delta}\langle 0|\phi_{\Delta}(z).
\end{equation}
and  CFT correlation functions can be regarded as bilinear forms in the Hilbert space. The
two-point function corresponds to the normalization choice of the inner product
\begin{equation}
\label{2point}
\langle\Delta|\Delta'\rangle=\delta_{\Delta,\Delta'}
\end{equation}
and the three-point function of scalar primary fields follows from the knowledge of the structure constants
\begin{equation}
\label{3point}
C(\Delta_1,\Delta_2,\Delta_3)=\lim_{z\rightarrow\infty}|z|^{4\Delta_3}\langle\Phi_{\Delta_3}(z,\bar{z})\Phi_{\Delta_2}(1)
\Phi_{\Delta_1}(0)\rangle. 
\end{equation}
Structure constants are not determined algebraically but are rather free parameters of the theory that must be fixed imposing
additional constraints. Once they are known all the correlation functions can be in principle computed,
starting  from the four-point function which will be at length discuss in this paper. Introducing the short-hand notation
$\{\Delta\}$ for the set of external dimensions and recalling \cite{BPZ} one has 
\begin{eqnarray}
\label{4point}
\lim_{z_4\rightarrow\infty}&& |z_4|^{4\Delta_4}\langle \Phi_{\Delta_4}(z_4,\bar{z_4})\Phi_{\Delta_3}(1)
\Phi_{\Delta_2}(x,\bar{x})\Phi_{\Delta_1}(0)\rangle=\nonumber \\
&& \sum_p |x|^{-2(\Delta_1-\Delta_2+\Delta_p)}
C(\Delta_1,\Delta_2,\Delta_p)C(\Delta_3,\Delta_4,\Delta_p)~|F(x|c,\Delta_p,\{\Delta\})|^2,
\end{eqnarray}
The summation in \eqref{4point} is over all the possible internal fields with conformal dimensions $\Delta_p$ and those can constitute a continuous set.  The CFT is well-defined when \eqref{4point} produces crossing symmetric correlation functions.
$F(x|c,\Delta_p,\{\Delta\})$ is the holomorphic conformal block,  usually represented by the diagram
\begin{equation}
\begin{tikzpicture}
\draw (-3,0.5) node[left]{$F(x|c,\Delta_p,\{\Delta\})=$}--(-3.0001,0.5);
 \draw  (-1,0.0) node[left]{$\Delta_1,0$}-- (0,0);
  \draw (0,1) node[above]{$\Delta_2,x$}--(0,0);
\draw (0,0)--(1,0) node[above]{${\small \Delta_p}$}--(2,0);
\draw  (2,0)-- (2,1) node[above]{$\Delta_3, 1$};
  \draw (2,0)--(3,0) node[right]{$\Delta_4, \infty$};
\end{tikzpicture}
\label{diag_cb}
\end{equation}
The holomorphic conformal block $F(x|c,\Delta_p,\{\Delta\})$ can be written in term of the following expansion
\begin{equation}
\label{conformal_block}
F(x|c,\Delta_p,\{\Delta\})=\sum_{D=0}^{\infty}x^D\sum_{\{N\},\{M\}}\Gamma_N H^{-1}_{NM} \tilde{\Gamma}_M, 
\end{equation}
 where 
 \begin{equation}
H_{NM}=\langle\{N\}|\{M\}\rangle, 
\end{equation}
is the Shapovalov matrix and the coefficients $\Gamma_N$ are given by $\Gamma_N=\prod_{j=1}^k(\Delta_p+\sum_{i=1}^{j-1} n_i+n_j\Delta_3-\Delta_4)$,
$\tilde{\Gamma}_M=\prod_{j=1}^l(\Delta_p+\sum_{i=1}^{j-1} m_i+m_j\Delta_2-\Delta_1)$.
Let us introduce the standard parameterizations for the central charge
\begin{equation}
\label{central}
 c=1-24\alpha_0^2,\quad 2\alpha_0=\beta-\beta^{-1}
\end{equation}
 and for  the conformal dimensions
\begin{equation}
\label{dimension}
\Delta(\alpha)=\alpha(\alpha-2\alpha_0),
\end{equation}
where $\beta$ and $\alpha$ (the charge) are reals. The equation \eqref{dimension} shows that the charges $\alpha$ and
$2\alpha_0-\alpha$ lead to the same conformal dimensions; since we are assuming a non-degenerate spectrum, through all the paper the primary fields with dimension
$\Delta(\alpha)$ and $\Delta(2\alpha_0-\alpha)$ will be identified.  
From the study of the representation theory of the Virasoro
algebra it follows that when the charges $\alpha$ belong
to the lattice\footnote{Notice that the symmetry $\alpha\rightarrow 2\alpha_0-\alpha$ of the
conformal dimensions is equivalent to identify field with charges $\alpha_{r,s}$ and $\alpha_{-r,-s}$.}
\begin{equation}
\label{kac_lattice}
\alpha_{r,s}=\frac{1-r}{2}\beta+\frac{s-1}{2\beta},\quad r,s\in\mathbb N
\end{equation}
the Verma module $\mathcal{V}_{rs}$ of the field $\phi_{r,s}$ with conformal dimension
$\Delta_{r,s}\equiv\Delta(\alpha_{r,s})$, contains a vector $|\chi_{r,s}\rangle$ at level $rs$
with vanishing norm. In general, the vector $|\chi_{r,s}\rangle$ has the form
\begin{equation}
|\chi_{r,s}\rangle=\bigl[L_{-1}^{rs}+d_1L_{-2}L_{-1}^{rs-2}+\dots]|\phi_{r,s}\rangle, 
\end{equation}
where a particular normalization has been chosen.
The null vector $|\chi_{r,s}\rangle$ renders singular the Shapovalov matrix $H$ and ill-defined the expansion
\eqref{conformal_block}. The fields  $\phi_{r,s}$ with $r,s>0$, are termed degenerate. The existence of null
vectors in Verma modules led  \cite{Zam_rec1, Zam_rec2} to the recursive formula for the conformal block
\begin{equation}
\label{rec_conf}
 F(x|c,\Delta_p,\{\Delta\})=g(x|c,\Delta_p,\{\Delta\})+
 \sum_{r,s > 0}\frac{S_{rs}x^{rs}}{\Delta_p-\Delta_{r,s}}F(x|c,\Delta_p+rs,\{\Delta\}).
\end{equation} 
The residue $S_{rs}$ is a  polynomial of degree $rs$ in each external charge that
vanishes when the fusion rules \cite{BPZ} 
$\Phi_{\Delta_1}\cdot\Phi_{\Delta_p}$  (resp. $\Phi_{\Delta_3}\cdot\Phi_{\Delta_p}$) allow the dimension $\Delta_2$
(resp. $\Delta_4$). Namely
\begin{equation}
\label{residue}
 S_{rs}=\frac{R_{rs}(\alpha_1,\alpha_2)R_{rs}(\alpha_4,\alpha_3)}{B_{rs}},
 \end{equation}
 where one has\cite{Zam_rec1, Zam_rec2} 
 \begin{equation}
 R_{rs}(\alpha_1,\alpha_2)=\prod_{\substack{p=-r+1,-r+3,\dots,r-1\\ q=-s+1,-s+3,\dots,s-1}}
 \left(-\alpha_1-\alpha_2+\alpha_{p-1,q-1}\right)  \left(-\alpha_1+\alpha_2+\alpha_{p+1,q+1}\right)
\label{deg_con}
\end{equation}
and, after defining $\lambda_{rs}=-r\beta+s\beta^{-1}$, \cite{Zam_rec2}
\begin{equation}
B_{rs}=-2\prod_{k=1-r}^{r}\prod_{l=1-s}^s \lambda_{kl},\quad\{k,l\}\not=\{0,0\},\{r,s\}. 
\label{deg_norm}
\end{equation}
The  $B_{rs}$  \cite{Zamhem, Yan} are related to the vanishing  norm of the null vector $|\chi_{rs}\rangle$
in the following way
\begin{equation}
\lim_{\Delta\rightarrow\Delta_{r,s}}\langle\chi_{rs}|\chi_{rs}\rangle=B_{rs}(\Delta-\Delta_{r,s})+O\bigl((\Delta-\Delta_{r,s})^2\bigr).
\end{equation} 
Finally, the  null  field $\chi_{r,s}(x)$ associated to the null vector $|\chi_{r,s}\rangle$
is said to decouple from the theory when all of its correlation functions vanish. Such condition
requires in particular
\begin{equation}
\label{null_vector_cond} 
\lim_{x\rightarrow\infty}x^{2rs}\frac{\langle\chi_{r,s}(x)\phi_{\alpha_1}(1)\phi_{\alpha_2}(0)\rangle}{\langle\phi_{r,s}(x)\phi_{\alpha_1}(1)\phi_{\alpha_2}(0)\rangle}=R_{rs}(\alpha_1,\alpha_2)=0,
\end{equation}
i.e. $R_{r,s}(\alpha_1,\alpha_2)=0$.

 \subsection{Structure constants and Coulomb gas representation}
As we discussed in the previous section, the structure constants \eqref{3point} are not fixed by conformal invariance
and additional dynamical constraints are needed. The request of associativity of the operator product expansion (OPE)
was shown in \cite{BPZ} to produce a set of functional equations for the structure constants
$C(\Delta_1,\Delta_2,\Delta_3)$, the so-called conformal bootstrap, whose solution would have completely solved the field theory. 
Under the assumptions of a non-degenerate spectrum and of the normalization choice \eqref{2point}, \cite{Za} solved the conformal
bootstrap, proving the existence of a unique solution for  $\beta^2\not \in\mathbb Q$. Such solution has the following form
\begin{equation}
\label{Zam_structure}
C(\alpha_1,\alpha_2,\alpha_3)=A_{\beta}\frac{\Upsilon_{\beta}(\beta-\alpha_{13}^2)\Upsilon_{\beta}(\beta-\alpha_{23}^1)\Upsilon_{\beta}(\beta-\alpha_{12}^3)\Upsilon_{\beta}(2\beta-\beta^{-1}-\alpha_{123})}
{\Bigl[\prod_{i=1}^3\Upsilon_{\beta}(\beta-2\alpha_i)\Upsilon_{\beta}(2\beta-\beta^{-1}-2\alpha_i)\Bigr]^{1/2}}, 
\end{equation}
where we chose to write the structure constants as a function of the charges $\alpha_i$ ($i=1,2,3$) related to the conformal dimensions by
\eqref{dimension} for the value of the central charge \eqref{central} and $\alpha_{ijk}=\alpha_i+\alpha_j+\alpha_k$,
$\alpha_{ij}^k=\alpha_i+\alpha_j-\alpha_k$. The special function $\Upsilon_{\beta}(x)$, first introduce
in \cite{ZaZa}, is briefly considered in the Appendix B; here we remind that it has integral representation
\begin{equation}
\label{Zam_int}
\log\Upsilon_{\beta}(x)=\int_{0}^{\infty}\frac{dt}{t}\left[\frac{(Q/2-x)^2}{e^t}-\frac{\sinh^2\frac{t}{2}(Q/2-x)}{\sinh\frac{\beta t}{2}
\sinh\frac{t}{2\beta}}\right],
\end{equation}
convergent for $0<x<Q$ ($Q=\beta+\beta^{-1}$) and that satisfies the shift relations
\begin{equation}
\label{rec_Zam}
\frac{\Upsilon_{\beta}(x+\beta)}{\Upsilon_{\beta}(x)}=\gamma(\beta x)\beta^{1-2\beta x} \quad 
\frac{\Upsilon_{\beta}(x+\beta^{-1})}{\Upsilon_{\beta}(x)}=\gamma(\beta^{-1} x)\beta^{-1+2\beta^{-1} x},  
\end{equation}
with $\gamma(x)=\Gamma(x)/\Gamma(1-x)$; notice $\gamma(x)\gamma(-x)=-x^{-2}$ and $\gamma(x+1)=-x^2\gamma(x)$. The constant $A_{\beta}$ is given by
\begin{equation}
 \label{A_const}
 A_{\beta}=\frac{\beta^{\beta^{-2}-\beta^2-1}\gamma(\beta^2)\gamma(\beta^{-2}-1)}{\Upsilon_{\beta}(\beta)}.
\end{equation}
In \cite{DPSV}, it has then been shown that the solution \eqref{Zam_structure} can actually be
recovered from analytic continuation of the three-point correlation functions of scalar vertex operator
in the CG formalism \cite{DoFa}. Also for $\beta^2$ rational, \eqref{Zam_structure} has intriguing physical
applications in statistical mechanics \cite{DPSV} which are well beyond the predictions of the minimal conformal models.

Let us briefly outline the CG approach to CFT \cite{DoFa}. Coulomb gas is a free boson theory 
with an additional background charge $2\alpha_0$
at infinity which produces the total central charge \eqref{central}. Primary fields are vertex operator
$V_{\alpha}=e^{i\alpha\varphi(z,\bar z)}$ with conformal dimension \eqref{dimension}, $\varphi(z,\bar z)$ is the free bosonic field.
The  double degeneracy of the scaling dimensions \eqref{dimension} is solved \cite{Doco} by assuming that the vertex operators $V_{\alpha}$
and $V_{2\alpha_0-\alpha}$ represent
the same primary field $\Phi_{\Delta(\alpha)}\equiv \Phi_{\alpha}=\Phi_{2\alpha_0-\alpha}$ but acquire non trivial
normalizations. Correlation functions in the 
CG approach are non-zero if the charges of the vertex operators $\alpha_i$ sum to the background charge
$2\alpha_0$. If such condition cannot be satisfied one is free to add vertex operators in the correlation function
of charges $\beta$ and $-\frac{1}{\beta}$
and integrate over them. This peculiar vertex operators (screening operators) do not spoil conformal invariance and
do not modify the value of the central charge. The
correlation function of four scalar primary fields $\Phi_{\alpha_i}$, $i=1,\dots,4$:
\begin{equation}
\langle \Phi_{\alpha_4}(\infty)\Phi_{\alpha_3}(1)
\Phi_{\alpha_2}(x,\bar{x})\Phi_{\alpha_1}(0)\rangle,
\end{equation}
 which  satisfy the neutrality condition
\begin{equation}
\sum_{i=1}^4 \alpha_i
+n \beta-m \beta^{-1}= 2\alpha_0 \quad n,m=0,1,2\dots
\label{ch_neu}
\end{equation} 
coincides, up to a normalization constant, to the following integral
\begin{align}
\label{4vertex}
I_{nm}=|1-x|^{4\alpha_2\alpha_3}|x|^{4\alpha_1\alpha_2}&\int\prod_{i=1}^nd\vec{w}_i~\prod_{i=1}^n|w_i-1|^{4\beta\alpha_3}|w_i-x|^{4\beta\alpha_2}|w_i|^{4\beta\alpha_1}
\prod_{l<k}|w_l-w_k|^{4\beta^2}\nonumber\\
&\hspace*{-3.5cm}\times\int\prod_{j=1}^m d\vec{u}_j~\prod_{j=1}^m|u_j-1|^{-4\alpha_3/\beta}|u_j-x|^{-4\alpha_2/\beta}|u_j|^{-4\alpha_1/\beta}
\prod_{l<k}|u_l-u_k|^{4/\beta^2}
\prod_{i,j}|w_i-u_j|^{-4}.
\end{align}

A very remarkable property of the integral \eqref{4vertex} is that it can be decomposed \cite{DoFa,DoFa2,Doco}
into the sum of $(n+1)(m+1)$ terms
\begin{equation}
\label{exp_int}
I_{nm}= \sum_{i=0}^{n}\sum_{j=0}^{m}
|x|^{-2(\Delta_1+\Delta_2-\hat{\Delta}_{i,j})} X_{i,j} |F(x|c(\beta),\hat{\Delta}_{i,j},\{\Delta\})|^2, 
\end{equation}
where the constants
$X_{i,j}$ are proportional to the structure constants
\begin{equation}
\label{struc_X}
\frac{X_{i,j}}{X_{i',j'}}=\frac{C(\alpha_1,\alpha_2,\hat{\alpha}_{i,j})C(\alpha_4,\alpha_3,\hat{\alpha}_{i,j})}
{C(\alpha_1,\alpha_2,\hat{\alpha}_{i',j'})C(\alpha_4,\alpha_3,\hat{\alpha}_{i',j'})} 
\end{equation}
and the function $F(x|c(\beta),\hat{\Delta}_{i,j},\{\Delta\})$ is the conformal block defined in \eqref{conformal_block}. 
For a correlation function that is represented
by the CG integral (\ref{exp_int}) the summation (\ref{4point}) is restricted to the discrete set of internal channels with the following dimensions and charges
\begin{equation}
 \hat{\Delta}_{i,j}=\hat{\alpha}_{i,j}(\hat{\alpha}_{i,j}-2\alpha_0), \quad \hat{\alpha}_{i,j}=\alpha_1+\alpha_2+i\beta-j\beta^{-1}.
 \end{equation}
 
  We observe that the radius of
converge of the series expansions (\ref{conformal_block}) is one. Conformal blocks can
be analytically continued in the region $|x|>1$ implementing the condition of associativity of the OPE \cite{BPZ}.
In our study of logarithmic singularities of the integrals \eqref{4vertex} we will however restrict to the
domain  $|x|<1$, even if  a characterization of such singularities in the limit $x\rightarrow 1$, could be
carried out analogously exchanging the role of $\alpha_1$ and $\alpha_3$.

\section{Logarithmic singularities in one-screening Coulomb gas integrals}
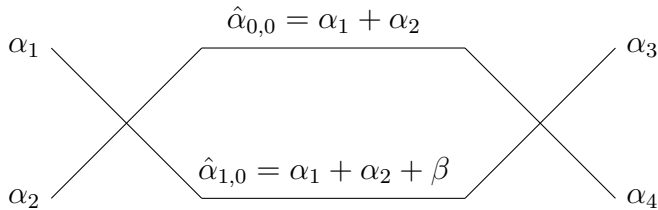
\begin{figure}[t]
\begin{center}
\begin{tikzpicture}
 \draw  (-1,1)node[left]{$\alpha_1$}-- (1,-1);
 \draw  (-1,-1)node[left]{$\alpha_2$}-- (1,1);
 \draw (1,1)node[above]{$\hspace*{3.25cm}\hat{\alpha}_{0,0}=\alpha_1+\alpha_2$}--(4.5,1);
 \draw (1,-1)node[above]{$\hspace*{3.25cm}\hat{\alpha}_{1,0}=\alpha_1+\alpha_2+\beta$}--(4.5,-1);
 \draw(4.5,-1)--(6.5,1)node[right]{$\alpha_3$};
 \draw(4.5,1)--(6.5,-1)node[right]{$\alpha_4$};
 \end{tikzpicture}
 \end{center}
\caption{Diagram representing the integral $I_{10}$ in \eqref{onescreanig}. The internal charges are
$\hat{\alpha}_{0,0}=\alpha_{r,s}$ and $\hat{\alpha}_{1,0}=\alpha_{r-2,s}$. For $r=1$, the corresponding conformal dimensions
\eqref{dimension} differ by a positive integer $s$ and the conformal block $F(x|c,\hat{\Delta}_{0,0},\{\Delta\})$ is singular.
The singularity is due to the null vector at level $s$ which does not decouple and is cured by
the contribution of the conformal block $F(x|c,\hat{\Delta}_{1,0},\{\Delta\})$. The integral $I_{1,0}$ has in this case a
logarithmic singularity as $x\rightarrow 0$.}
\label{log_pic}
\end{figure}

In this section we will discuss the appearance of logarithmic singularities in the integral \eqref{4vertex}, focusing
on the simplest example: the case where the neutrality condition \eqref{ch_neu} is satisfied with $n=1$ and $m=0$; the
case $n=0$ and $m=1$ follows through the replacement $\beta\rightarrow-1/\beta$.
For simplicity we will take $\beta^2\not\in\mathbb Q$ and $\alpha_4$  fixed
by \eqref{ch_neu} as $\alpha_4=-(\beta+\alpha_1+\alpha_2+\alpha_3)+2\alpha_0$. We will also assume 
$\alpha_1+\alpha_2=\alpha_{r,s}$, with $r,s>0$ ensuring that the internal channel with charge
$\hat{\alpha}_{00}$ is a degenerate field. For
$n=1$ and $m=0$, the integral \eqref{4vertex} has the following form \cite{Doco}
\begin{equation}
 \label{onescreanig}
 I_{10}=|x|^{-2(\Delta_1+\Delta_2-\hat{\Delta}_{0,0})}
 \Bigl[X_{0,0}~|F(x|c,\hat{\Delta}_{0,0},\{\Delta\})|^2+X_{1,0}~|x|^{2(\hat{\Delta}_{1,0}-\hat{\Delta}_{0,0})}
 |F(x|c,\hat{\Delta}_{1,0},\{\Delta\})|^2\Bigr],
\end{equation}
where $\hat{\alpha}_{0,0}=\alpha_{r,s}$, $\hat{\alpha}_{1,0}=\alpha_{r,s}+\beta$. The  corresponding
conformal blocks are expressed through hypergeometric functions as
\begin{align}
\label{F00}
&F(x|c,\hat{\Delta}_{0,0},\{\Delta\})=(1-x)^{-2\alpha_2\alpha_3}{}_2F_1\bigl(-2\beta\alpha_2, -1-2\beta\alpha_{123},
-2\beta(\alpha_1+\alpha_2),x\bigr),\\
\label{F10}
&F(x|c,\hat{\Delta}_{1,0},\{\Delta\})=(1-x)^{-2\alpha_2\alpha_3}{}_2F_1\bigl(-2\beta\alpha_3, 1+2\beta\alpha_1,
2+2\beta(\alpha_1+\alpha_2),x\bigr);
\end{align}
after defining $s(x)=\sin(\pi x)$, the coefficients $X$ read
\begin{align}
\label{X00}
 &X_{0,0}=\frac{s(2\beta\alpha_3)s(2\beta\alpha_{123})}{s(2\beta(\alpha_1+\alpha_2))}\left[\frac{\Gamma(1+2\beta\alpha_3)
 \Gamma(-1-2\beta\alpha_{123})}{\Gamma(-2\beta(\alpha_1+\alpha_2))}\right]^2,\\
 \label{X10}
 &X_{1,0}=\frac{s(2\beta\alpha_1)s(2\beta\alpha_{2})}{s(2\beta(\alpha_1+\alpha_2))}\left[\frac{\Gamma(1+2\beta\alpha_1)
 \Gamma(1+2\beta\alpha_{2})}{\Gamma(2+2\beta(\alpha_1+\alpha_2))}\right]^2.
 \end{align}
Inside the circle $|x|<1$, the hypergeometric function ${}_2F_1(\alpha,\beta,\gamma,x)$ is represented by the power series
\begin{equation}
\label{power_series}
{}_2F_1(\alpha,\beta,\gamma,x)=\sum_{k=0}^{\infty}\frac{[\alpha]_k[\beta]_k}{[\gamma]_k}\frac{x^k}{k!},
\end{equation}
where $[q]_k=\Gamma(q+k)/\Gamma(q)$ is the Pochhammer symbol, see for example \cite{Lebedev}. As a function of the variable
$\gamma$, \eqref{power_series}  has simple poles for $\gamma=(1-s)$ with $s$ a positive integer. The residue is
proportional to the hypergeometric function ${}_2F_1(\alpha+s,\beta+s,s+1,x)$ 
\begin{equation}
\label{hyper_exp}
\text{Res}_{\gamma=1-s}~{}_2F_1(\alpha,\beta,\gamma,x)=
\frac{(-1)^{s-1}[\alpha]_s[\beta]_s}{\Gamma(s)\Gamma(1+s)}x^{s}~{}_2F_1(\alpha+s,\beta+s,s+1,x).
\end{equation}
If the sum $\alpha_1+\alpha_2=\alpha_{1,s}=\frac{s-1}{2\beta}$  the conformal
block $F(x|c,\hat{\Delta}_{00},\{\Delta\})$ in \eqref{F00} is then  singular\footnote{If 
$\alpha_1+\alpha_2+\beta=\alpha_{-1,-s}$, the conformal block $F(x|c,\hat{\Delta}_{1,0},\{\Delta\})$ is singular and
$\hat{\Delta}_{1,0}=\Delta_{1,s}$ now.} with residue on the pole
at $\hat\Delta_{0,0}=\Delta_{1,s}$ proportional to the conformal block $F(x|c,\hat{\Delta}_{10},\{\Delta\})$, see
eq. \eqref{F10}. The singularity of the conformal block $F(x|\hat{\Delta}_{00},\{\Delta\})$ and its residue
could have been computed directly form the recursive formula \eqref{rec_conf}. Indeed when $\alpha_1+\alpha_2=\alpha_{1,s}$
and, by the neutrality condition, $\alpha_3+\alpha_4=\alpha_{1,-s}$, eq. \eqref{residue} predicts
\begin{equation}
\label{residue_one_s}
S_{1s}=\frac{(-1)^{s-1}\lambda_{1,s}[-2\beta\alpha_2]_s[-s-2\beta\alpha_3]_s}{2\beta~\Gamma(s)\Gamma(1+s)},
\end{equation} 
which is non-vanishing\footnote{The eq. \eqref{residue_one_s} shows that the analysis we carried out in this section extends to
the case $\alpha_2\not=\alpha_{1,j}$ and $\alpha_3\not=\alpha_{1,j-s}$, $j=1,\dots,s$.}.
Expanding the sum of the charges $\alpha_1$ and $\alpha_2$ near the singular value $\alpha_{1,s}$ as
\begin{equation}
\label{reg}
\alpha_1+\alpha_2=\alpha_{1,s}+\delta,
\end{equation}
and observing that $\Delta-\Delta_{r,s}=\lambda_{r,s}(\alpha-\alpha_{r,s})+O\bigl((\alpha-\alpha_{r,s}\bigr)^2)$, one concludes that
\eqref{hyper_exp} is consistent with \eqref{rec_conf}. This singularity is associated
to the presence of a null vector at level $s$  in the Verma
module of the degenerate primary $\phi_{\hat{\Delta}_{0,0}}=\phi_{1,s}$ which cannot be decoupled since $S_{1,s}\not=0$.
Following a general scheme in LCFT, the divergence is cured by the mixing of the null-vector in the Verma
module $\mathcal{V}_{\hat\Delta_{00}}$  with the highest weight of the Verma module
$\mathcal{V}_{\hat{\Delta}_{1,0}}$ corresponding to the other internal channel in \eqref{onescreanig}. When
$\alpha_{1}+\alpha_2=\alpha_{1,s}$, the two states have indeed the same conformal dimension $\Delta_{1,s}+s=\Delta_{-1,s}$.
A schematic picture is presented in Fig. \ref{log_pic}.\\
In the remaining part of the section, we  point out  that the existence of logarithmic singularities in the four-point
function requires moreover a precise relation between the values of the
structure constants \eqref{Zam_structure} and the residue \eqref{residue}. This mechanism has been outlined for some special cases, see for instance \cite{HaYa,Ya,CrRi}.
In order to show this we regularize the CG integral by introducing a parameter $\delta$ as in \eqref{reg} and to
ensure that the neutrality condition \eqref{ch_neu} will be always satisfied we also take
\begin{equation}
\alpha_3+\alpha_4=\alpha_{1,-s}-\delta.
\label{reg2}
\end{equation}
In the limiting procedure the charges $\alpha_1$ and $\alpha_4$ will depend on the $\delta$, through the
(\ref{reg}, \ref{reg2}) whereas the charges $\alpha_2$ and $\alpha_3$ are fixed; we will  take $\delta\to 0$ at the end.  Note that the regularization prescriptions (\ref{reg}, \ref{reg2}) do not
modify the central charge, i.e.  $\beta$ is independent from $\delta$. 
The conformal blocks (\ref{F00}, \ref{F10})  and the coefficients (\ref{X00}, \ref{X10}) have the series expansion
in $\delta$ 
\begin{eqnarray}
F(x|c,\hat{\Delta}_{0,0},\{\Delta\})&=& \frac{1}{\delta} G^{(-1)}_{00}(x)+ G^{(0)}_{00}(x) +O(\delta),
\\
F(x|c,\hat{\Delta}_{0,0},\{\Delta\})&=& G^{(0)}_{10}(x)+\delta\; G^{(1)}_{10}(x)+O(\delta^2),
\label{exp_cb}\\
X_{0,0}=X^{(1)}_{00} \delta + X_{0,0}^{(2)} \delta^2+O(\delta^3),& & X_{1,0}=\frac{1}{\delta}X^{(-1)}_{1,0}+
X_{1,0}^{(0)} +O(\delta),
\end{eqnarray}
explicitly computed in Appendix A. The leading singularity of order $\delta^{-1}$ in the correlation
functions $I_{10}$ disappears due to the identity $X^{(1)}_{0,0}|G^{(-1)}_{0,0}(x)|^2+X_{1,0}^{(-1)}|x|^{2s}|G^{0}_{(1,0)}(x)|^2=0$,
which can be recast in the more expressive form
\begin{equation}
\label{rel_log}
\frac{C(\alpha_1,\alpha_2,\alpha_{1,s}+\delta)C(\alpha_3,\alpha_4,\alpha_{1,s}+\delta)}
{C(\alpha_1,\alpha_2,\alpha_{-1,s}+\delta)C(\alpha_3,\alpha_4,\alpha_{-1,s}+\delta)}=
-\delta^2\left[\frac{\lambda_{1,s}}{S_{1,s}}\right]^2+O(\delta^3).
\end{equation}
Notice that due to the form of the structure constants \eqref{Zam_structure}, eq. (\ref{rel_log}) can be simplified into
\begin{equation}
\label{simp_rel_log}
\frac{C(\alpha_1,\alpha_2,\alpha_{1,s}+\delta)}{C(\alpha_1,\alpha_2,\alpha_{-1,s}+\delta)}=i\delta\frac{\lambda_{1,s}B_{1,s}}{
R^2_{1,s}(\alpha_1,\alpha_2)}+O(\delta^2),
\end{equation}
valid for any value of the $\alpha_i$; the formula \eqref{simp_rel_log}  can be also verified by using the shift relation
\eqref{rec_Zam} and its generalization in the multi-screening case will be the object of the next section. It is interesting
to observe that the vanishing in the $\delta\rightarrow 0$ limit of the ratio of the structure constants in
\eqref{simp_rel_log} is due, see Appendix B, to zeros 	in the function $\Upsilon_{\beta}(x)$ at $x=\beta-2\alpha_{-1,s}$
and $x=2\beta-\beta^{-1}-2\alpha_{-1,s}$. Such zeros are responsible  \cite{DPSV} for the singular behavior of the normalization
factors of the field $\phi_{-1,s}$ in the CG representation. As it is well known, the cancellation of the leading $\delta^{-1}$ singularity in the integral
\eqref{onescreanig} leads to a logarithmic divergence and one has
\begin{align}
I_{1,0}=|x|^{-2(\Delta_1+\Delta_2-\hat{\Delta}_{00})}&\left[a\log|x|^2|x|^{2s} X_{1,0}^{-1}|G_{10}^{(0)}(x)|^2+
X_{0,0}^{(1)}|G_{00}^{(-1)}(x)|^2+X_{1,0}^{(0)}|x|^{2s}|G_{10}^{(0)}(x)|^2\right.\nonumber\\
&\hspace*{-1cm}\Bigl.+X_{0,0}^{(1)}\bigr(G_{00}^{(-1)}(x)G_{00}^{(0)}(\bar{x})+c.c.\bigl)+
X_{1,0}^{(-1)}|x|^{2s}\bigl(G_{10}^{(0)}(x)G_{10}^{(1)}(\bar{x})+c.c.\bigr)\Bigr],
\label{cg_1s}
\end{align}
where the logarithmic prefactor is
\begin{equation}
a=\lim_{\delta\to 0}\left(s-\hat{\Delta}_{1,0}(\delta)+\hat{\Delta}_{0,0}(\delta)\right)=-2\beta.
\end{equation}
Let us mention that some examples of bulk correlation functions of the kind (\ref{cg_1s})  appeared
in the study of classical statistical systems at criticality. For instance, the function (\ref{cg_1s}) with
$\alpha_{1}=2\alpha_{0}-\alpha_4=\alpha_{1,0} \quad \alpha_2=\alpha_{3}=\alpha_{1,2}$ and $1/2\leq \beta^2 \leq 1$, was argued to be relevant in the study of  interfaces in  the random bond Potts model
\cite{JDPSW}. 
It determines, in particular, the effects of quenched bond disorder on the fractal dimension of FK clusters boundaries.
These functions are, at our knowledge, the only bulk logarithmic correlation functions to have found a numerical
test\cite{JDPSW}. Similarly, the bulk correlation function  (\ref{cg_1s})
with $\alpha_{1}=2\alpha_0-\alpha_4=\alpha_{2,0}$, determines the disorder effects on 
the fractal dimensions of the pivotal bonds and  has found numerical verification \cite{JDPSW_un}.
Finally, the correlation function (\ref{cg_1s}) with 
$\alpha_{1}=2\alpha_{0}-\alpha_4=\alpha_{0,1} \quad \alpha_2=\alpha_{3}=\alpha_{2,1}$
has been argued in  \cite{CaSi} to  
describe the propagation of a loop in the dilute phase of the $O(n)$ model in the presence of two twist operators.

\section{Logarithmic singularities in multi-screening Coulomb gas integrals}
We now extend the results obtained in previous section to the case where four-point correlation functions of vertex operators
are computed from \eqref{4vertex} with an arbitrary number $n$ and $m$ of screening operators. We assume
$\beta^2\not\in \mathbb Q$ and defer the discussion of the peculiarity of $\beta^2$ rational to the next section.  
Analogously to the one-screening case we consider 
\begin{equation}
\alpha_1+\alpha_2=\alpha_{r,s} \quad  \alpha_3+\alpha_4=\alpha_{2n-r,2m-s} \quad r,s\in\mathbb N.
\label{sum_12}
\end{equation}
Under this assumption the internal charges take the form
$\hat{\alpha}_{ij}=\alpha_{r-2i,s-2j}$. 
The condition (\ref{sum_12}) satisfies the neutrality condition (\ref{ch_neu}) and 
assures us that the integral (\ref{exp_int}) 
contains conformal block of degenerate fields. 
Indeed,among the $(n+1)(m+1)$ internal charges $\hat{\alpha}_{ij}$, $i=0,..,n$ and $j=0,..,m$, the ones 
with $(r-2i\geq 1 \wedge s-2j \geq 1)$ or $(r-2i\leq -1 \wedge s-2j \leq -1)$ belongs to the lattice \eqref{kac_lattice}. 
It is also useful to observe that the channels  $\hat{\alpha}_{ij}$ and $\hat{\alpha}_{r-i,s-j}$
correspond to the same conformal block and to the same structure constants
\begin{equation}
F(x|c,\hat{\Delta}_{ij},\{\Delta\})=F(x|c,\hat{\Delta}_{r-i,s-j},\{\Delta\});\quad X_{i,j}=X_{r-i,s-j}, 
\label{tr_sym}
\end{equation}
due to the identity $\hat{\Delta}_{i,j}=\hat{\Delta}_{r-i,s-j}$. 

\subsection{Singularities of the internal channels}
Logarithmic terms in the four-point correlation function may arise when some of the 
conformal blocks are singular, i.e. when the dimension $\hat{\Delta}_{i,j}$ in \eqref{exp_int}
corresponds to that of a degenerate field  and  the condition
 \begin{equation}
R_{r-2i,s-2j}(\alpha_{r,s}-\alpha_2,\alpha_2)R_{r-2i,s-2j}(\alpha_{2n-r,2m-s}-\alpha_3,\alpha_3) \neq 0,
\label{cond_sing2}
 \end{equation}
is satisfied. From (\ref{rec_conf}) it follows that the conformal block  has in this case a
singularity at the level $(r-2i)(s-2j)$ and, as we have previously noticed, the condition (\ref{cond_sing2}) is equivalent
to consider a CFT where the null-vector $|\chi_{r-2i,s-2j}\rangle$ cannot be set to zero. Using (\ref{deg_con}), one can check that the requirement of
the degeneracy of the internal channel and  the condition (\ref{cond_sing2}) are met by the charges
$\hat{\alpha}_{i,j}$ such that
\begin{align}
&(r-2i\geq 1) \;\wedge\; (s-2j\geq 1)\;\wedge \;(i\geq r-n \vee j\geq s-m),\\
&\hspace*{5cm}\text{or}\nonumber\\
&(r-2i\leq -1) \;\wedge\; (s-2j\leq -1)\;\wedge \;(i<n+1 \vee j< m+1\;).
\label{ij_deg_sing}
\end{align}
Note that we are always assuming $\alpha_2$ and $\alpha_3$ to take general real values. In particular, 
the condition (\ref{cond_sing2}) also requires
 \begin{eqnarray}
  \alpha_{2}&\neq& \alpha_{i+k,j+l}\quad \mbox{and}     \nonumber \\
\alpha_3&\neq& \alpha_{n-r+i+k,m-s+j+l}, \quad \mbox{for}  \;  k=1,\dots,r-2i,\;l=1,\dots,s-2j.
\end{eqnarray} 
Now that we have identified the possible sources of singularities of the conformal blocks, we pass to analyze those of
the coefficients  $X_{i,j}$ defined in (\ref{struc_X}).  Taking into account (\ref{sum_12}) one has
\begin{equation}
\label{X_ij_dege}
X_{i,j}=
\mathcal{N}(c,\{\alpha\})~C(\alpha_{r,s}-\alpha_2,\alpha_2,\hat{\alpha}_{i,j})C(\alpha_{2n-r,2m-s}-\alpha_3,\alpha_2,\hat{\alpha}_{i,j}).
\end{equation}
where $\mathcal{N}$ is a normalization constant which depends only on the external charges and the central charge.
For arbitrary real values of $\alpha_2$ and $\alpha_3$,
we can single out the factors  $X_{i,j}|_{\text{sing}}$ present in $X_{i,j}$ which are zero or diverge
\begin{align}
 X_{i,j}|_{\text{sing}}&= \frac{1}
{\underbrace{\Upsilon_{\beta}(\beta-2\alpha_{r-2 i,s-2 j})}_{\equiv A}\underbrace{\Upsilon_{\beta}(2\beta-\beta^{-1}-2\alpha_{r-2 i,s-2 j})}_{\equiv B}}\times \nonumber \\
&\times\underbrace{\Upsilon_{\beta}(\beta-\alpha_{r,s}+\alpha_{r-2i,s-2j})\Upsilon_{\beta}(2\beta-\beta^{-1}-\alpha_{r,s}-
\alpha_{r-2i,s-2j})}_{\equiv C}\times \nonumber \\
&\times \underbrace{\Upsilon_{\beta}(\beta-\alpha_{2n-r,2m-s}+\alpha_{r-2i,s-2j})\Upsilon_{\beta}(2\beta-\beta^{-1}-\alpha_{2n-r,2m-s}-
\alpha_{r-2i,s-2j})}_{\equiv D}.
\label{exp_sing}
\end{align}
From the location of the zeros of the function $\Upsilon_{\beta}(x)$ defined in \eqref{Zam_int}, see also Appendix B, 
one finds
\begin{align}
&A=0:~ (r-2 i\geq 0 \;\wedge\; s-2j\leq -1) \quad \text{or} \quad (r-2 i\leq -1\; \wedge\; s-2j\geq 0), 
\label{infies1}
\\
&B=0:~  (r-2 i\geq 1 \;\wedge\; s-2j\leq 0) \quad \text{or}\quad  (r-2 i\leq 0 \;\wedge\;s-2j\geq 1),
\label{infies2}
 \\
&C=0:~
 (i\leq r \;\wedge\; j> s)\quad \text{or}\quad ( i> r\; \wedge\; j\leq s),
 \label{zeros1} \\
&D=0:~ (i\geq r-n \;\wedge\; j< s-m)\quad \text{or}\quad (i< r-n \;\wedge\; j\geq s-m).
\label{zeros2}
\end{align}
We need now to regularize the sum (\ref{exp_int}). This can be done 
by shifting the internal channels by
\begin{equation}
\hat{\alpha}_{i,j}\to \hat{\alpha}_{i,j}(\delta)=\alpha_{r,s}+i\beta -j \beta^{-1}+\delta.
\label{regul}
\end{equation}
Note that on may use different regularization procedures. We can 
regularize the sum (\ref{exp_int}) by keeping the neutrality condition satisfied.
This is obtained by shifting
$\alpha_1,\alpha_2\to\alpha_1+\delta/2,\alpha_2+\delta/2$; $\alpha_3,\alpha_4\to\alpha_3-\delta/2,\alpha_4-\delta/2$,
as we have done in the one-screening case.  In principle,  one can also study the integral (\ref{exp_int})  with 
the internal charges shifted as in (\ref{regul}) by keeping the external charges fixed: the sum (\ref{exp_int}) remains well defined also in this case
even if it does not correspond to a CG integral, which is only 
recovered in the limit $\delta\to 0$. If $i,j$ satisfies the condition (\ref{ij_deg_sing}), the corresponding 
term $|F(x|c,\hat{\Delta}_{ij},\{\Delta\})|^2$  has a singularity $ O(\delta^{-2})$. 
The coefficient $X_{i,j}$ is $O(\delta)$  (resp. $O(\delta^{-1})$) anytime one of the conditions   
(\ref{zeros1}) and  (\ref{zeros2}) (resp. (\ref{infies1}) and  (\ref{infies2})) is satisfied.

We find useful to consider in the following, the points  $(x,y)$ with integer coordinates, where 
$(x,y)$ label the charges $\alpha_{x,y}=1/2(1-x)\beta+1/2(y-1)\beta^{-1}$.
We can then visualize the set of channels entering in (\ref{exp_int}) as a grid of points of size $2n 
\times 2m$ whose  upper right corner has coordinates $(r,s)$.
The symmetry (\ref{tr_sym}) relates the 
pair of points symmetric with respect to the origin.
Points in the lattice representing the internal channels with charge $\hat{\alpha}_{ij}$ will have
coordinates 
\begin{equation}
(r-2i,s-2j)\equiv[i,j],
\end{equation}
 and to each of them we associate the order in the expansion in powers of $\delta$ 
of the corresponding coefficient $X_{i,j}$ as well as of the  conformal block
$|F(x|\hat{\Delta}_{ij},\{\Delta\})|^2$. Let us consider first the behavior of  the conformal blocks.
The points $[i,j]$ with positive coordinates, i.e. $r-2 i\geq 1$ and $s-2j\geq 1$
(belonging to the first quadrant of the lattice), and the ones with negative coordinates,
$r-2 i\leq -1$ and $s-2j\leq -1$ (in the third quadrant),
represent degenerate operators.
If the point  $[i,j]$ and at least one of the two points $[r-i,j]$, $[i,s-j]$, obtained by reflection with respect
to the horizontal and vertical axis, belong to the grid of internal channels,
then the pair $i,j$ satisfies the condition (\ref{ij_deg_sing}) and is associated to a singular conformal block. 
On the other hand, given a point   $[i,j]$, for what concerns the behavior of the coefficients $X_{i,j}$ we should distinguish
three cases.
\begin{itemize}
\item[\textit{(i)}] The point $[r-i,s-j]$ symmetric with respect to the origin belongs to the grid.
In this case, the (\ref{zeros1})-(\ref{zeros2})  are not satisfied. If $[i,j]$ or  $[r-i,s-j]$
belong to the first or third quadrant,  $X_{i,j}=O(\delta^{0})$
otherwise  $X_{i,j}=O(\delta^{-2})$ (we remind the identification (\ref{tr_sym})).
For those points sitting on the two axis but excluding the origin,
i.e. $2i=r$ and $2j\neq s$ or  $2i\neq r$ and $2j= s$ , $X_{i,j}= O(\delta^{-1})$
while for the origin, we have  $X_{r/2,s/2}=O(\delta^0)$;
\item[\textit{(ii)}] One among the axis-symmetric points $[r-i,j]$ or $[r,s-j]$ belong to the grid.
This implies that one of the two conditions (\ref{zeros1}) and (\ref{zeros2}) is verified. If $[i,j]$ belong to the first
or third quadrant,  $X_{i,j}= O(\delta)$, if they are sitting on one of two axis, $X_{i,j}=O(\delta^0)$,
otherwise  $X_{i,j}=O(\delta^{-1})$;
\item [\textit{(iii)}] The two-axis symmetric points  $[r-i,j]$ or $[r,s-j]$  do not belong to the grid:
in this case  $X_{i,j}=O(\delta^0)$. 
\end{itemize}
The terms in (\ref{exp_int}) can have amplitudes  of different order in $\delta$, 
in particular singularities of order $\delta^{-2}$ or $\delta^{-1}$. 
The important observation is that, in any case,
to any singular term  $X_{i,j}F(x|\hat{\Delta}_{i,j},\{\Delta\})|^2$  of $O(\delta^{-2})$ or $O( \delta^{-1})$, corresponds
an other term  $X_{r-i,j}|F(x|c,\hat{\Delta}_{r-i,j},x)|^2$ or 
$X_{i,s-j}F(x|x,\hat{\Delta}_{i,s-j},\{\Delta\})|^2$  of the same order. 
This pair of singular channels is represented by points which can 
(if $i=r/2$ or $s=j/2$)  or not belong to one of the two axis.  
In this last case,  the singular vector $|\chi_{r-2i,s-2j}\rangle$ of the degenerate primary $\phi_{r-2i,s-2j}$
enters in collision with the highest weight $|\phi_{2i-r,s-2j}\rangle$ (or $|\phi_{r-2i,2j-s}\rangle$)
while in the first case one has a collision between the two primaries with the same dimension $\Phi_{0,s-2j}$
and $\Phi_{0,2j-s}$ (or $\Phi_{r-2i,0}$ and $\Phi_{2i -r,0}$).
In both situations,  the leading order singularity in  the integral cancels
and the subleading order in the power series in $\delta$ contains logarithms. Before showing such cancellation explicitly,
let us discuss some examples.
 
\begin{itemize}
\item $r>2n$ and $s>2 m$ 

 All the points $[i,j]$ are in the first quadrant.
Since (\ref{ij_deg_sing}) cannot be satisfied, the ratio of the all  coefficients $X_{i,j}$ is finite 
and there are not singular conformal blocks. In the geometrical interpretation we proposed, the grid does not
contain points obtained by reflection with respect to the horizontal or vertical axis.

In Fig. \ref{grid_ch} we show as an example the case $n=2,m=3$ and $r=5,s=7$. 

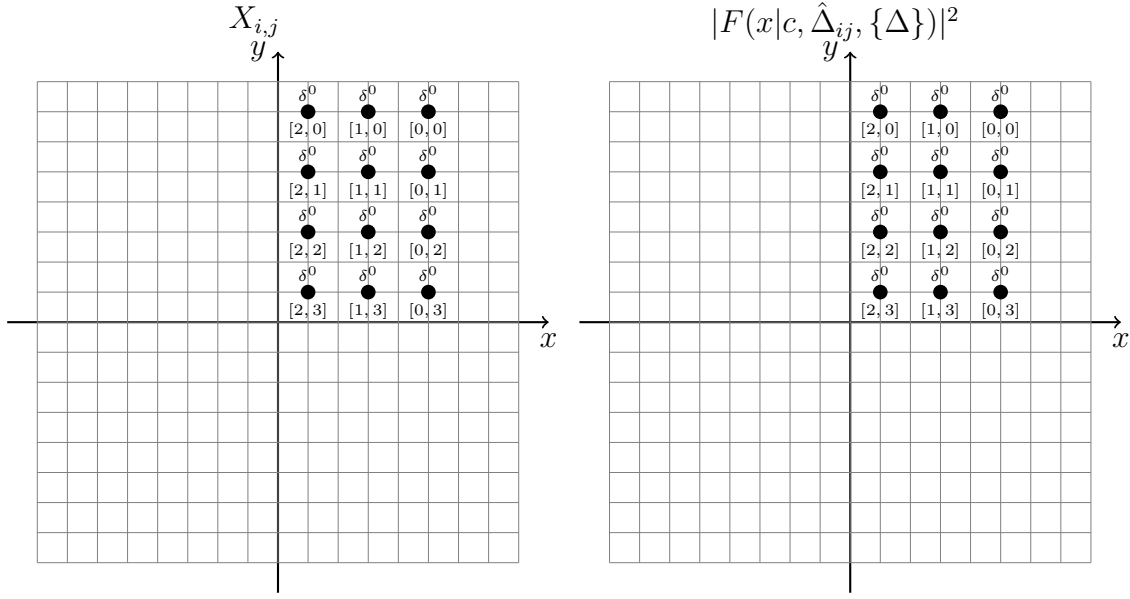
\begin{figure}[t]
\begin{tikzpicture}[scale=0.4]
\draw (-2,10) node[right]{$X_{i,j}$};
    \draw[->,thick] (-9,0) -- (9,0) node[below]{$x$};
    \draw[->,thick] (0,-9) --(0,9)node[left]{$y$};
\foreach \x in {-8,...,8}
                \draw[very thin, gray] (\x,-8)--(\x,8);
\foreach \y in {-8,...,8}
                \draw[very thin, gray] (-8,\y)--(8,\y);
\draw (5,7) node[circle,fill=black,minimum size=1pt,scale=0.5] {};
\draw (5,7) node[below]{{\tiny $[0,0]$}};
\draw (5,7) node[above]{{\tiny $\delta^0$}};

\draw (5,5) node[circle,fill=black,minimum size=1pt,scale=0.5] {};
\draw (5,5) node[below]{{\tiny $[0,1]$}};
\draw (5,5) node[above]{{\tiny $\delta^0$}};

\draw (5,3) node[circle,fill=black,minimum size=1pt,scale=0.5] {};
\draw (5,3) node[below]{{\tiny $[0,2]$}};
 \draw (5,3) node[above]{{\tiny $\delta^0$}};

\draw (5,1) node[circle,fill=black,minimum size=1pt,scale=0.5] {};
\draw (5,1) node[below]{{\tiny $[0,3]$}};
\draw (5,1) node[above]{{\tiny $\delta^0$}};

\draw (3,7) node[circle,fill=black,minimum size=1pt,scale=0.5] {};
\draw (3,7) node[below]{{\tiny $[1,0]$}};
\draw (3,7) node[above]{{\tiny $\delta^0$}};

\draw (3,5) node[circle,fill=black,minimum size=1pt,scale=0.5] {};
\draw (3,5) node[below]{{\tiny $[1,1]$}};
\draw (3,5) node[above]{{\tiny $\delta^0$}};

\draw (3,3) node[circle,fill=black,minimum size=1pt,scale=0.5] {};
\draw (3,3) node[below]{{\tiny $[1,2]$}};
 \draw (3,3) node[above]{{\tiny $\delta^0$}};

\draw (3,1) node[circle,fill=black,minimum size=1pt,scale=0.5] {};
\draw (3,1) node[below]{{\tiny $[1,3]$}};
\draw (3,1) node[above]{{\tiny $\delta^0$}};

\draw (1,7) node[circle,fill=black,minimum size=1pt,scale=0.5] {};
\draw (1,7) node[below]{{\tiny $[2,0]$}};
\draw (1,7) node[above]{{\tiny $\delta^0$}};

\draw (1,5) node[circle,fill=black,minimum size=1pt,scale=0.5] {};
\draw (1,5) node[below]{{\tiny $[2,1]$}};
\draw (1,5) node[above]{{\tiny $\delta^0$}};

\draw (1,3) node[circle,fill=black,minimum size=1pt,scale=0.5] {};
\draw (1,3) node[below]{{\tiny $[2,2]$}};
\draw (1,3) node[above]{{\tiny $\delta^0$}};

\draw (1,1) node[circle,fill=black,minimum size=1pt,scale=0.5] {};
\draw (1,1) node[below]{{\tiny $[2,3]$}};
\draw (1,1) node[above]{{\tiny $\delta^0$}};
\end{tikzpicture}
\begin{tikzpicture}[scale=0.4]
\draw (-5,10) node[right]{$|F(x|c,\hat{\Delta}_{ij},\{\Delta\})|^2$};
   \draw[->,thick] (-9,0) -- (9,0) node[below]{$x$};
    \draw[->,thick] (0,-9) --(0,9)node[left]{$y$};
\foreach \x in {-8,...,8}
                \draw[very thin, gray] (\x,-8)--(\x,8);
\foreach \y in {-8,...,8}
                \draw[very thin, gray] (-8,\y)--(8,\y);
\draw (5,7) node[circle,fill=black,minimum size=1pt,scale=0.5] {};
\draw (5,7) node[below]{{\tiny $[0,0]$}};
\draw (5,7) node[above]{{\tiny $\delta^0$}};

\draw (5,5) node[circle,fill=black,minimum size=1pt,scale=0.5] {};
\draw (5,5) node[below]{{\tiny $[0,1]$}};
\draw (5,5) node[above]{{\tiny $\delta^0$}};

\draw (5,3) node[circle,fill=black,minimum size=1pt,scale=0.5] {};
\draw (5,3) node[below]{{\tiny $[0,2]$}};
 \draw (5,3) node[above]{{\tiny $\delta^0$}};

\draw (5,1) node[circle,fill=black,minimum size=1pt,scale=0.5] {};
\draw (5,1) node[below]{{\tiny $[0,3]$}};
\draw (5,1) node[above]{{\tiny $\delta^0$}};

\draw (3,7) node[circle,fill=black,minimum size=1pt,scale=0.5] {};
\draw (3,7) node[below]{{\tiny $[1,0]$}};
\draw (3,7) node[above]{{\tiny $\delta^0$}};

\draw (3,5) node[circle,fill=black,minimum size=1pt,scale=0.5] {};
\draw (3,5) node[below]{{\tiny $[1,1]$}};
\draw (3,5) node[above]{{\tiny $\delta^0$}};

\draw (3,3) node[circle,fill=black,minimum size=1pt,scale=0.5] {};
\draw (3,3) node[below]{{\tiny $[1,2]$}};
 \draw (3,3) node[above]{{\tiny $\delta^0$}};

\draw (3,1) node[circle,fill=black,minimum size=1pt,scale=0.5] {};
\draw (3,1) node[below]{{\tiny $[1,3]$}};
\draw (3,1) node[above]{{\tiny $\delta^0$}};

\draw (1,7) node[circle,fill=black,minimum size=1pt,scale=0.5] {};
\draw (1,7) node[below]{{\tiny $[2,0]$}};
\draw (1,7) node[above]{{\tiny $\delta^0$}};

\draw (1,5) node[circle,fill=black,minimum size=1pt,scale=0.5] {};
\draw (1,5) node[below]{{\tiny $[2,1]$}};
\draw (1,5) node[above]{{\tiny $\delta^0$}};

\draw (1,3) node[circle,fill=black,minimum size=1pt,scale=0.5] {};
\draw (1,3) node[below]{{\tiny $[2,2]$}};
\draw (1,3) node[above]{{\tiny $\delta^0$}};

\draw (1,1) node[circle,fill=black,minimum size=1pt,scale=0.5] {};
\draw (1,1) node[below]{{\tiny $[2,3]$}};
\draw (1,1) node[above]{{\tiny $\delta^0$}};

\end{tikzpicture}

\caption{Example of a grid with $r>2n$ and $s>2m$. According to our analysis all the internal channels have finite
coefficients $X_{i,j}$ and all the conformal blocks $F(x|c,\hat{\Delta}_{ij},\{\Delta\})$ are non-singular.}
\label{grid_ch}
\end{figure}

\item $r>2n$ and $s\leq2m$

In this case the grid contains points which are obtained by reflection with respect to the $x$ axis.
Correspondingly,  we expect singularities $O(\delta^{-1})$ coming from the pair
$X_{i,j}|F(x|c,\hat{\Delta}_{ij},\{\Delta\})|^2$ and $X_{i,s-j}|F(x|c,\hat{\Delta}_{i,s-j},\{\Delta\})|^2$. 
 As we previously remarked,
 the sum of these two terms is $O(\delta^0)$ and produces a logarithmic term
 which adds with the other (non-logarithmic) contributions when considering the $\delta\to 0$ limit. 
Note that there are no singularities $O(\delta^{-2})$, because there are no points
of the grid which are symmetric with respect to the origin. Fig. \ref{grid_ch2} and Fig \ref{grid_ch3} show the examples 
$n=2,m=3$ with $r=5,s=4$ and  $n=2,m=3$ with $r=5,s=2$

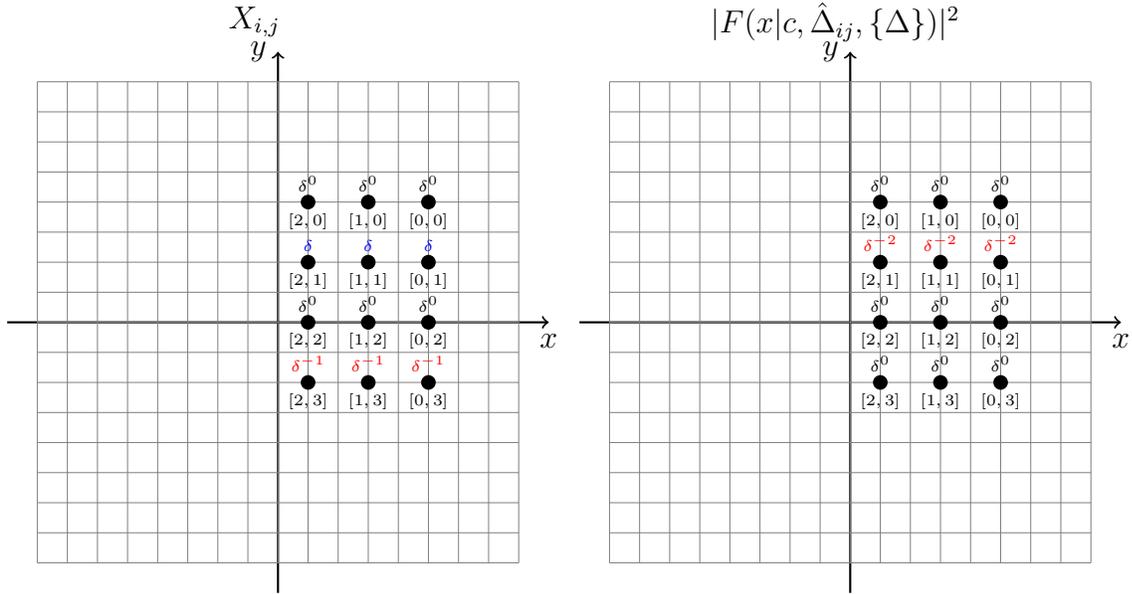
\begin{figure}[t]
\begin{tikzpicture}[scale=0.4]
\draw (-2,10) node[right]{$X_{i,j}$};
    \draw[->,thick] (-9,0) -- (9,0) node[below]{$x$};
    \draw[->,thick] (0,-9) --(0,9)node[left]{$y$};
\foreach \x in {-8,...,8}
                \draw[very thin, gray] (\x,-8)--(\x,8);
\foreach \y in {-8,...,8}
                \draw[very thin, gray] (-8,\y)--(8,\y);
\draw (5,4) node[circle,fill=black,minimum size=1pt,scale=0.5] {};
\draw (5,4) node[below]{{\tiny $[0,0]$}};
\draw (5,4) node[above]{{\tiny $\delta^0$}};

\draw (5,2) node[circle,fill=black,minimum size=1pt,scale=0.5] {};
\draw (5,2) node[below]{{\tiny $[0,1]$}};
\draw (5,2) node[above]{{\tiny ${\color{blue} \delta}$}};

\draw (5,0) node[circle,fill=black,minimum size=1pt,scale=0.5] {};
\draw (5,0) node[below]{{\tiny $[0,2]$}};
 \draw (5,0) node[above]{{\tiny $\delta^0$}};

\draw (5,-2) node[circle,fill=black,minimum size=1pt,scale=0.5] {};
\draw (5,-2) node[below]{{\tiny $[0,3]$}};
\draw (5,-2) node[above]{{\tiny ${\color{red} \delta^{-1}}$}};

\draw (3,4) node[circle,fill=black,minimum size=1pt,scale=0.5] {};
\draw (3,4) node[below]{{\tiny $[1,0]$}};
\draw (3,4) node[above]{{\tiny $\delta^0$}};

\draw (3,2) node[circle,fill=black,minimum size=1pt,scale=0.5] {};
\draw (3,2) node[below]{{\tiny $[1,1]$}};
\draw (3,2) node[above]{{\tiny ${\color{blue}\delta}$}};

\draw (3,0) node[circle,fill=black,minimum size=1pt,scale=0.5] {};
\draw (3,0) node[below]{{\tiny $[1,2]$}};
 \draw (3,0) node[above]{{\tiny $\delta^0$}};

\draw (3,-2) node[circle,fill=black,minimum size=1pt,scale=0.5] {};
\draw (3,-2) node[below]{{\tiny $[1,3]$}};
\draw (3,-2) node[above]{{\tiny ${\color{red}\delta^{-1}}$}};

\draw (1,4) node[circle,fill=black,minimum size=1pt,scale=0.5] {};
\draw (1,4) node[below]{{\tiny $[2,0]$}};
\draw (1,4) node[above]{{\tiny $\delta^0$}};

\draw (1,2) node[circle,fill=black,minimum size=1pt,scale=0.5] {};
\draw (1,2) node[below]{{\tiny $[2,1]$}};
\draw (1,2) node[above]{{\tiny ${\color{blue}\delta}$}};

\draw (1,0) node[circle,fill=black,minimum size=1pt,scale=0.5] {};
\draw (1,0) node[below]{{\tiny $[2,2]$}};
\draw (1,0) node[above]{{\tiny $\delta^0$}};

\draw (1,-2) node[circle,fill=black,minimum size=1pt,scale=0.5] {};
\draw (1,-2) node[below]{{\tiny $[2,3]$}};
\draw (1,-2) node[above]{{\tiny ${\color{red} \delta^{-1}}$}};
\end{tikzpicture}
\begin{tikzpicture}[scale=0.4]
\draw (-5,10) node[right]{$|F(x|c,\hat{\Delta}_{ij},\{\Delta\})|^2$};
   \draw[->,thick] (-9,0) -- (9,0) node[below]{$x$};
    \draw[->,thick] (0,-9) --(0,9)node[left]{$y$};
\foreach \x in {-8,...,8}
                \draw[very thin, gray] (\x,-8)--(\x,8);
\foreach \y in {-8,...,8}
                \draw[very thin, gray] (-8,\y)--(8,\y);
\draw (5,4) node[circle,fill=black,minimum size=1pt,scale=0.5] {};
\draw (5,4) node[below]{{\tiny $[0,0]$}};
\draw (5,4) node[above]{{\tiny $\delta^0$}};

\draw (5,2) node[circle,fill=black,minimum size=1pt,scale=0.5] {};
\draw (5,2) node[below]{{\tiny $[0,1]$}};
\draw (5,2) node[above]{{\tiny ${\color{red}\delta^{-2}}$}};

\draw (5,0) node[circle,fill=black,minimum size=1pt,scale=0.5] {};
\draw (5,0) node[below]{{\tiny $[0,2]$}};
 \draw (5,0) node[above]{{\tiny $\delta^0$}};

\draw (5,-2) node[circle,fill=black,minimum size=1pt,scale=0.5] {};
\draw (5,-2) node[below]{{\tiny $[0,3]$}};
\draw (5,-2) node[above]{{\tiny $\delta^0$}};

\draw (3,4) node[circle,fill=black,minimum size=1pt,scale=0.5] {};
\draw (3,4) node[below]{{\tiny $[1,0]$}};
\draw (3,4) node[above]{{\tiny $\delta^0$}};

\draw (3,2) node[circle,fill=black,minimum size=1pt,scale=0.5] {};
\draw (3,2) node[below]{{\tiny $[1,1]$}};
\draw (3,2) node[above]{{\tiny ${\color{red}\delta^{-2}}$}};

\draw (3,0) node[circle,fill=black,minimum size=1pt,scale=0.5] {};
\draw (3,0) node[below]{{\tiny $[1,2]$}};
 \draw (3,0) node[above]{{\tiny $\delta^0$}};

\draw (3,-2) node[circle,fill=black,minimum size=1pt,scale=0.5] {};
\draw (3,-2) node[below]{{\tiny $[1,3]$}};
\draw (3,-2) node[above]{{\tiny $\delta^0$}};

\draw (1,4) node[circle,fill=black,minimum size=1pt,scale=0.5] {};
\draw (1,4) node[below]{{\tiny $[2,0]$}};
\draw (1,4) node[above]{{\tiny $\delta^0$}};

\draw (1,2) node[circle,fill=black,minimum size=1pt,scale=0.5] {};
\draw (1,2) node[below]{{\tiny $[2,1]$}};
\draw (1,2) node[above]{{\tiny ${\color{red}\delta^{-2}}$}};

\draw (1,0) node[circle,fill=black,minimum size=1pt,scale=0.5] {};
\draw (1,0) node[below]{{\tiny $[2,2]$}};
\draw (1,0) node[above]{{\tiny $\delta^0$}};

\draw (1,-2) node[circle,fill=black,minimum size=1pt,scale=0.5] {};
\draw (1,-2) node[below]{{\tiny $[2,3]$}};
\draw (1,-2) node[above]{{\tiny $\delta^0$}};
\end{tikzpicture}
\caption{In this figure we show the correlation function with $r=5,~s=4$ and $n=2,~m=3$. Notice that the degenerate
primaries represented by the points $[2,1],~[1,1]$ and $[0,1]$ have a singular conformal block but their coefficient $X_{i,j}$
is $O(\delta)$. On the other hand the reflected points $[2,3],~[1,3]$,and $[0,3]$
corresponds to non-degenerate primaries with a
finite conformal block but with a coefficient $X_{i,j}$ which is now $O(\delta^{-1})$. This situation is completely analogous to the
one we encountered in the one-screening case of Sect. 3.}
\label{grid_ch2}
\end{figure}

\begin{figure}[t]
\begin{tikzpicture}[scale=0.4]
\draw (-2,10) node[right]{$X_{i,j}$};
    \draw[->,thick] (-9,0) -- (9,0) node[below]{$x$};
    \draw[->,thick] (0,-9) --(0,9)node[left]{$y$};
\foreach \x in {-8,...,8}
                \draw[very thin, gray] (\x,-8)--(\x,8);
\foreach \y in {-8,...,8}
                \draw[very thin, gray] (-8,\y)--(8,\y);
\draw (5,2) node[circle,fill=black,minimum size=1pt,scale=0.5] {};
\draw (5,2) node[below]{{\tiny $[0,0]$}};
\draw (5,2) node[above]{{\tiny ${\color{blue}\delta}$}};

\draw (5,0) node[circle,fill=black,minimum size=1pt,scale=0.5] {};
\draw (5,0) node[below]{{\tiny $[0,1]$}};
\draw (5,0) node[above]{{\tiny $\delta^0$}};

\draw (5,-2) node[circle,fill=black,minimum size=1pt,scale=0.5] {};
\draw (5,-2) node[below]{{\tiny $[0,2]$}};
 \draw (5,-2) node[above]{{\tiny ${\color{red} \delta^{-1}}$}};

\draw (5,-4) node[circle,fill=black,minimum size=1pt,scale=0.5] {};bookos
\draw (5,-4) node[below]{{\tiny $[0,3]$}};
\draw (5,-4) node[above]{{\tiny $\delta^0$}};

\draw (3,2) node[circle,fill=black,minimum size=1pt,scale=0.5] {};
\draw (3,2) node[below]{{\tiny $[1,0]$}};
\draw (3,2) node[above]{{\tiny ${\color{blue}\delta}$}};

\draw (3,0) node[circle,fill=black,minimum size=1pt,scale=0.5] {};
\draw (3,0) node[below]{{\tiny $[1,1]$}};
\draw (3,0) node[above]{{\tiny $\delta^0$}};

\draw (3,-2) node[circle,fill=black,minimum size=1pt,scale=0.5] {};
\draw (3,-2) node[below]{{\tiny $[1,2]$}};
 \draw (3,-2) node[above]{{\tiny ${\color{red} \delta^{-1}}$}};

\draw (3,-4) node[circle,fill=black,minimum size=1pt,scale=0.5] {};
\draw (3,-4) node[below]{{\tiny $[1,3]$}};
\draw (3,-4) node[above]{{\tiny $\delta^0$}};

\draw (1,2) node[circle,fill=black,minimum size=1pt,scale=0.5] {};
\draw (1,2) node[below]{{\tiny $[2,0]$}};
\draw (1,2) node[above]{{\tiny ${\color{blue}\delta}$}};

\draw (1,0) node[circle,fill=black,minimum size=1pt,scale=0.5] {};bookos
\draw (1,0) node[below]{{\tiny $[2,1]$}};
\draw (1,0) node[above]{{\tiny $\delta^0$}};

\draw (1,-2) node[circle,fill=black,minimum size=1pt,scale=0.5] {};
\draw (1,-2) node[below]{{\tiny $[2,2]$}};
\draw (1,-2) node[above]{{\tiny ${\color{red} \delta^{-1}}$}};

\draw (1,-4) node[circle,fill=black,minimum size=1pt,scale=0.5] {};
\draw (1,-4) node[below]{{\tiny $[2,3]$}};
\draw (1,-4) node[above]{{\tiny $\delta^{0}$}};
\end{tikzpicture}
\begin{tikzpicture}[scale=0.4]
\draw (-5,10) node[right]{$|F(x|c,\hat{\Delta}_{ij},\{\Delta\})|^2$};
   \draw[->,thick] (-9,0) -- (9,0) node[below]{$x$};
   \draw[->,thick] (0,-9) --(0,9)node[left]{$y$};
\foreach \x in {-8,...,8}
                \draw[very thin, gray] (\x,-8)--(\x,8);
\foreach \y in {-8,...,8}
                \draw[very thin, gray] (-8,\y)--(8,\y);
\draw (5,2) node[circle,fill=black,minimum size=1pt,scale=0.5] {};
\draw (5,2) node[below]{{\tiny $[0,0]$}};
\draw (5,2) node[above]{{\tiny ${\color{red} \delta^{-2}}$}};

\draw (5,0) node[circle,fill=black,minimum size=1pt,scale=0.5] {};
\draw (5,0) node[below]{{\tiny $[0,1]$}};
\draw (5,0) node[above]{{\tiny $\delta^0$}};

\draw (5,-2) node[circle,fill=black,minimum size=1pt,scale=0.5] {};
\draw (5,-2) node[below]{{\tiny $[0,2]$}};
 \draw (5,-2) node[above]{{\tiny $\delta^0$}};

\draw (5,-4) node[circle,fill=black,minimum size=1pt,scale=0.5] {};
\draw (5,-4) node[below]{{\tiny $[0,3]$}};
\draw (5,-4) node[above]{{\tiny $\delta^0$}};

\draw (3,2) node[circle,fill=black,minimum size=1pt,scale=0.5] {};
\draw (3,2) node[below]{{\tiny $[1,0]$}};
\draw (3,2) node[above]{{\tiny ${\color{red} \delta^{-2}}$}};

\draw (3,0) node[circle,fill=black,minimum size=1pt,scale=0.5] {};
\draw (3,0) node[below]{{\tiny $[1,1]$}};
\draw (3,0) node[above]{{\tiny $\delta^0$}};

\draw (3,-2) node[circle,fill=black,minimum size=1pt,scale=0.5] {};
\draw (3,-2) node[below]{{\tiny $[1,2]$}};
 \draw (3,-2) node[above]{{\tiny $\delta^0$}};

\draw (3,-4) node[circle,fill=black,minimum size=1pt,scale=0.5] {};
\draw (3,-4) node[below]{{\tiny $[1,3]$}};
\draw (3,-4) node[above]{{\tiny $\delta^0$}};

\draw (1,2) node[circle,fill=black,minimum size=1pt,scale=0.5] {};
\draw (1,2) node[below]{{\tiny $[2,0]$}};
\draw (1,2) node[above]{{\tiny ${\color{red} \delta^{-2}}$}};

\draw (1,0) node[circle,fill=black,minimum size=1pt,scale=0.5] {};
\draw (1,0) node[below]{{\tiny $[2,1]$}};
\draw (1,0) node[above]{{\tiny $\delta^0$}};

\draw (1,-2) node[circle,fill=black,minimum size=1pt,scale=0.5] {};
\draw (1,-2) node[below]{{\tiny $[2,2]$}};
\draw (1,-2) node[above]{{\tiny $\delta^0$}};

\draw (1,-4) node[circle,fill=black,minimum size=1pt,scale=0.5] {};
\draw (1,-4) node[below]{{\tiny $[2,3]$}};
\draw (1,-4) node[above]{{\tiny $\delta^0$}};
\end{tikzpicture}
\caption{Figure representing the case $r=5~,s=2$ and $n=2,~m=3$. Analogous considerations to the ones of Fig. \ref{grid_ch2}
apply.}
\label{grid_ch3}
\end{figure}
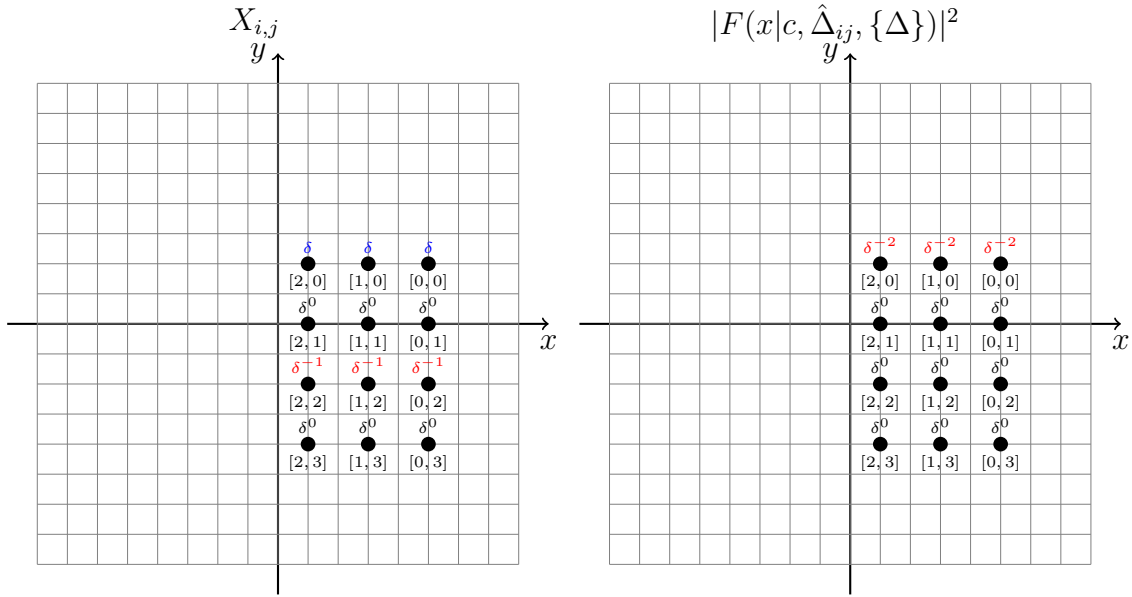

\item $r \leq 2n$ and $m\leq s \leq 2m$

In this case there are also points of the grid which are symmetric with respect to the origin.
These points are associated
to contributions $O(\delta^{-2})$. In Fig. \ref{grid_ch4}, we show the example $n=2$, $m=3$ and $r=3$, $s=4$.
In this case we expect that the correlation function gets dominant contributions of  order $\delta^{-1}$.
To get a finite correlation function, one has to choose a vanishing normalization constant $\mathcal{N}(c,\{\Delta\})$ in
\eqref{X_ij_dege}, of $O(\delta)$. Note also that in this case there are terms in the (\ref{exp_int})
whose contribution is $0(\delta^0)$ and therefore disappear in the limit $\delta\to 0$.

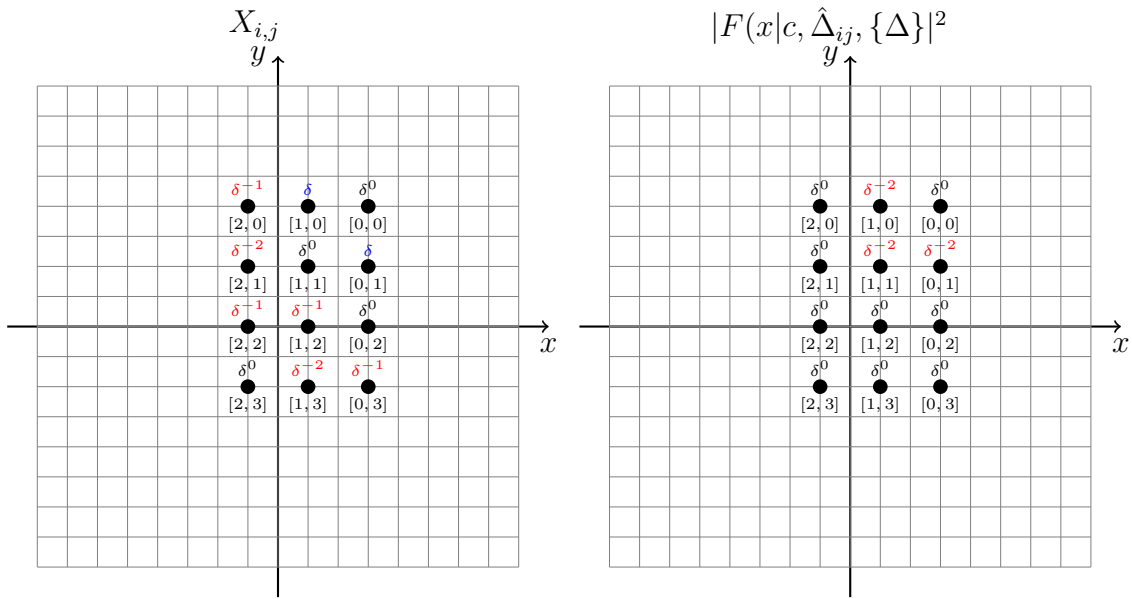
\begin{figure}[t]
\begin{tikzpicture}[scale=0.4]
\draw (-2,10) node[right]{$X_{i,j}$};
    \draw[->,thick] (-9,0) -- (9,0) node[below]{$x$};
    \draw[->,thick] (0,-9) --(0,9)node[left]{$y$};
\foreach \x in {-8,...,8}
                \draw[very thin, gray] (\x,-8)--(\x,8);
\foreach \y in {-8,...,8}
                \draw[very thin, gray] (-8,\y)--(8,\y);
\draw (3,4) node[circle,fill=black,minimum size=1pt,scale=0.5] {};
\draw (3,4) node[below]{{\tiny $[0,0]$}};
\draw (3,4) node[above]{{\tiny $\delta^0$}};

\draw (3,2) node[circle,fill=black,minimum size=1pt,scale=0.5] {};
\draw (3,2) node[below]{{\tiny $[0,1]$}};
\draw (3,2) node[above]{{\tiny ${\color{blue}\delta}$}};

\draw (3,0) node[circle,fill=black,minimum size=1pt,scale=0.5] {};
\draw (3,0) node[below]{{\tiny $[0,2]$}};
 \draw (3,0) node[above]{{\tiny $\delta^{0}$}};

\draw (3,-2) node[circle,fill=black,minimum size=1pt,scale=0.5] {};
\draw (3,-2) node[below]{{\tiny $[0,3]$}};
\draw (3,-2) node[above]{ {\tiny ${\color{red} \delta^{-1}}$}};

\draw (1,4) node[circle,fill=black,minimum size=1pt,scale=0.5] {};
\draw (1,4) node[below]{{\tiny $[1,0]$}};
\draw (1,4) node[above]{{\tiny ${\color{blue}\delta}$}};

\draw (1,2) node[circle,fill=black,minimum size=1pt,scale=0.5] {};
\draw (1,2) node[below]{{\tiny $[1,1]$}};
\draw (1,2) node[above]{{\tiny $\delta^0$}};

\draw (1,0) node[circle,fill=black,minimum size=1pt,scale=0.5] {};
\draw (1,0) node[below]{{\tiny $[1,2]$}};
 \draw (1,0) node[above]{{\tiny ${\color{red} \delta^{-1}}$}};

\draw (1,-2) node[circle,fill=black,minimum size=1pt,scale=0.5] {};
\draw (1,-2) node[below]{{\tiny $[1,3]$}};
\draw (1,-2) node[above]{{\tiny ${\color{red} \delta^{-2}}$}};

\draw (-1,4) node[circle,fill=black,minimum size=1pt,scale=0.5] {};
\draw (-1,4) node[below]{{\tiny $[2,0]$}};
\draw (-1,4) node[above]{{\tiny ${\color{red} \delta^{-1}}$}};

\draw (-1,2) node[circle,fill=black,minimum size=1pt,scale=0.5] {};
\draw (-1,2) node[below]{{\tiny $[2,1]$}};
\draw (-1,2) node[above]{{\tiny ${\color{red} \delta^{-2}}$}};

\draw (-1,0) node[circle,fill=black,minimum size=1pt,scale=0.5] {};
\draw (-1,0) node[below]{{\tiny $[2,2]$}};
\draw (-1,0) node[above]{{\tiny ${\color{red} \delta^{-1}}$}};

\draw (-1,-2) node[circle,fill=black,minimum size=1pt,scale=0.5] {};
\draw (-1,-2) node[below]{{\tiny $[2,3]$}};
\draw (-1,-2) node[above]{{\tiny $\delta^{0}$}};

\end{tikzpicture}
\begin{tikzpicture}[scale=0.4]
\draw (-5,10) node[right]{$|F(x|c,\hat{\Delta}_{ij},\{\Delta\}|^2$};
   \draw[->,thick] (-9,0) -- (9,0) node[below]{$x$};
    \draw[->,thick] (0,-9) --(0,9)node[left]{$y$};
\foreach \x in {-8,...,8}
                \draw[very thin, gray] (\x,-8)--(\x,8);
\foreach \y in {-8,...,8}
                \draw[very thin, gray] (-8,\y)--(8,\y);
                \draw (3,4) node[circle,fill=black,minimum size=1pt,scale=0.5] {};
\draw (3,4) node[below]{{\tiny $[0,0]$}};
\draw (3,4) node[above]{{\tiny $\delta^0$}};

\draw (3,2) node[circle,fill=black,minimum size=1pt,scale=0.5] {};
\draw (3,2) node[below]{{\tiny $[0,1]$}};
\draw (3,2) node[above]{{\tiny ${\color{red} \delta^{-2}}$}};

\draw (3,0) node[circle,fill=black,minimum size=1pt,scale=0.5] {};
\draw (3,0) node[below]{{\tiny $[0,2]$}};
 \draw (3,0) node[above]{{\tiny $\delta^0$}};

\draw (3,-2) node[circle,fill=black,minimum size=1pt,scale=0.5] {};
\draw (3,-2) node[below]{{\tiny $[0,3]$}};
\draw (3,-2) node[above]{{\tiny $\delta^0$}};

\draw (1,4) node[circle,fill=black,minimum size=1pt,scale=0.5] {};
\draw (1,4) node[below]{{\tiny $[1,0]$}};
\draw (1,4) node[above]{{\tiny ${\color{red} \delta^{-2}}$}};

\draw (1,2) node[circle,fill=black,minimum size=1pt,scale=0.5] {};
\draw (1,2) node[below]{{\tiny $[1,1]$}};
\draw (1,2) node[above]{{\tiny ${\color{red} \delta^{-2}}$}};

\draw (1,0) node[circle,fill=black,minimum size=1pt,scale=0.5] {};
\draw (1,0) node[below]{{\tiny $[1,2]$}};
 \draw (1,0) node[above]{{\tiny $\delta^0$}};

\draw (1,-2) node[circle,fill=black,minimum size=1pt,scale=0.5] {};
\draw (1,-2) node[below]{{\tiny $[1,3]$}};
\draw (1,-2) node[above]{{\tiny $\delta^0$}};

\draw (-1,4) node[circle,fill=black,minimum size=1pt,scale=0.5] {};
\draw (-1,4) node[below]{{\tiny $[2,0]$}};
\draw (-1,4) node[above]{{\tiny $\delta^0$}};

\draw (-1,2) node[circle,fill=black,minimum size=1pt,scale=0.5] {};
\draw (-1,2) node[below]{{\tiny $[2,1]$}};
\draw (-1,2) node[above]{{\tiny $\delta^0$}};

\draw (-1,0) node[circle,fill=black,minimum size=1pt,scale=0.5] {};
\draw (-1,0) node[below]{{\tiny $[2,2]$}};
\draw (-1,0) node[above]{{\tiny $\delta^0$}};

\draw (-1,-2) node[circle,fill=black,minimum size=1pt,scale=0.5] {};
\draw (-1,-2) node[below]{{\tiny $[2,3]$}};
\draw (-1,-2) node[above]{{\tiny $\delta^{0}$}};
\end{tikzpicture}
\caption{Example of a grid with $r=3,~s=4$ and $n=2,~m=3$. The associated correlation function will be $O(\delta^{-1})$, due to
the cancellation of leading $O(\delta^{-2})$ singularity. Such kind of correlation function can be render finite by the
choice $\mathcal{N}(c,\{\Delta\})=O(\delta)$.}
\label{grid_ch4}
\end{figure}

\end{itemize}
\clearpage
\subsection{Origin of logarithmic singularities in the four-point function}

We have seen that the correlation function (\ref{exp_int}) has singularities  $O(\delta^{-1})$ 
when only one pair of colliding channels $\hat{\alpha}_{ij}$, $\hat{\alpha}_{r-i,j}$
 or $\hat{\alpha}_{i,j}$, $\hat{\alpha}_{i,s-j}$ appears in its decomposition. The singularity is $O(\delta^{-2})$
 when all the four channels $\hat{\alpha}_{i,j}$, $\hat{\alpha}_{r-i,j}$,
$\hat{\alpha}_{i,s-j}$ and $\hat{\alpha}_{r-i,s-j}$ appear in its decomposition. Here we prove a  generalization of  \eqref{simp_rel_log} and show that the leading order in $\delta$ to the integral cancels,
leaving us with a logarithmic singularity. We consider first the leading order of the ratios $X_{i,j}/X_{r-i,j}$
and $X_{i,j}/X_{i,s-j}$ in the $\delta\to 0$ limit. 
By using the recursion relation (\ref{rec_Zam}), we obtain  for general values of
$\alpha_a$ and $\alpha_b$, with $\alpha_{a}^{b}\equiv \alpha_a-\alpha_b$ and $\alpha_{ab}\equiv \alpha_a+\alpha_b$ 
the following relations
\begin{align}
&\hspace*{-2cm}\frac{\Upsilon_{\beta}(\beta-\alpha_{a}^b-\alpha_{P,Q})\Upsilon_{\beta}(\beta-\alpha_{b}^a-\alpha_{P,Q})
\Upsilon_{\beta}(\beta-\alpha_{ab}+\alpha_{P,Q})}
{\Upsilon_{\beta}(\beta-\alpha_{a}^b-\alpha_{-P,Q})\Upsilon_{\beta}(\beta-\alpha_{b}^a-\alpha_{-P,Q})
\Upsilon_{\beta}(\beta-\alpha_{ab}+\alpha_{-P,Q})}\times \nonumber \\
&\hspace*{2.5cm}\times\frac{\Upsilon_{\beta}(2\beta-\beta^{-1}-\alpha_{ab}-\alpha_{P,Q})}
{\Upsilon_{\beta}(2\beta-\beta^{-1}-\alpha_{ab}-\alpha_{-P,Q})}= \frac{1}{R_{P,Q}(\alpha_a,\alpha_b)^2}, 
\end{align}
as well as
\begin{equation}
 \sqrt{\frac{\Upsilon_{\beta}(\beta-2\alpha_{-P,Q}-2\delta)\Upsilon_{\beta}(2\beta-\beta^{-1}-2\alpha_{-P,Q}-2\delta)}
{\Upsilon_{\beta}(\beta-2\alpha_{P,Q})
\Upsilon_{\beta}(2\beta-\beta^{-1}-2\alpha_{P,Q})}}=  i \;\delta \; B_{P,Q}\lambda_{P,Q}+O(\delta^{2})
\end{equation}
where $P$ and $Q$ are two positive integers. The ratios of the structure 
constant $C(\alpha_a,\alpha_b,\alpha_{P,Q})$, $C(\alpha_a,\alpha_b,\alpha_{-P,Q}+\delta)$ and $C(\alpha_a,\alpha_b,\alpha_{P,-Q}+\delta)$ 
take then the form 
\begin{eqnarray}
 \frac{C(\alpha_a,\alpha_b,\alpha_{P,Q}+\delta)}
{C(\alpha_a,\alpha_b,\alpha_{-P,Q}+\delta)}
&=&i \;\delta\; \frac{B_{P,Q}\lambda_{P,Q}}{R_{P,Q}(\alpha_a,\alpha_b)^2}+O(\delta^2) \nonumber \\
 \frac{C(\alpha_a,\alpha_b,\alpha_{P,Q}+\delta)}
{C(\alpha_a,\alpha_b,\alpha_{P,-Q}+\delta)}
&=&i \;\delta\; \frac{B_{P,Q}\lambda_{P,Q}}{R_{P,Q}(\alpha_a,\alpha_b)^2}+O(\delta^2).
\label{logza1}
\end{eqnarray} 
Notice that the  leading order in $\delta$ of the ratios \eqref{logza1}  are
the same. This is expected since $\Delta_{-P,Q}=\Delta_{P,-Q}$. 
We stress that a very similar 
equation relating the Liouville structure 
constants with the so-called logarithmic primaries has been found in \cite{Zamhem}. Using 
the property of the function $\Upsilon_{\beta}(x)$  and  
in particular its invariance with respect to the transformation $x\to\beta+\beta^{-1}-x$, 
we computed 
\begin{equation}
\frac{C(\alpha_a,\alpha_b,\alpha_{0,Q}+\delta)}
{C(\alpha_a,\alpha_b,\alpha_{0,-Q}+\delta)}=i;\quad \quad \frac{C(\alpha_a,\alpha_b,\alpha_{P,0}+\delta)}
{C(\alpha_a,\alpha_b,\alpha_{-P,0}+\delta)}=i.
\end{equation}
From these results,  
we can finally write the relation between
the coefficient $X_{i,j}$ associated to the point $[i,j]$ and to its symmetric under reflection $[r-i,j]$, $[i.s-j]$
\begin{eqnarray}
\label{canc_sing}
\frac{X_{i,j}(\delta)}{X_{r-i,j}(\delta)}&=&-\delta^{-2}\left(\frac{B_{r-2i,s-2j}\lambda_{r-2i,s-2j}}{R_{r-2i,s-2j}
(\{\alpha_{i}\})}\right)^2 +O\left(\delta^{-1}\right) \\
\label{canc_sing0}
\frac{X_{i,s/2}(\delta)}{X_{r-i,s/2}(\delta)}&=&-1 +O\left(\delta\right);\quad  \frac{X_{r/2,j}(\delta)}{X_{r/2,s-j}(\delta)}
=-1 +O\left(\delta\right)
\end{eqnarray}
Clearly, one can verify that the results 
(\ref{canc_sing}) and (\ref{canc_sing0}) are consistent with the previous findings based 
on the analysis of the poles and zeros of the structure constants.   

Let us focus on the situation in which  one pair of colliding channels is present,
for example the operators with charge $\hat{\alpha}_{i,j}(\delta)$ and $\hat{\alpha}_{r-i,j}(\delta)$. 
In the case ($i\neq r/2 \;\wedge\; j\neq s/2$), we have seen that the  
primary operator $\phi_{2i-r,s-2j}$ interferes with the null field $\chi_{r-2i,s-2j}$
responsible for the pole in the conformal block of $\phi_{r-2i,s-2j}$. 
We can fix the global normalization in (\ref{X_ij_dege}) such that
  \begin{equation}
  X_{i,j}(\delta)
= X_{i,j}^{(1)}\delta+X^{(2)}_{ij}\delta^2+O(\delta^3) \quad X_{r-i~j}(\delta)
= X_{r-i,j}^{(-1)}\delta^{-1}+X^{(0)}_{r-i,j}+O(\delta). 
  \end{equation} 

On the other hand, the
expansions in powers of $\delta$ of the singular conformal block 
$F(x|c,\hat{\Delta}_{ij},\{\Delta\})$ and of 
the non-singular conformal block $F(x|c,\hat{\Delta}_{r-i,j},\{\Delta\})$ can be written as
\begin{eqnarray}
F(x|c,\hat{\Delta}_{ij},\{\Delta\})&=&\delta^{-1}G^{(-1)}_{i,j}(z) +G^{(0)}_{i,j}(x)+O(\delta)\\
F(x|c,\hat{\Delta}_{r-i,j},\{\Delta\})&=&G^{(0)}_{r-i,j}+\delta G^{(1)}_{r-i,j}(z)+O(\delta^2), 
\end{eqnarray}
where as it follow from \eqref{rec_conf}
\begin{equation}
\label{delta_conf}
G^{(0)}_{r-i,j}(x)=F(x|c,\hat{\Delta}_{r-i,j},\{\Delta\}), \quad 
G^{(-1)}_{i,j}(x)=\frac{x^{P Q}S_{P,Q}}{\lambda_{P,Q}}F(x|c,\hat{\Delta}_{ij}+PQ,\{\Delta\}).
\end{equation}
In the equations \eqref{delta_conf}, $P= r-2i \geq 1$, and $Q=s-2j \geq 1$  and we recall
that $\hat{\Delta}_{i,j}(\delta)=\hat{\Delta}_{i,j}+\lambda_{P,Q}\delta +O(\delta^2)$. 
The functions 
\begin{equation}
 G^{(0)}_{i,j}(x)=1+\sum_{k=1}^{\infty}a_k~x^k;\quad G^{(1)}_{r-i,j}(x)=\sum_{k=1}^{\infty}b_k~x^k 
\end{equation}
are obtained by differentiating the conformal blocks with respect to $\delta$. In  general, compact analytic
expressions for the expansion coefficients $a_k$ and $b_k$  are not available. Nevertheless
one could derived them 
from the recursion formula \cite{Zam_rec1,Zam_rec2} or applying the AGT correspondence \cite{AGT} which 
relates a conformal block to the Nekrasov instantons partition function of a given $\mathcal N=2$ supersymmetric
four-dimensional gauge theory, see for instance \cite{SaTa10}. 
Applying the identity (\ref{canc_sing}), we can finally write the finite contribution $I^{PQ}_{nm}$ to the integral \eqref{exp_int} 
of the two colliding channels $\hat{\alpha}_{i,j}\equiv\alpha_{P,Q}$ and $\hat{\alpha}_{r-i,j}\equiv\alpha_{-P,Q}$
\begin{align}
&I_{nm}^{PQ}\equiv|x|^{2\Delta_{PQ}(\delta)}~X_{i,j}|F(x|c,\Delta_{PQ}(\delta), \{\Delta\})|^2+
|x|^{2\Delta_{-P,Q}(\delta)}~X_{r-i,j}|F(x|c,\Delta_{-P,Q}(\delta),\{\Delta\})|^2\nonumber\\
&=|x|^{2\Delta_{PQ}}\left[a^{(1)}_{P,Q}\log|x|^2|x|^{2PQ} X_{r-i,j}^{(-1)}|G_{r-i,j}^{(0)}(x)|^2+
X_{i,j}^{(1)}|G_{i,j}^{(-1)}(x)|^2+X_{r-i,j}^{(0)}|x|^{2PQ}|G_{r-i,j}^{(0)}(x)|^2\right.\nonumber\\
&\Bigl.\hspace*{0.5cm}+X_{i,j}^{(1)}\bigr(G_{i,j}^{(-1)}(x)G_{i,j}^{(0)}(\bar{x})+c.c.\bigl)+
X_{r-i,j}^{(-1)}|x|^{2PQ}\bigl(G_{r-i,j}^{(0)}(x)G_{r-i,j}^{(1)}(\bar{x})+c.c.\bigr)\Bigr],
\label{cg_ms}
\end{align}
where the factor $a_{P,Q}^{(1)}$ is defined by
\begin{equation}
a_{P,Q}^{(1)}\equiv\lim_{\delta \to 0}\frac{1}{\delta}\left(PQ-\Delta_{-P,Q}(\delta)+ \Delta_{P,Q}\right)=2P\beta.
\label{blog}
\end{equation}
If the colliding channels are   $\hat{\alpha}_{i,j}$ and $\hat{\alpha}_{i,s-j}$,
a formula identical to \eqref{cg_1s} will hold, but with the substitutions $r-i\to i $, $j\to s-j$ and of course the
parameter $a_{P,Q}^{(1)}$ will have to be replaced by 
\begin{equation}
\label{bdef}
a_{P,Q}^{(2)}
=\lim_{\delta \to 0}\frac{1}{\delta}\left(PQ-\Delta_{P,-Q}(\delta)+ \Delta_{PQ}(\delta)\right)=-2Q\beta^{-1}.
\end{equation}
Notice that the two coefficients $a_{P,Q}^{(1)}$ and $a_{P,Q}^{(2)}$ have always different signs.
We briefly outlined now the case where the channel with charge $\hat{\alpha}_{i,s/2}$ collides with $\hat{\alpha}_{r-i,s/2}$.
Using now the relation (\ref{canc_sing0}), we can write                                                                                                           
\begin{equation}
 X_{i,j}(\delta)
= X_{i,s/2}^{(-1)}\delta^{-1}+X^{(1)}_{(i,s/2)}+O(\delta) \quad X_{r-i,s/2}(\delta)
= -X_{i,s/2}^{(-1)}\delta^{-1}+X^{(1)}_{(r-i,s/2)}+O(\delta)
\label{canc0}
\end{equation}
while the expansions of the corresponding conformal blocks, both non-singular, are
\begin{align}
&F(x|c,\hat{\Delta}_{i,s/2},\{\Delta\})=G^{(0)}_{i,s/2}(x)+\delta G^{(1)}_{i,s/2}(x)+O(\delta^2)\\
&F(x|c,\hat{\Delta}_{i,s/2},\{\Delta\})=G^{(0)}_{(i,s/2)}(x)+\delta G^{(1)}_{(r-i,s/2)}(x)+O(\delta^2).
\end{align}
The finite  contribution in the correlation function can be then easily derived and the
coefficient of the logarithm is $a_{P,0}^{(1)}$, see \eqref{bdef}.

From the result (\ref{cg_1s}) one can infer an Operator Product Expansion (OPE), compatible with the cancellation of the leading singularities in
the four point function.
Consider two scalar operators
$\Phi_{\alpha_a}$ and $\Phi_{\alpha_b}$ which fuse into a third operator
$\Phi_{P,Q}^{\delta}(x,\bar{x})=\phi_{P,Q}^{\delta}(x)
\otimes \bar{\phi}_{P,Q}^{\delta}(\bar{x})$, having charge $\alpha_{P,Q}(\delta)=\alpha_{P,Q}+\delta$
and dimension $\Delta_{P,Q}(\delta)=\Delta_{P,Q}+\lambda_{P,Q}\delta+O(\delta^2)$
\begin{align}
\Phi_{\alpha_a}(x,\bar{x})\Phi_{\alpha_b}(0)=\frac{C(\alpha_a,\alpha_b,\alpha_{P,Q}(\delta))}
{|x|^{2\Delta_{\alpha_a}+2\Delta_{\alpha_b}-
2\Delta_{P,Q}(\delta)}
}&\left(\phi^{\delta}_{P,Q}(0)+\dots+\eta_{P,Q} x^{PQ}\chi^{\delta}_{P,Q}(0)+\dots\right) \times \nonumber \\
\times
&\left(\bar{\phi}^{\delta}_{P,Q}(0)+\dots+\eta_{P,Q} \bar{x}^{PQ} \bar{\chi}^{\delta}_{P,Q}(0)+\dots\right),
\label{ope_s}
\end{align}
where we have considered both the holomorphic and anti-holomorphic parts of the OPE.
In the limit $\delta\to 0$ the descendant $\chi_{P,Q}^{\delta}$ is the null-field  $\chi_{P,Q}$ of the representation
$\mathcal{V}_{\Delta_{P,Q}}$ of the Virasoro algebra.
From (\ref{deg_con}), the coefficient $\eta_{P,Q}$ is given by
\begin{equation}
\eta_{P,Q}=\frac{\langle \chi^{\delta}_{P,Q}(\infty) \phi_{\alpha_a}(1) \phi_{\alpha_b}(0)\rangle}
{\langle \chi_{P,Q}(\infty) \chi_{P,Q}(0) \rangle}
=\frac{R_{P,Q}(\alpha_a,\alpha_b)}{\delta \lambda_{P,Q} B_{P,Q}}+O(\delta^0),
\end{equation}
and is singular in the limit $\delta\to 0$ if $R_{P,Q}(\alpha_a,\alpha_b)\neq 0$.
As we have seen before local correlation functions may have  singularities $O(\delta^{-1})$ and $O(\delta^{-2})$.
The first one is always eliminated by the vanishing of the structure constant
$C(\alpha_a,\alpha_b,\alpha_{P,Q}(\delta))$ which is $0(\delta)$.
The collision between $\Lambda_{P,Q}^{\delta}\equiv \chi^{\delta}_{P,Q}\otimes\bar{\chi}^{\delta}_{P,Q}$
and the operator $\Phi^{\delta}_{-P,Q}$ (or $\Phi^{\delta}_{P,-Q}$)
which enters in the correlation function and is produced in the  OPE
\begin{equation}
 \Phi_{\alpha_a}(x,\bar{x})\Phi_{\alpha_b}(0)=\frac{C(\alpha_a,\alpha_b,\alpha_{-P,Q}(\delta))}
{|x|^{2\Delta_{\alpha_a}+2\Delta_{\alpha_b}-
2\Delta_{-P,Q}(\delta)}}\left(\Phi_{-P,Q}^{\delta}+\dots\right),
\label{reg_ter}
\end{equation}
cancels instead the remaining $\delta^{-1}$ singularity. 
Consider for instance the mixing between the fields $ \Lambda_{P,Q}^{\delta}$ and $\Phi^{\delta}_{-P,Q}$.
The formula \eqref{logza1} suggests to define a pair of fields $(\mathcal{C},\mathcal{D})$
\begin{eqnarray}
\label{log_field1}
\mathcal{D}&\equiv&\frac{1}{\sqrt{\delta}}\left(\Phi_{-P,Q}^{\delta}+\frac{i\Lambda_{P,Q}^{\delta}}{\delta \lambda_{P,Q}B_{P,Q}}\right)
\\
\label{log_field2}
\mathcal{C}&\equiv&-\left(PQ -\Delta_{-P,Q}(\delta)+\Delta_{P,Q}(\delta) \right)\frac{\Phi^{\delta}_{-P,Q}}{\sqrt{\delta}}
\end{eqnarray}
such that in the limit $\delta\to 0$ one has the finite correlation functions 
\begin{eqnarray}
\label{log_pair1}
\langle \mathcal{D}(x,\bar{x})\mathcal{D}(0)\rangle &=&\frac{-2a^{(1)}_{P,Q}(\ln{|x|^2}+O(1))}{|x|^{4(\Delta_{PQ}+ PQ)}}  \\
\label{log_pair2}
\langle \mathcal{C}(x,\bar{x})\mathcal{D}(0)\rangle &=&\frac{a^{(1)}_{P,Q}}{|x|^{4(\Delta_{PQ}+ PQ)}}  \\
\label{log_pair3}
\langle \mathcal{C}(x,\bar{x})\mathcal{C}(0)\rangle&=&0
\end{eqnarray}
The fields $\mathcal{C},\mathcal{D}$ are called a logarithmic pair, see for example \cite{Ca_rev}. Similar
conclusions in the chiral case have been reached in \cite{VaJaSa}, see in particular eq. (1.22). 

\section{Appearance of logarithms for rational central charges}
In the previous section we have shown
 that logarithms in the CG integrals  (\ref{exp_int}) are necessarily generated by singular conformal blocks.  
The results we have presented so far 
are based on the assumption that  $\beta^2\not\in\mathbb Q$.  
Indeed, when $\beta$ (and therefore the central charge $c$) takes the value
\begin{equation}
\beta=\beta_c(p,p')\equiv\sqrt{\frac{p'}{p}};\quad c(p',p)=1-6\frac{(p-p')^2}{p p'},
\label{scc}
\end{equation}
with $p$ and $p'$  two coprime positive integers, one has to take into account that 
\begin{equation}
p \beta_c(p,p')-p' \beta_c(p,p')^{-1}=0.
\label{spcco}
\end{equation} 
Accordingly, the charges $\alpha_{r,s}$ satisfy the additional symmetry 
\begin{equation}
\alpha_{r,s}=\alpha_{r\pm p,s\pm p'} 
\label{nsym}
\end{equation}
which implies new identifications between operators. 
The minimal models $\mathcal{M}_{p,p'}$ are CFT with central charge $c(p,p')$ 
which are constructed from the finite set of  $\frac{1}{2}(p-1)(p'-1)$ irreducible representations 
$\Phi_{r,s}$, $1\leq r\leq p-1$ and $1\leq s\leq p'-1$.  The algebraic structure of extensions of minimal models, 
which contain in general indecomposable representations has been intensively studied, see for instance \cite{PeRaZu,FeGaSeTi} 
and references therein. 
Here we are interested in obtaining local correlation functions 
for theories with central charge (\ref{scc}) 
by taking the limits $\beta\to \beta_c$  of integrals of 
type (\ref{exp_int}), associated to a set of external charges $\alpha_i$ obeying the condition (\ref{sum_12}). 

As we previously showed, the presence of zeros and poles in the structures constants, 
as well as the singularities in the conformal block, depend upon
the position of the points $[i,j]$, representing the internal charges $\hat{\alpha}_{ij}=\alpha_1+\alpha_2+i\beta-j\beta^{-1}$ on the grid of Sec.4.
In order to classify every type of singularity entering in the correlation function \eqref{exp_int}, we can use 
the expansion
\begin{equation}
p\beta-p'\beta^{-1}=2 p(\beta-\beta_c)+O\left((\beta-\beta_c)^2\right),
\end{equation}
to define new charges $\tilde{\alpha}_{i}$ such as
\begin{eqnarray}
\alpha_{i}&=& \underbrace{\alpha_i+\frac{k p}{4}\beta-\frac{k p'}{4}\beta^{-1}-\frac{k p}{2}
(\beta-\beta_c)}_{\tilde{\alpha}_{i}}+O\left((\beta-\beta_c)^2\right)\quad i=1,2\nonumber \\
\alpha_{i}&=& \underbrace{\alpha_i-\frac{k p}{4}\beta+\frac{k p'}{4}\beta^{-1}
+\frac{k p}{2}(\beta-\beta_c)}_{\tilde{\alpha}_{i}}+O\left((\beta-\beta_c)^2\right) \quad i=3,4
\label{ncr}
\end{eqnarray}
with $k$ a positive integer. 
In this way the correlation function (\ref{exp_int}) with given initial external charges $\alpha_i$ 
can be also written in terms of 
an integral with the same number of screening $n$ and $m$ 
but with new external charges $\tilde{\alpha_{i}}=\alpha_i+O\left((\beta-\beta_c)^2\right)$ and
with new values of $r,s$: $r\to r- k p, s\to s-k p'$.
The grid of points representing the internal channels is therefore obtained by translating
the initial grid by a multiple $k$ of $(p,p')$
\begin{equation}
\hat{\alpha}_{i,j}=\alpha_{r-kp -2i,s-k p'-2j}+k p(\beta-\beta_c).
\end{equation}
Note that  the distance from the central charge 
(\ref{scc}) play the role of a small regularization parameter $\delta$
\begin{equation}
\delta=-k p(\beta-\beta_c).
\label{delta_c}
\end{equation}
In the case of special values of $\beta$, we can compare the order of the different contribution of the integral using the 
\eqref{logza1}, (\ref{canc_sing}) and (\ref{canc_sing0}) applied to any translated grid, in particular when the translated 
grid intersects different quadrants. It is important to stress that, in the limit $\beta\to \beta_c(p,p')$,
the new symmetry (\ref{spcco}) affect the analysis of the zeros of the function (\ref{deg_con}) and of the function
$\Upsilon_{\beta}(x)$.
Therefore the conditions (\ref{ij_deg_sing}) and the (\ref{infies1})-(\ref{zeros2}) fail to take into account new zeros.
Moreover, in the limit $\beta\to \beta_c(p,p')$, the norm $B_{PQ}$, see (\ref{deg_con}), can vanish due to the (\ref{spcco}).
For these reasons, differently from the case $\beta^2 \notin \mathbb{Q}$, we did not find a simple graphical rule 
to determine the order of contribution of each channel.  Nevertheless, 
the analysis can be done case by case and we will show some concrete examples below.

Finally, note that for $\beta^2\not\in\mathbb Q$, we have regularized 
the expressions by introducing a parameter 
 $\delta$ which does not depend on $\beta$.  Here, instead,   we consider $\delta=O(\beta-\beta_c)$ and we study the limit $\beta\to \beta_c$.    
The parameters $a_{P,Q}^{(1)}$ and $a_{P,Q}^{(2)}$, which are the prefactors of term  $\log{|x|^2}$, are given by
\begin{equation}
a_{P,Q}^{(1)}=2P \beta_c\quad a_{P,Q}^{(2)}=-2 Q \beta_c. 
\label{bspc}
\end{equation}

\subsection{$c=-2$, $\beta_c^2=1/2$}

The theories with $c=-2$  occupy a very particular place as paradigms of LCFTs. Indeed,
 local $c=-2$ theories  can be constructed as models of free simplectic fermions and correlation functions can be
 computed exactly, see for instance  \cite{Salog,GaKa,Ka}. The $c=-2$ models have been intensively studied in
 polymer theory as they describe the dense phase of the $n\rightarrow 0$ of the $O(n)$ model known to  capture
 the critical properties of dense polymers and spanning trees\cite{DuLERW,DuSa3,Maj}.  In \cite{Salog}, 
 the CG formalism has been used to determine local four-point correlation functions of leg operators.
 In particular the correlation function of four $\Phi_{2,1}$ operators with conformal
 weight $\Delta_{2,1}=-1/8$ has been shown to contain logarithmic terms. As a physically relevant example
 we discuss the behavior of correlation functions
\begin{equation}
 F^{l}(x)\equiv\langle \Phi_{2+2l,1}(0)\Phi_{2+2l,1}(1)\Phi_{2+2l,1}(x)\Phi_{2+2l,1}(\infty) \rangle.
\end{equation}
 which for $l\in\mathbb N$ is  the four-point function of the  $4l$ leg operator for dense polymers \cite{Salog}. By choosing 
  \begin{equation}
  \alpha_1=\alpha_2=\alpha_{3}=\alpha_{2+2l,1}\quad \alpha_4=\alpha_{-2-2l,-1}
  \label{extchargecm2}
  \end{equation} 
the function $F^{l}$ can be expressed by the CG integral with $n=2 l+1$ and $m=0$ screenings. 
In the following we define:
\begin{equation}
\delta=-2 \left(\beta-\sqrt{\frac{1}{2}}\right)
\end{equation}

\underline{Four-point function with l=0}

\begin{figure}[t]
\begin{center}
\begin{tikzpicture}[scale=0.6]
\draw (-5,10) node[right]{$X_{i,j}|F(x|c,\hat{\Delta}_{i,j},\{\Delta\}|^2$};
    \draw[->,thick] (-9,0) -- (9,0) node[below]{$x$};
    \draw[->,thick] (0,-9) --(0,9)node[left]{$y$};
\foreach \x in {-8,...,8}
                \draw[very thin, gray] (\x,-8)--(\x,8);
\foreach \y in {-8,...,8}
                \draw[very thin, gray] (-8,\y)--(8,\y);

\draw[->,dashed] (5,2)--(3,1);
\draw[->,dashed] (3,2)--(1,1);
\draw (3,1) node[circle,fill=black,minimum size=1pt,scale=0.5] {};
\draw (3,1) node[above]{{\tiny $\delta^0$}};

\draw (1,1) node[circle,fill=black,minimum size=1pt,scale=0.5] {};
\draw (1,1) node[above]{{\tiny $\delta^0$}};

\draw[->] (3,1)--(1,0);
\draw[->] (1,1)--(-1,0);

\draw (1,0) node[circle,fill=black,minimum size=1pt,scale=0.5] {};
\draw (1,0) node[above]{{\tiny $\delta^{0}$}};

\draw (-1,0) node[circle,fill=black,minimum size=1pt,scale=0.5] {};
\draw (-1,0) node[above]{{\tiny $\delta^{0}$}};

\draw[->] (-1,0)--(-3,-1);
\draw[->] (1,0)--(-1,-1);
\draw (-1,-1) node[circle,fill=black,minimum size=1pt,scale=0.5] {};
\draw (-1,-1) node[above]{{\tiny $\delta^{0}$}};

\draw (-3,-1) node[circle,fill=black,minimum size=1pt,scale=0.5] {};
\draw (-3,-1) node[above]{{\tiny $\delta^{0}$}};
\draw[->,dashed] (-3,-1)--(-5,-2);
\draw[->,dashed] (-1,-1)--(-3,-2);
\end{tikzpicture}
\end{center}
\caption{The two internal channels $[0,0]$ and $[1,0]$ and the points obtained by  translation of the vector $(2,1)$. The
analysis of Sec.~4 can be carried out analogously with the \textit{caveat} that new sources of zeros and poles may
arise at $\beta^2$ rational in the structure constants.}
\label{l0cm2}
\end{figure}
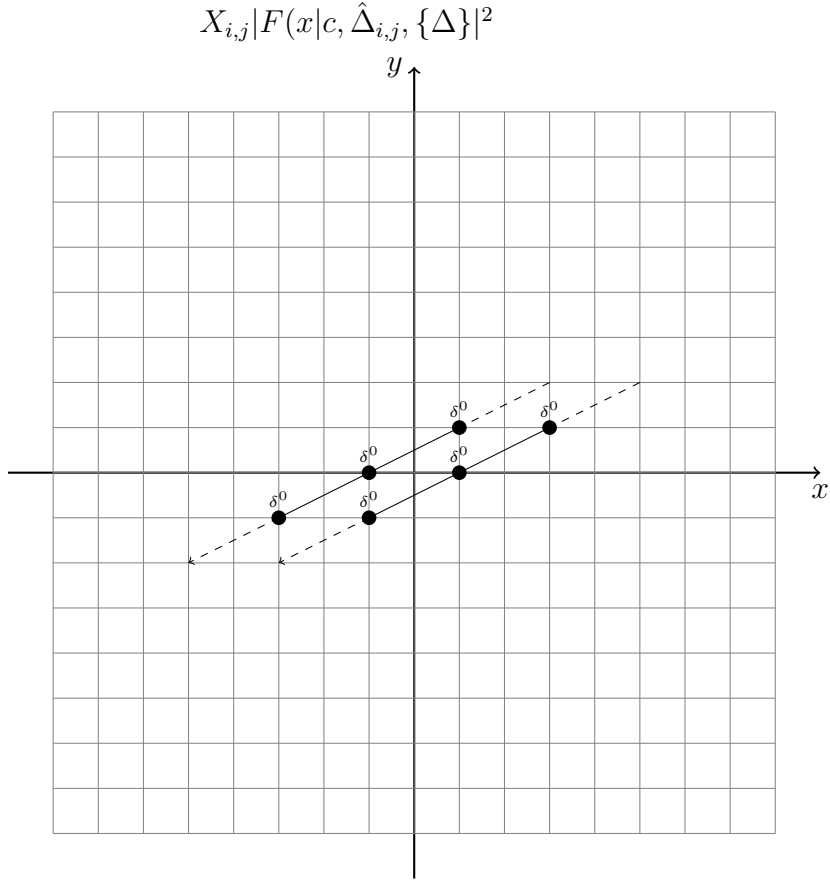

The case $l=0$ is particularly simple as it corresponds to the one-screening case, i.e.
it has only two internal fusion channels, see Fig. \ref{l0cm2}.  As can be seen by translating the grid, these two fusion channels
can be related to the fields $\Phi_{1,0}$ and $\Phi_{-1,0}$ with the same conformal dimension.
The (\ref{canc_sing}) tells us that the coefficients $X_{1,0}$ and $X_{-1,0}$ are of the same order in
$\delta$ and in particular that  $X_{1,0}=-X_{-1,0}$. The  function $F^{0}(x)$ takes therefore
the form (\ref{cg_ms}). This was pointed out in detail in \cite{Gu}, where the conformal block
corresponding to the case $l=0$  was shown to satisfy  a degenerate hypergeometric differential equation.
Note that the divergence $\delta^{-1}$ of the coefficients $X_{i,j}$ associate to points on the axis is here canceled by
the vanishing of the global normalization $A_\beta$ in \eqref{Zam_structure} for $\beta=\sqrt{1/2}$.   

By choosing the normalization factor $\mathcal{N}=O(\delta^{-1})$, the CG integral gives a finite $F^{0}$.
In order to familiarize with our notations, it may be useful to write explicitly  the result for  $l=0$ by setting 
the (\ref{extchargecm2}) into the equations (\ref{F00},\ref{F10}) and (\ref{X00},\ref{X10}) and take
the limit $\beta \to \sqrt{1/2}$.
One obtains
\begin{align}
 &G^{(0)}_{0,0}(x)=G^{(0)}_{1,0}(x)=\; _2F_1(\frac{1}{2},\frac{1}{2},1,x) \nonumber \\
&G^{(1)}_{0,0}=\sqrt{2}(\beta-\sqrt{1/2})\sum_{k=1}^{\infty} \frac{[1/2]_k^2}{(k!)^2}
\left(4\psi(k+1/2)-4\psi(1/2)-2\psi(k+1)+2\psi(1)\right)\nonumber \\
&G^{(1)}_{1,0}=\sqrt{2}(\beta-\sqrt{1/2})\sum_{k=1}^{\infty} \frac{[1/2]_k^2}{(k!)^2}
\left(2\psi(k+1)-2\psi(1)\right) \nonumber \\
&X^{(0)}_{0,0}=-X^{(0)}_{(1,0)}=-\frac{\pi}{2\sqrt{2} (\beta-\sqrt{1/2})}+\frac{\pi}{2}\left(1-4\gamma_{E}-4\psi(1/2)\right)+O((\beta-\sqrt{1/2})^2)\nonumber\\
&X_{1,0}=\frac{\pi}{2\sqrt{2} (\beta-\sqrt{1/2})}-\frac{\pi}{2}\left(1-4\gamma_{E}-4\psi(1/2)\right)+O((\beta-\sqrt{1/2})^2),
\end{align}
 where $\gamma_{E}$ is the Euler constant and $\psi(z)$ the digamma function. 
Using the above results in the (\ref{cg_1s}) one obtains
\begin{equation}
 F^0(x)=|x|^{1/2}|x-1|^{1/2}\left[\sqrt{2}\ln{|x|^2} |_2F_1(1/2,1/2,1,x)|^2
+\left(_2F_1(1/2,1/2,1,x) M(\bar{x})+c.c.\right)\right],
\label{cm2l0}
\end{equation}
with
\begin{equation}
 M(x)=\sum_{k=1}^{\infty} \frac{[1/2]_k^2}{(k!)^2} \left(\psi(k+1)-\psi(1)-\psi(k+1/2)-\psi(1/2)\right).
\end{equation}
Note that the factor of the logarithmic term corresponds to $a^{(1)}_{1,0}$ of the equation (\ref{bspc}).
The correlation function (\ref{cm2l0}) was also studied in \cite{Gu} and discussed in \cite{GaKa,HaYa}. 

\underline{Four-point function with l=1/2}
\begin{figure}
\begin{center} 
  \begin{tikzpicture}[scale=0.6]
\draw (-5,10) node[right]{$X_{i,j}|F(x|c,\hat{\Delta}_{i,j},\{\Delta\}|^2$};
    \draw[->,thick] (-9,0) -- (9,0) node[below]{$x$};
    \draw[->,thick] (0,-9) --(0,9)node[left]{$y$};
\foreach \x in {-8,...,8}
                \draw[very thin, gray] (\x,-8)--(\x,8);
\foreach \y in {-8,...,8}
                \draw[very thin, gray] (-8,\y)--(8,\y);
\draw[->,dashed] (7,2)--(5,1);
\draw[->,dashed] (5,2)--(3,1);
\draw[->,dashed] (3,2)--(1,1);

\draw (5,1) node[circle,fill=black,minimum size=1pt,scale=0.5] {};
\draw (5,1) node[above]{{\tiny $\delta^0$}};
                
\draw (3,1) node[circle,fill=black,minimum size=1pt,scale=0.5] {};
\draw (3,1) node[above]{{\tiny $\delta^0$}};

\draw (1,1) node[circle,fill=black,minimum size=1pt,scale=0.5] {};
\draw (1,1) node[above]{{\tiny $\delta^0$}};

\draw[->] (5,1)--(3,0);
\draw[->] (3,1)--(1,0);
\draw[->] (1,1)--(-1,0);

\draw (3,0) node[circle,fill=black,minimum size=1pt,scale=0.5] {};
\draw (3,0) node[above]{{\tiny $\delta^{0}$}};

\draw (1,0) node[circle,fill=black,minimum size=1pt,scale=0.5] {};
\draw (1,0) node[above]{{\tiny $\delta^{0}$}};

\draw (-1,0) node[circle,fill=black,minimum size=1pt,scale=0.5] {};
\draw (-1,0) node[above]{{\tiny $\delta^{0}$}};

\draw[->] (3,0)--(1,-1);
\draw[->] (1,0)--(-1,-1);
\draw[->] (-1,0)--(-3,-1);

\draw (-3,-1) node[circle,fill=black,minimum size=1pt,scale=0.5] {};
\draw (-3,-1) node[above]{{\tiny $\delta^{0}$}};

\draw (-1,-1) node[circle,fill=black,minimum size=1pt,scale=0.5] {};
\draw (-1,-1) node[above]{{\tiny $\delta^{0}$}};

\draw (1,-1) node[circle,fill=black,minimum size=1pt,scale=0.5] {};
\draw (1,-1) node[above]{{\tiny $\delta^{0}$}};

\draw[->,dashed] (1,-1)--(-1,-2);
\draw[->,dashed] (-1,-1)--(-3,-2);
\draw[->,dashed] (-3,-1)--(-5,-2);
\end{tikzpicture}
\end{center}
\caption{This figure shows the points associated to the three internal channels in the case  $l=1/2$ and 
the corresponding translated grids.}
\label{l1cm2}
\end{figure}
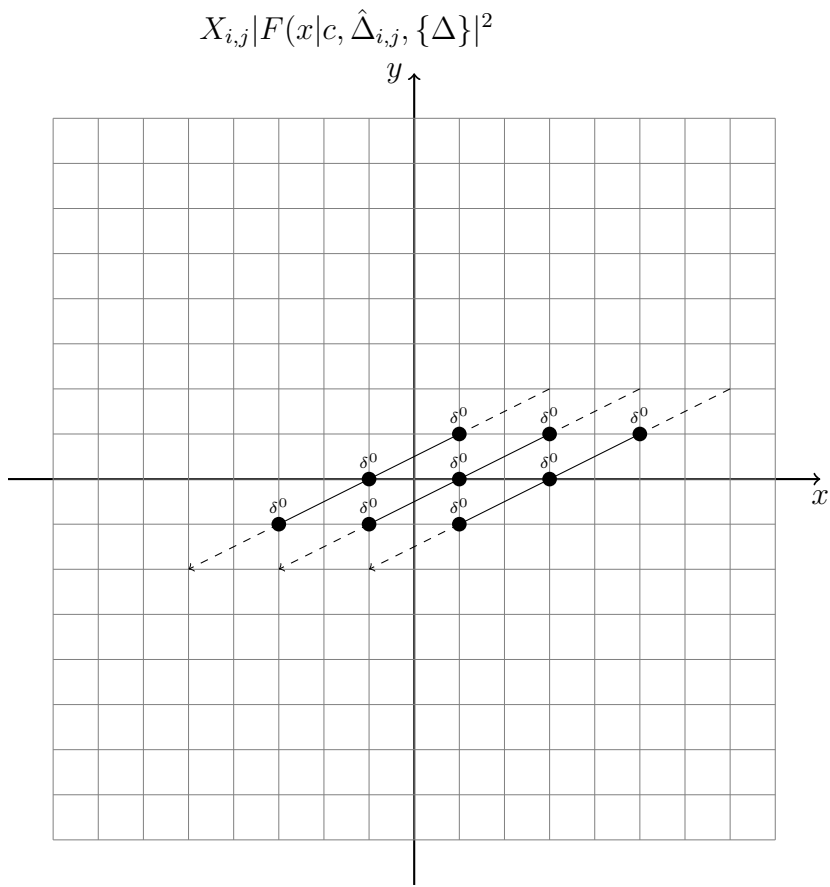

Analogously to the case $l=0$, by fixing $\mathcal{N}=O(\delta^{-1})$, the  contributions of the channels
$\Phi_{3,1}$ and $\Phi_{1,1}$  sum to produce a finite result.   As for the remaining third channel $\Phi_{5,1}=\Phi_{1,-1}$,
we can determine its order in $\delta$, using (\ref{logza1}) and compare it with the one of the field $\Phi_{3,1}=\Phi_{-1,-1}$. 
For external charges as in (\ref{extchargecm2}) with $l=1$, one obtains
\begin{equation}
S_{1,1}(\alpha_1,\alpha_2,\alpha_3,\alpha_4)= O(\delta^2).
\end{equation}
The ratio of the coefficients $X_{1,0}$ and $X_{0,0}$ is therefore given by
\begin{equation}
\frac{X_{1,0}}{X_{0,0}}=
\frac{C(\alpha_{1},\alpha_2,\alpha_{1,1}+\delta)C(\alpha_{1},\alpha_2,\alpha_{1,1}+\delta)}
{C(\alpha_{1},\alpha_2,\alpha_{-1,1}+\delta)C(\alpha_{1},\alpha_2,\alpha_{-1,1}+\delta)}\sim
\delta^2\frac{\lambda_{1,1}^2}{S_{1,1}(\{\alpha_i\})^2}=O(\delta^{-2}).
\end{equation}
Therefore we can conclude that in the limit $\delta\to 0$, i.e. $c\to -2$,  $F^{1}(x)$ is given 
by the contributions of the two channels $\Phi_{3,1}$ and $\Phi_{1,1}$, which produce 
a logarithmic term  $\propto \frac{1}{\sqrt{2}} \ln|x|^2$. 
Notice that we could have considered the correlation function involving more general external 
charges $\alpha_i$ (for instance $\alpha_1\neq \alpha_2\neq \alpha_3$) 
satisfying the (\ref{sum_12}) with $n=2,m=0$ and $(r=5,s=1)$. 
In this case the coefficient $S_{1,1}(\{\alpha_i \})$ 
vanishes linearly in $\delta$, $S_{1,1}(\{\alpha_i \})\sim \delta$. 
This would imply that the contribution of the channel $\Phi_{5,1}$ 
would be of  the same order of the other two. 
As the the dominant term of the channels $\Phi_{3,1}$ and $\Phi_{1,1}$ cancels, 
this would imply that,  in the limit $\beta\to 1/\sqrt{2}$, the correlation $F^{1}(x)$ is given by the channel 
$\Phi_{5,1}$ alone, without logarithmic terms.

 \underline{Four-point function with l=1}
\\
 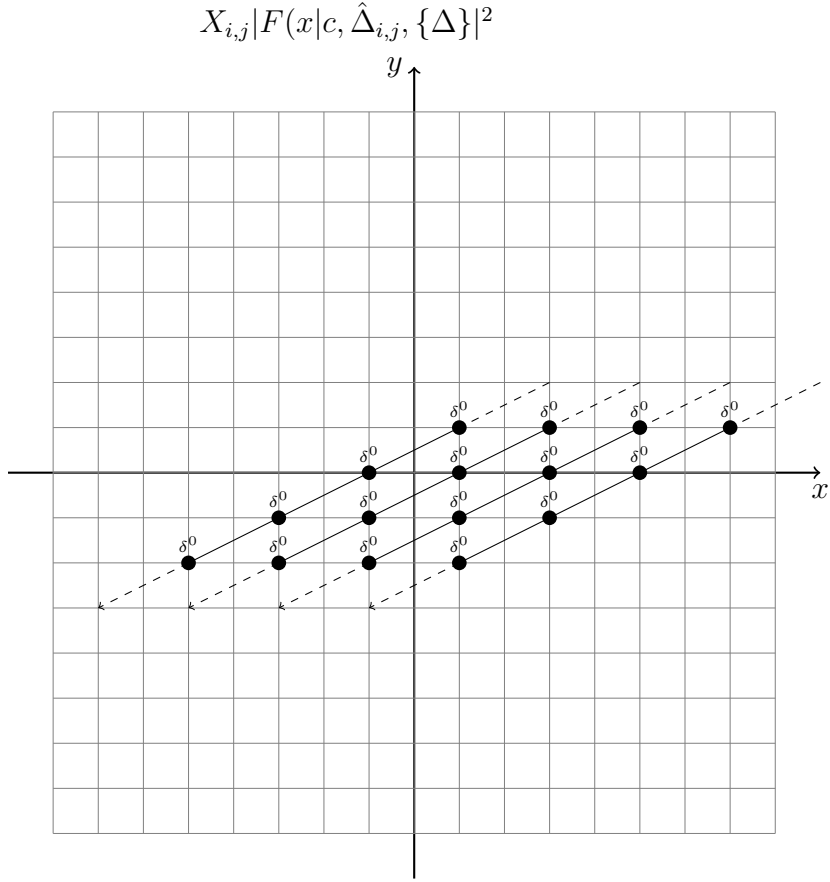
\begin{figure}[t]
 \begin{center} 
  \begin{tikzpicture}[scale=0.6]
\draw (-5,10) node[right]{$X_{i,j}|F(x|c,\hat{\Delta}_{i,j},\{\Delta\}|^2$};
    \draw[->,thick] (-9,0) -- (9,0) node[below]{$x$};
    \draw[->,thick] (0,-9) --(0,9)node[left]{$y$};
\foreach \x in {-8,...,8}
                \draw[very thin, gray] (\x,-8)--(\x,8);
\foreach \y in {-8,...,8}
                \draw[very thin, gray] (-8,\y)--(8,\y);
\draw[->,dashed] (9,2)--(7,1);
\draw[->,dashed] (7,2)--(5,1);
\draw[->,dashed] (5,2)--(3,1);
\draw[->,dashed] (3,2)--(1,1);
\draw (7,1) node[circle,fill=black,minimum size=1pt,scale=0.5] {};
\draw (7,1) node[above]{{\tiny $\delta^0$}};

\draw (5,1) node[circle,fill=black,minimum size=1pt,scale=0.5] {};
\draw (5,1) node[above]{{\tiny $\delta^0$}};
                
\draw (3,1) node[circle,fill=black,minimum size=1pt,scale=0.5] {};
\draw (3,1) node[above]{{\tiny $\delta^0$}};

\draw (1,1) node[circle,fill=black,minimum size=1pt,scale=0.5] {};
\draw (1,1) node[above]{{\tiny $\delta^0$}};

\draw[->] (7,1)--(5,0);
\draw[->] (5,1)--(3,0);
\draw[->] (3,1)--(1,0);
\draw[->] (1,1)--(-1,0);

\draw (5,0) node[circle,fill=black,minimum size=1pt,scale=0.5] {};
\draw (5,0) node[above]{{\tiny $\delta^0$}};

\draw (3,0) node[circle,fill=black,minimum size=1pt,scale=0.5] {};
\draw (3,0) node[above]{{\tiny $\delta^{0}$}};

\draw (1,0) node[circle,fill=black,minimum size=1pt,scale=0.5] {};
\draw (1,0) node[above]{{\tiny $\delta^{0}$}};

\draw (-1,0) node[circle,fill=black,minimum size=1pt,scale=0.5] {};
\draw (-1,0) node[above]{{\tiny $\delta^{0}$}};

\draw[->] (5,0)--(3,-1);
\draw[->] (3,0)--(1,-1);
\draw[->] (1,0)--(-1,-1);
\draw[->] (-1,0)--(-3,-1);

\draw (3,-1) node[circle,fill=black,minimum size=1pt,scale=0.5] {};
\draw (3,-1) node[above]{{\tiny $\delta^{0}$}};

\draw (-3,-1) node[circle,fill=black,minimum size=1pt,scale=0.5] {};
\draw (-3,-1) node[above]{{\tiny $\delta^{0}$}};

\draw (-1,-1) node[circle,fill=black,minimum size=1pt,scale=0.5] {};
\draw (-1,-1) node[above]{{\tiny $\delta^{0}$}};

\draw (1,-1) node[circle,fill=black,minimum size=1pt,scale=0.5] {};
\draw (1,-1) node[above]{{\tiny $\delta^{0}$}};

\draw[->] (3,-1)--(1,-2);
\draw[->] (1,-1)--(-1,-2);
\draw[->] (-1,-1)--(-3,-2);
\draw[->] (-3,-1)--(-5,-2);
\draw (1,-2) node[circle,fill=black,minimum size=1pt,scale=0.5] {};
\draw (1,-2) node[above]{{\tiny $\delta^{0}$}};

\draw (-5,-2) node[circle,fill=black,minimum size=1pt,scale=0.5] {};
\draw (-5,-2) node[above]{{\tiny $\delta^{0}$}};

\draw (-3,-2) node[circle,fill=black,minimum size=1pt,scale=0.5] {};
\draw (-3,-2) node[above]{{\tiny $\delta^{0}$}};

\draw (-1,-2) node[circle,fill=black,minimum size=1pt,scale=0.5] {};
\draw (-1,-2) node[above]{{\tiny $\delta^{0}$}};

\draw[->,dashed] (1,-2)--(-1,-3);
\draw[->,dashed] (-1,-2)--(-3,-3);
\draw[->,dashed] (-3,-2)--(-5,-3);
\draw[->,dashed] (-5,-2)--(-7,-3);
\end{tikzpicture}
\end{center}
\caption{Grids of internal channels for the correlation function $F^{2}(x)$.}
\label{l2cm2}
\end{figure}
We consider finally the case $l=2$, see Fig. \ref{l2cm2} 
In order to evaluate  the order of its contribution, 
we compare the channel $\Phi_{7,1}\equiv\Phi_{1,-2}$ with the channel  $\Phi_{5,1}\equiv\Phi_{-1,-2}$ 
by using the (\ref{logza1}). 
Assuming the values of the external charges as in (\ref{extchargecm2}) with $l=2$,one can verify that
\begin{equation}
S_{12}(\alpha_1,\alpha_2,\alpha_3,\alpha_4)\neq 0.
\end{equation} 
This means that the conformal block associated to $\Phi_{5,1}$ 
has a singularity $O(\delta^{-1})$ at level 2. 
As we have said before, this singularity is canceled by the collision with the field 
$\Phi_{7,1}$. Taking into account that $S_{1,1}(\{\alpha\})$ is of order $\delta^{-1}$ we can conclude that  $F^{2}(x)$
has two logarithmic terms coming from the pairs $\Phi_{1,1}$, $\Phi_{3,1}$ and $\Phi_{5,1}$, $\Phi_{7,1}$ with prefactors 
 $a_{1,0}^{(1)}=\sqrt{2}$ and $a_{2,1}^{(1)}=2\sqrt{2}$.

The correlation function $F^{l}(x)$ for general $l$, 
as well as other correlation functions, 
such as for instance the four point correlation function $\langle \Phi_{1, l}(0)\Phi_{1,l}(1)\Phi_{1,l}(x)\Phi_{1,l}(\infty) \rangle$
of $L=4l+2$ operators \cite{Salog} can be studied case by case with the same techniques without any additional 
technical complication.
\subsection{$c=0$, $\beta^2=2/3$}

CFTs at central charge $c=0$ are of particular interest as 
they play a crucial role in the study of critical systems with
quenched disorder or in the description of dilute self-avoiding walk and critical percolation theory.   
For $\beta=\sqrt{2/3}$, the vanishing of the dimension $\Delta_{2,1}$ 
is at the origin of the so-called $c=0$ catastrophe \cite{GuLu1, GuLu2}. Such
terminology refers to the fact that OPE
\begin{equation}
\Phi_{\alpha}(x,\bar{x})
\Phi_{\alpha}(0)=\frac{C_{\Delta_{\alpha}}}{|z|^{4\Delta_{\alpha}}}
\left(1+x^2\frac{2\Delta_{\alpha}}{c}T(0)\right)\left(1+\bar{x}^2\frac{2\Delta_{\alpha}}{c}\bar{T}(0)\right), 
\label{id_fusion}
\end{equation}
is singular in the $c\rightarrow 0$ limit. The $c=0$ catastrophe can be rephrased in the following terms:
if we consider the identity as internal channel with dimension $\Delta_p=0$ in \eqref{rec_conf} then the
conformal block $F(x|c=0,\Delta_p=0,\{\Delta\})$, with $\Delta_1=\Delta_2=\Delta_{\alpha}$
does not have a singularity at level 1 since $R_{1,1}(\alpha,\alpha)=0$
but is divergent at the  level two because  $R_{2,1}(\alpha,\alpha)\neq 0$ for $\Delta_{\alpha}\neq 0$. 
As we have already seen in the general case, 
the singularities are  $O(c^{-2})$ and  $O(c^{-1})$, and they 
are canceled inside correlation functions by the vanishing 
of the structure constant $C_{\Delta_{\alpha}}$ which is $O(c)$ and by 
the collision with one of the operators $\Phi_{-2,1}$ and $\Phi_{2,-1}$.

It may be useful to compare  the expression (\ref{id_fusion}) with the expression (\ref{ope_s}).
Let us define $\chi^{\delta}_{2,1}=(L_{-1}^2-\beta^2 L_{-2}) V_{1,1}^{\delta}$, where $V_{1,1}^{\delta}$ is the 
primary with charge $\alpha_{-1,-1}(\delta)=\alpha_{-1,-1}+\delta$ and  
\begin{equation}
\delta=-3(\beta-\sqrt{2/3}).
\label{delta_c_0}
\end{equation}
From the (\ref{nsym}) we have $\alpha_{2,1}=\alpha_{-1,-1}(\delta)+O(\delta^2)$. 
In the limit $\beta\to \sqrt{2/3}$, i.e. $c\to0$, the field $\chi_{2,1}^{\delta}(x)$ is 
therefore proportional to
the stress-energy tensor,
\begin{equation}
 \lim_{\delta\to 0} \chi_{2,1}^{\delta}=-2/3\;T
\end{equation}
which becomes a null-vector at level $2$ in the Verma module of the identity. Setting $T=0$ would imply to
consider a trivial theory that contains in its spectrum only  fields invariant under all conformal transformations, i.e. only
the identity field.  The coefficient
\begin{equation}
\frac{3}{2}\eta_{2,1}=\frac{3}{2}\frac{\langle \chi_{2,1}^{\delta}(\infty) V_{\alpha}(1) V_{\alpha}(0)\rangle}{\langle \chi_{2,1}^{\delta}(\infty) \chi_{2,1}^{\delta}(0) \rangle}
\sim \frac{1}{2}\frac{R_{2,1}(\alpha,\alpha)}{(\beta-\sqrt{2/3})(2\beta-\beta^{-1})B_{2,1}}\sim\frac{2\Delta_{\alpha}}{c}
\end{equation}
is consistent with the expansion (\ref{ope_s}) where the regularization parameter 
is played by $\delta$.

In order to make connection with previous results,
let us consider for instance the correlation function
\begin{equation}
\langle \Phi_{n+1,m+1}(0)\Phi_{n+1,m+1}(x) \Phi_{n+1,m+1}(1)\Phi_{-n-1,-m-1}(\infty)
\rangle
\label{cfco}
\end{equation}
of four operators with the same dimension $\Delta=\Delta_{n+1,m+1}$. 
The neutrality condition (\ref{ch_neu}) is satisfied with $n$ and $m$ screenings 
respectively of type $\beta$ and $\beta^{-1}$. 

Let us consider in the following some illustrative cases

\underline{$n=2, m=0$}

\begin{figure}
\begin{center}
\begin{tikzpicture}[scale=0.6]
\draw (-5,10) node[right]{$X_{i,j}|F(x|c,\hat{\Delta}_{i,j},\{\Delta\}|^2$};
    \draw[->,thick] (-9,0) -- (9,0) node[below]{$x$};
    \draw[->,thick] (0,-9) --(0,9)node[left]{$y$};
\foreach \x in {-8,...,8}
                \draw[very thin, gray] (\x,-8)--(\x,8);
\foreach \y in {-8,...,8}
                \draw[very thin, gray] (-8,\y)--(8,\y);

\draw[->,dashed] (8,3)--(5,1);
\draw[->,dashed] (6,3)--(3,1);
\draw[->,dashed] (4,3)--(1,1);

\draw (5,1) node[circle,fill=black,minimum size=1pt,scale=0.5] {};
\draw (5,1) node[above]{{\tiny $\delta^{-2}$}};

\draw (3,1) node[circle,fill=black,minimum size=1pt,scale=0.5] {};
\draw (3,1) node[above]{{\tiny $\delta^{-1}$}};

\draw (1,1) node[circle,fill=black,minimum size=1pt,scale=0.5] {};
\draw (1,1) node[above]{{\tiny $\delta^{-2}$}};

\draw[->] (5,1)--(2,-1);
\draw[->] (3,1)--(0,-1);
\draw[->] (1,1)--(-2,-1);

\draw (2,-1) node[circle,fill=black,minimum size=1pt,scale=0.5] {};
\draw (2,-1) node[above]{{\tiny $\delta^{-2}$}};

\draw (0,-1) node[circle,fill=black,minimum size=1pt,scale=0.5] {};
\draw (0,-1) node[above]{{\tiny $\delta^{-1}$}};

\draw (-2,-1) node[circle,fill=black,minimum size=1pt,scale=0.5] {};
\draw (-2,-1) node[above]{{\tiny $\delta^{-2}$}};

\draw[->] (5,1)--(2,-1);
\draw[->] (3,1)--(0,-1);
\draw[->] (1,1)--(-2,-1);

\draw (2,-1) node[circle,fill=black,minimum size=1pt,scale=0.5] {};
\draw (2,-1) node[above]{{\tiny $\delta^{-2}$}};

\draw (0,-1) node[circle,fill=black,minimum size=1pt,scale=0.5] {};
\draw (0,-1) node[above]{{\tiny $\delta^{-1}$}};

\draw (-2,-1) node[circle,fill=black,minimum size=1pt,scale=0.5] {};
\draw (-2,-1) node[above]{{\tiny $\delta^{-2}$}};

\draw[->] (2,-1)--(-1,-3);
\draw[->] (0,-1)--(-3,-3);
\draw[->] (-2,-1)--(-5,-3);

\draw (-1,-3) node[circle,fill=black,minimum size=1pt,scale=0.5] {};
\draw (-1,-3) node[above]{{\tiny $\delta^{-2}$}};

\draw (-3,-3) node[circle,fill=black,minimum size=1pt,scale=0.5] {};
\draw (-3,-3) node[above]{{\tiny $\delta^{-1}$}};

\draw (-5,-3) node[circle,fill=black,minimum size=1pt,scale=0.5] {};
\draw (-5,-3) node[above]{{\tiny $\delta^{-2}$}};

\draw[->,dashed] (-1,-3)--(-4,-5);
\draw[->,dashed] (-3,-3)--(-6,-5);
\draw[->,dashed] (-5,-3)--(-8,-5);
\end{tikzpicture}
\end{center}
\caption{We show an example of a correlation functions at $c=0$, with two screening charges. Notice that the case $n=1$ does
not contain any logarithm.}
\end{figure}
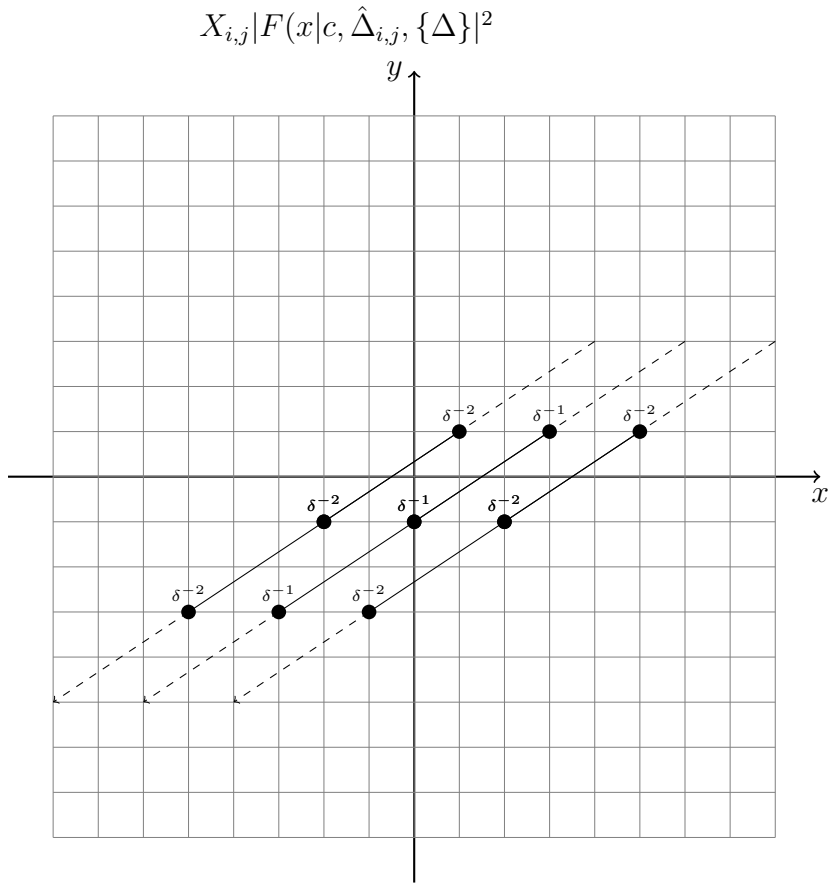

Using the analysis of the previous sections, 
one observes that the most singular terms are of order $\delta^{-2}$ and correspond
to the charges  $\alpha_{-2,-1}$ and $\alpha_{2,-1}$, associated respectively to 
the identity $\Phi_{1,1}$ and to the $\Phi_{5,1}$ fusion channels. 
The dominant contribution $\delta^{-2}$ is canceled in the sum, 
providing an $O(\delta^{-1})$ contribution with logarithms to the correlation function. 
Choosing a global normalization constant of $O(\delta)$, the correlation function (\ref{cfco}) 
with $n=2$ and $m=0$ shows a a logarithmic contribution
generated by  the colliding channels $\Phi_{1,1}$ and $\Phi_{5,1}$ and a non-logarithmic contribution from
the channel $\Phi_{3,1}$. The prefactor of the logarithmic terms  is
\begin{equation}
a_{2,1}^{(1)}=4\sqrt{6}.
\end{equation}
As we previously noticed, see (\ref{log_field1}, \ref{log_field2}), the value $a_{2,1}^{(1)}$ can be seen
as the mixing parameter 
between the \textit{scalar} field $\Phi_{5,1}(x,\bar{x})$ and the field $T(x) \bar{T}(\bar{x})$.
Let us comment on the fact that in the chiral $c=0$ CFT \cite{GuLu1,GuLu2} which describes 
boundary percolation, one is general interested in the mixing parameter $b_{\tiny\mbox{perco}}$, see \cite{Ri} for a recent
survey,   between the \textit{chiral} operator $T(x)$ and $\phi_{5,1}(x)$.
It is therefore not surprising that such parameter $b_{\tiny\mbox{perco}}=-5/8$,   numerically coincides with 
\begin{equation}
-\left(\left.\frac{d \beta_c}{d c}\right|_{c=0}~ 2 a_{2,1}^{(1)}\right)^{-1}=-\frac{5}{8}.
\end{equation} 
The factor $-\frac{d \beta_c}{d c}|_{c=0}$ takes into account that, in the definition of
the parameter $b_{{\tiny\mbox{perco}}}$ one uses the limit  $c\to0$ instead of $\beta\to \sqrt{2/3}$, while the factor $2$
comes from the fact that we are considering a non-chiral theory where the chiral and anti-chiral degrees of freedom are coupled
symmetrically.

\underline{$n=0,m=2$}

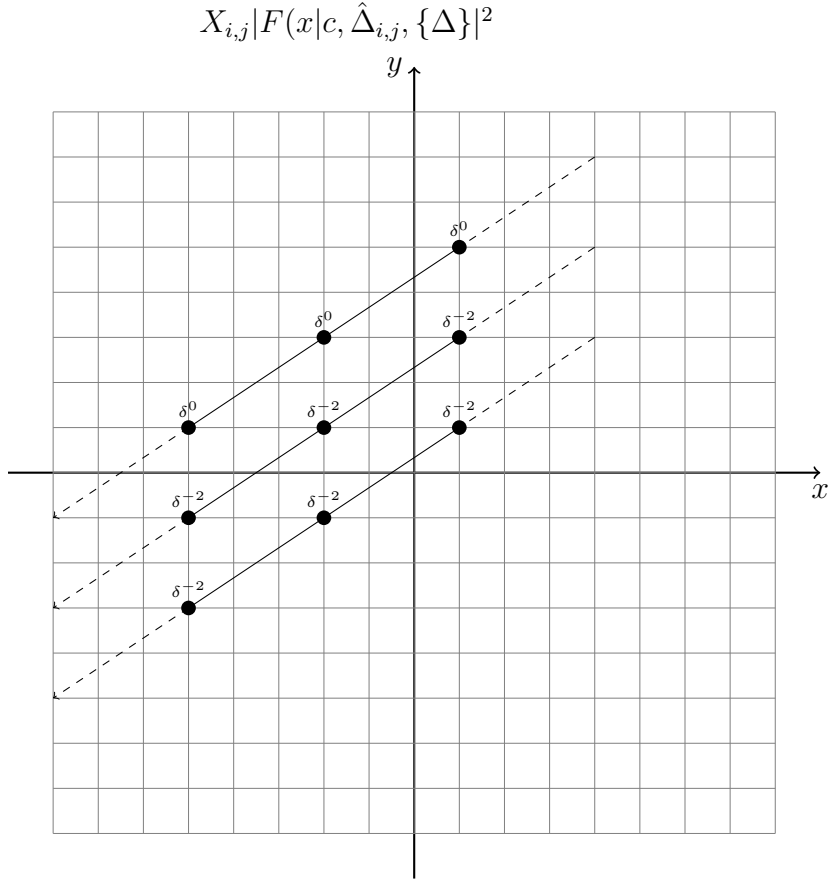
\begin{figure}[t]
\begin{center}
\begin{tikzpicture}[scale=0.6]
\draw (-5,10) node[right]{$X_{i,j}|F(x|c,\hat{\Delta}_{i,j},\{\Delta\}|^2$};
    \draw[->,thick] (-9,0) -- (9,0) node[below]{$x$};
    \draw[->,thick] (0,-9) --(0,9)node[left]{$y$};
\foreach \x in {-8,...,8}
                \draw[very thin, gray] (\x,-8)--(\x,8);
\foreach \y in {-8,...,8}
                \draw[very thin, gray] (-8,\y)--(8,\y);

\draw[->,dashed] (4,7)--(1,5);
\draw[->,dashed] (4,5)--(1,3);
\draw[->,dashed] (4,3)--(1,1);

\draw (1,5) node[circle,fill=black,minimum size=1pt,scale=0.5] {};
\draw (1,5) node[above]{{\tiny $\delta^{0}$}};

\draw (1,3) node[circle,fill=black,minimum size=1pt,scale=0.5] {};
\draw (1,3) node[above]{{\tiny $\delta^{-2}$}};

\draw (1,1) node[circle,fill=black,minimum size=1pt,scale=0.5] {};
\draw (1,1) node[above]{{\tiny $\delta^{-2}$}};

\draw[->] (1,5)--(-2,3);
\draw[->] (1,3)--(-2,1);
\draw[->] (1,1)--(-2,-1);

\draw (-2,3) node[circle,fill=black,minimum size=1pt,scale=0.5] {};
\draw (-2,3) node[above]{{\tiny $\delta^{0}$}};

\draw (-2,1) node[circle,fill=black,minimum size=1pt,scale=0.5] {};
\draw (-2,1) node[above]{{\tiny $\delta^{-2}$}};

\draw (-2,-1) node[circle,fill=black,minimum size=1pt,scale=0.5] {};
\draw (-2,-1) node[above]{{\tiny $\delta^{-2}$}};

\draw[->] (-2,3)--(-5,1);
\draw[->] (-2,1)--(-5,-1);
\draw[->] (-2,-1)--(-5,-3);

\draw (-5,1) node[circle,fill=black,minimum size=1pt,scale=0.5] {};
\draw (-5,1) node[above]{{\tiny $\delta^{0}$}};

\draw (-5,-1) node[circle,fill=black,minimum size=1pt,scale=0.5] {};
\draw (-5,-1) node[above]{{\tiny $\delta^{-2}$}};

\draw (-5,-3) node[circle,fill=black,minimum size=1pt,scale=0.5] {};
\draw (-5,-3) node[above]{{\tiny $\delta^{-2}$}};

\draw[->,dashed] (-5,1)--(-8,-1);
\draw[->,dashed] (-5,-1)--(-8,-3);
\draw[->,dashed] (-5,-3)--(-8,-5);

\end{tikzpicture}
  \end{center}
\caption{The figure shows a correlation function at $c=0$ with $m=2$ and $n=0$.}
\label{c0_m2}
\end{figure}

Using  $S_{2,1}(\{\alpha_i\})\neq 0$ and $S_{5,1}(\{\alpha_i\})=O(\delta^2)$ together with (\ref{logza1}), one obtains
the contributions shown in Fig. \ref{c0_m2}. Setting the global normalization constant $\mathcal{N}$ of order $\delta$,
the function (\ref{cfco}) for $n=0,m=2$ is given by the \eqref{cg_ms} with  colliding channels
$\Phi_{1,1}=\Phi_{-2,-1}$ and $\Phi_{1,3}=\Phi_{-2,1}$ with  
\begin{equation}
a_{2,1}^{(2)}=-3\sqrt{6}.
\end{equation}
The above factor $a_{2,1}^{(2)}$ is formally related to the  parameter $b_{{\tiny\mbox{poly}}}=5/6$ 
\cite{GuLu1,GuLu2} which is defined as the mixing factor between the chiral fields $\phi_{1,3}(x)$ and $T(x)$ in the 
$c=0$  theory describing dilute polymers with a boundary. Analogously to the previous case, we have
\begin{equation}
-\left(\left.\frac{d \beta_c}{d c}\right|_{c=0}~ 2 a_{2,1}^{(2)}\right)^{-1}=\frac{5}{6}.
\end{equation} 
Although their numerical values are simply related, we stress again that the two mixing factors $b_{{\tiny\mbox{poly}}}$ and $a_{2,1}^{(2)}$ corresponds to different physical
situations: in the first case the LCFT is chiral and the chiral field $\phi_{1,3}$ is the logarithmic partner of the
stress-energy tensor while in the second case, the scalar field $\Phi_{3,1}(x,\bar{x})$
is the logarithmic partner of the operator $T\bar{T}$.  

\underline{$n=2, m=1$}

\begin{figure}
\begin{center}
\begin{tikzpicture}[scale=0.6]
\draw (-5,10) node[right]{$X_{i,j}|F(x|c,\hat{\Delta}_{i,j},\{\Delta\}|^2$};
    \draw[->,thick] (-9,0) -- (9,0) node[below]{$x$};
    \draw[->,thick] (0,-9) --(0,9)node[left]{$y$};
\foreach \x in {-8,...,8}
                \draw[very thin, gray] (\x,-8)--(\x,8);
\foreach \y in {-8,...,8}
                \draw[very thin, gray] (-8,\y)--(8,\y);


\draw (5,3) node[circle,fill=red,minimum size=1pt,scale=0.5] {};

\draw (5,1) node[circle,fill=red,minimum size=1pt,scale=0.5] {};

\draw (3,3) node[circle,fill=red,minimum size=1pt,scale=0.5] {};

\draw (3,1) node[circle,fill=red,minimum size=1pt,scale=0.5] {};

\draw (1,3) node[circle,fill=red,minimum size=1pt,scale=0.5] {};

\draw (1,1) node[circle,fill=red,minimum size=1pt,scale=0.5] {};

\draw[->] (5,3)--(2,1);
\draw[->] (5,1)--(2,-1);
\draw[->] (3,3)--(0,1);
\draw[->] (3,1)--(0,-1);
\draw[->] (1,3)--(-2,1);
\draw[->] (1,1)--(-2,-1);

\draw (2,1) node[circle,fill=black,minimum size=1pt,scale=0.5] {};
\draw (2,1) node[above]{{\tiny $\delta^{-2}$}};

\draw (2,-1) node[circle,fill=black,minimum size=1pt,scale=0.5] {};
\draw (2,-1) node[above]{{\tiny $\delta^{-2}$}};

\draw (0,1) node[circle,fill=black,minimum size=1pt,scale=0.5] {};
\draw (0,1) node[above]{{\tiny $\delta^{-1}$}};

\draw (0,-1) node[circle,fill=black,minimum size=1pt,scale=0.5] {};
\draw (0,-1) node[above]{{\tiny $\delta^{-1}$}};

\draw (-2,1) node[circle,fill=black,minimum size=1pt,scale=0.5] {};
\draw (-2,1) node[above]{{\tiny $\delta^{-2}$}};

\draw (-2,-1) node[circle,fill=black,minimum size=1pt,scale=0.5] {};
\draw (-2,-1) node[above]{{\tiny $\delta^{-2}$}};

\end{tikzpicture}
  \end{center}
\caption{The figure shows a  correlation function computed at $c=0$ in which both the operators $\Phi_{1,3}$ and
$\Phi_{5,1}$ enters in collision with the null-field $T\bar{T}$.
The others two channels can be shown to give a subleading contribution.}
  \label{c0_n3}  
\end{figure}
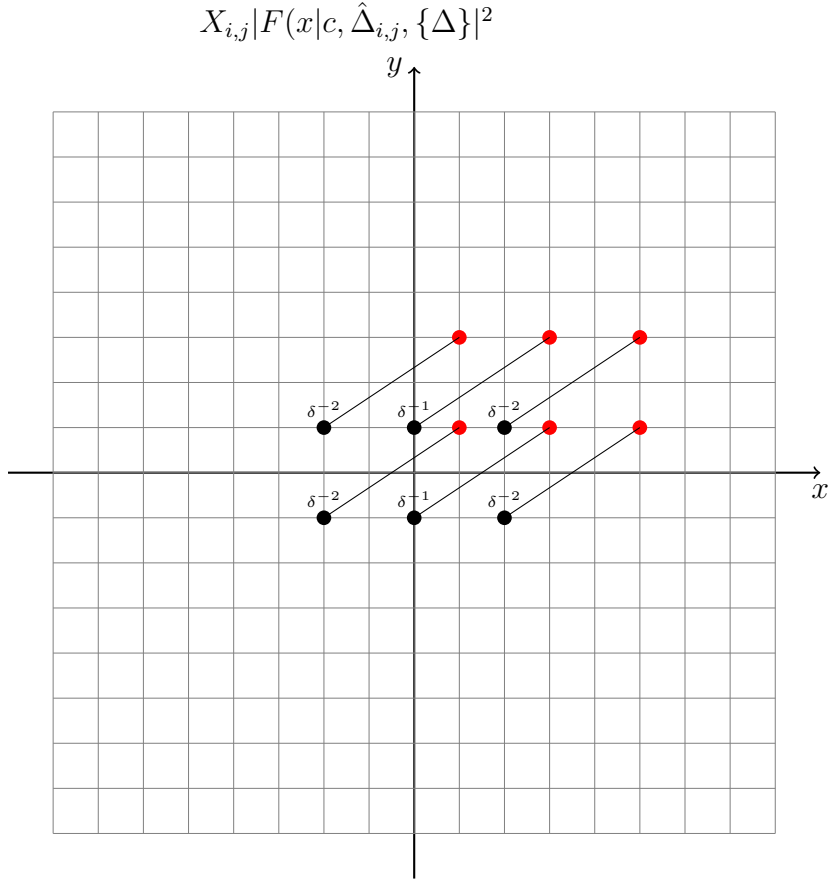

Until now we have seen correlation functions 
of the $c=0$ theory given by the 
collision of the channel $\Phi_{1,1}$ with $\Phi_{5,1}$ or 
with $\Phi_{1,3}$. The case $n=2,m=1$ is particularly interesting because, fixing $\mathcal{N}=O(\delta)$ both the channels
$\Phi_{5,1}$ and $\Phi_{1,3}$ provide a finite contribution to the integral, as shown in Fig. \ref{c0_n3}.

In this case, the logarithmic prefactor $a$ turns out to be
\begin{equation}
a=a_{2,1}^{(1)}+a_{2,1}^{(2)}=\sqrt{6},
\label{sum_a}
\end{equation}

This value is again formally related to the mixing parameter $b_{{\tiny\mbox{bulk}}}=-5$, recently conjectured \cite{VaGaJaSa, Ri}
for the bulk percolation and dilute polymer theory  indeed
\begin{equation}
\label{bulk}
-\left(\left.\frac{d \beta_c}{d c}\right|_{c=0} a\right)^{-1}=-5.
\end{equation} 
We emphasize that our result is obtained in a very different setting respect to the one in \cite{VaGaJaSa}.
We recall that CG integrals are necessarily  associated to a diagonal CFT whereas in \cite{VaGaJaSa},
the $b_{{\tiny\mbox{bulk}}}$ emerged as the mixing parameter between the non-diagonal field
$\phi_{-2,1}(x)\phi_{2,1}(\bar{x})$ and the stress energy tensor $T(x)$. Moreover,
since the structure constant $C_{\Delta_{\alpha}}$ in the OPE \eqref{id_fusion} is $O(c)$ the introduction of a logarithmic partner for $T(x)$, usually called $t(x)$, does not appear
necessary in our treatment; the numerical coincidence \eqref{bulk} is however quite interesting. 

Summarizing, we have shown three different possible situations occurring for the $c=0$ logarithmic 
CG integral, which correspond respectively to the mixing of $T\bar{T}$ with the scalar fields
$\Phi_{5,1}$,  $\Phi_{1,3}$ or with both. In this latter case, the operator $T\bar{T}$ has two logarithmic partners which
get identified, see eq. \eqref{sum_a}, in the $\beta^2\rightarrow 2/3$ limit, a circumstance reminiscent of the findings of
\cite{Ri}.  
We point out that the existence of three different scenarios is not in contradiction with the unicity of the $b$ parameter. Indeed, in order to construct a 
consistent theory, one has to fix the normalization of the two point functions for 
all the operators entering in the spectrum.
For instance, one could imagine that a consistent choice of these normalization factors would force all 
the correlation functions of the first two kinds to vanish, as it has also been argued in \cite{VaGaJaSa}.

\subsection{$\mathcal{M}_{p,p'}$ minimal models}
It is well known that the minimal models $\mathcal{M}_{p,p'}$ do not show logarithmic behavior.
Indeed the conformal blocks  are not singular as the Kac operators $\Phi_{r,s}$ satisfy
two differential equations, one of 
order $rs$ and the other of order $(p-r)(p'-s)$. 
The divergences associated to the corresponding two null-vector states are therefore compensated by the
vanishing of the associated matrix elements (\ref{deg_con}).

Maybe less known is the  fact that  the property of  $\mathcal{M}_{p,p'}$ to be a rational theory, does not imply the vanishing of the CG
structure constants $C(\alpha_{r_1,s_1},\alpha_{r_2,s_2},\alpha_{r_3,s_3})$
\cite{DoFa,DoFa2,DoFa3} when  they contain operators inside and outside the Kac table.
This is well explained in \cite{Doco}, where the $\mathcal{M}_{5,4}$ minimal model is taken as an example.
We review the argument given in \cite{Doco} following our scheme.
Let us consider for instance the correlation function 
\begin{equation}
\langle \Phi_{3,1}(0)\Phi_{3,1}(x) \Phi_{3,1}(1)\Phi_{-3,-1}(\infty)
\rangle,
\label{cfci}
\end{equation}
for $\beta=\sqrt{3/4}$ ($c=1/2$), i.e. the four-point energy correlation function in the Ising model.
The computation of \eqref{cfci} in the CG approach has been also proposed by \cite{DiFra} as an exercise.
This example shows that, even if the structure constant
$C(\alpha_{31},\alpha_{31},\alpha_{51})\neq 0$ does not vanish, the field $\Phi_{5,1}$, which is out side the Kac table,
do not enter in the computation of the four point correlation function.

\begin{figure}
\begin{center}
\begin{tikzpicture}[scale=0.6]
\draw (-5,10) node[right]{$X_{i,j}|F(x|c,\hat{\Delta}_{i,j},\{\Delta\}|^2$};
    \draw[->,thick] (-9,0) -- (9,0) node[below]{$x$};
    \draw[->,thick] (0,-9) --(0,9)node[left]{$y$};
\foreach \x in {-8,...,8}
                \draw[very thin, gray] (\x,-8)--(\x,8);
\foreach \y in {-8,...,8}
                \draw[very thin, gray] (-8,\y)--(8,\y);

\draw[->,dashed] (9,4)--(5,1);
\draw[->,dashed] (7,4)--(3,1);
\draw[->,dashed] (5,4)--(1,1);

\draw (5,1) node[circle,fill=black,minimum size=1pt,scale=0.5] {};
\draw (5,1) node[above]{{\tiny $\delta^{0}$}};

\draw (3,1) node[circle,fill=black,minimum size=1pt,scale=0.5] {};
\draw (3,1) node[above]{{\tiny $\delta^{0}$}};

\draw (1,1) node[circle,fill=black,minimum size=1pt,scale=0.5] {};
\draw (1,1) node[above]{{\tiny $\delta^{0}$}};

\draw[->] (5,1)--(1,-2);
\draw[->] (3,1)--(-1,-2);
\draw[->] (1,1)--(-3,-2);

\draw (1,-2) node[circle,fill=black,minimum size=1pt,scale=0.5] {};
\draw (1,-2) node[above]{{\tiny $\delta^{0}$}};

\draw (-1,-2) node[circle,fill=black,minimum size=1pt,scale=0.5] {};
\draw (-1,-2) node[above]{{\tiny $\delta^{0}$}};

\draw (-3,-2) node[circle,fill=black,minimum size=1pt,scale=0.5] {};
\draw (-3,-2) node[above]{{\tiny $\delta^{0}$}};

\draw[->,dashed] (1,-2)--(-3,-5);
\draw[->,dashed] (-1,-2)--(-5,-5);
\draw[->,dashed] (-3,-2)--(-7,-5);

\end{tikzpicture}
  \end{center}
 \caption{The grid of internal channel for the four-point correlation function of the energy operator
 $\Phi_{3,1}$ in the Ising model.} 
 \label{Ising}
\end{figure}
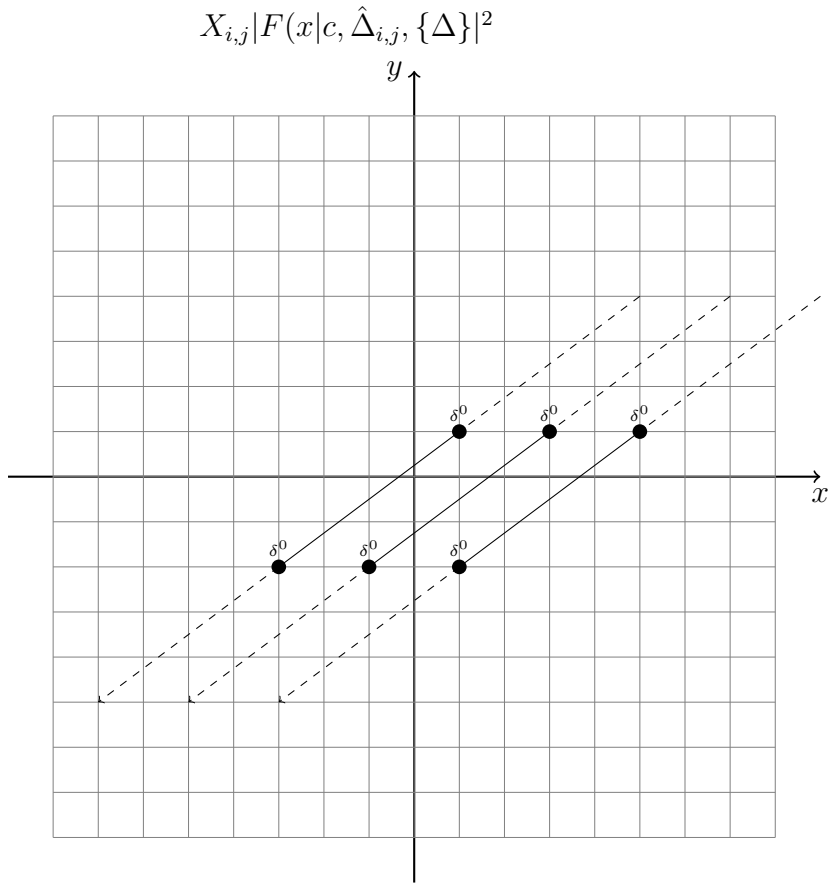

From $R_{12}(\alpha_{31},\alpha_{31})\neq 0$ one sees that the conformal block with $\Phi_{3,1}=\Phi_{1,2}$ 
has a singularity at level two. On the other hand, using the equations (\ref{logza1}), one obtains 
$C(\alpha_{31},\alpha_{31},\alpha_{31})=O(\delta)$ and
$C(\alpha_{31},\alpha_{31},\alpha_{51})=O(\delta^0)$ where $\delta\propto (\beta-\sqrt{3/4})$.
The internal translated channels of the correlator (\ref{cfci}) as well as their contributions in $\delta\propto
(\beta-\sqrt{3/4})$ are shown in Fig. \ref{Ising}.

  The dominant term of the colliding channel $\Phi_{3,1}$ and $\Phi_{5,1}$ cancels.
  The only contribution to the correlation function (\ref{cfci}) is therefore  given by the identity field
  $\Phi_{1,1}=\Phi_{-3,-2}$. One can therefore verify that only  operators inside the Kac table contribute to the correlation functions.

\section{Conclusions}
 We considered  local monodromy-invariant four-point
correlation functions
which are built from Virasoro algebra conformal blocks and which
posses a CG integral representation (\ref{4vertex}). In
particular we focused our attention on the singularities of the
integrals  (\ref{exp_int})
which generate, with an appropriate limit procedure, logarithmic terms.
We have shown that the analysis of these singularities can be
efficiently carried out by considering the analytic behavior of the
time-like Liouville structure constants (\ref{Zam_structure}) and of
Virasoro conformal blocks (\ref{rec_conf}).  In particular, we found
the key equations (\ref{logza1}, \ref{canc_sing}, \ref{canc_sing0}) which relate
the structure constants  (\ref{Zam_structure}) to combinatorial factor
coming from the Virasoro algebra representation theory. We remark that  in the
cases under consideration, the structure constants
(\ref{Zam_structure}) correspond to the Dotsenko-Fateev structure
constants \cite{DoFa,DoFa2,DoFa3}. Nevertheless, the use of the
properties of the function $\Upsilon_{\beta}(x)$, the building block for the formula 
(\ref{Zam_structure}),  greatly simplifies the
analysis, especially when multi-screening integrals are involved.

The relations   (\ref{logza1}),(\ref{canc_sing}) and
(\ref{canc_sing0}) assure us that the dominant contributions coming from
singular conformal block cancel and that the remaining finite terms
contain logarithmic singularities. In particular they express in a compact form the
fact  that, for a general central charge, the logarithms are generated
by the collision of a singular null field
$\Lambda_{P,Q}(x,\bar{x})=\chi_{P,Q}(x)\otimes\chi_{P,Q}(\bar{x})$ with one
of the non-degenerate primaries $\Phi_{-P,Q}(x,\bar{x})$ or $\Phi_{P,-Q}(x,\bar{x})$.

Using our approach, we considered in detail the CG
correlation functions when $\beta^2 \notin \mathbb {Q}$,  see
(\ref{central}). When the external charges take general values, we
provided a direct graphical method to predict the contributions of the
expansion (\ref{exp_int}). We adapted our approach to the case of
$\beta^2=\sqrt{p'/p}$, with $p'$ and $p$ positive and coprime.  We
applied our method to the study of correlation functions of leg
operators in dense polymers and spanning trees ($c=-2$)  and to correlation functions of $c=0$ theories.

We showed in particular
that the situation in which the CG integral gets
contributions from both the colliding fields
$\Phi_{-P,Q}(x,\bar{x})$ and  $\Phi_{P,-Q}(x,\bar{x})$ can occur.  In
the $c=0$ theory, this suggests the possibility that the operator
$T(x)\bar{T}(\bar{x})$ has two logarithmic partner which get
identified in the limit $c\to0$. In this particular case the prefactor of the logarithmic term is
simply related to the value $b_{{\tiny\text{bulk}}}=-5$, determined in \cite{Ri, VaGaJaSa} for bulk $c=0$ LCFTs. 

The motivation behind this work is  the construction of consistent
CFTs to study geometric properties of critical phases.
Although the correlation functions, for which the CG integral furnishes an integral representation necessarily come from
a diagonal CFT, theoretical
arguments and numerical investigations on Potts models \cite{DPSV}
recently suggested that the diagonal theories based on the time-like Liouville structure constants
are physically relevant to construct  new CFTs capable of
describing random conformal fractals.  The results presented
here show also that the analytic properties of time-like Liouville
structure constants encode logarithmic features of Virasoro CFTs, and
turn out to be very useful to determine the singularities of  CG
integrals as well as to compute local logarithmic correlation
functions. This was shown by evaluating the logarithmic behavior of
physically relevant correlation functions which satisfy differential
equations of order higher than two, i,e. which are given by
multi-screening CG integrals.
\vspace{1cm}

\noindent\textbf{ACKNOWLEDGMENTS.} 
We thank R. Vasseur for having share with us his knowledge of LCFTs and
for showing us some of his results before
publication. We are grateful to V. Dotsenko and H. Saleur for suggestions and comments and we also benefit from discussions
with J. Jacobsen, Y. Ikhlef, V. Schomerus and J. Teschner.

\begin{appendices}
 
 \section{Explicit expressions  in the one-screening case}
 
 From the identities
\begin{equation}
\frac{d}{d x} [x]_s=[x]_s \sum_{j=0}^{s-1}\frac{1}{x+j},\quad \Gamma(x+n)=[x]_n \Gamma(x),
\quad \sin{\pi x}=\frac{1}{\Gamma(1-x)\Gamma(x)}
\end{equation}
with $[x]_n=\Gamma(x+n)/\Gamma(x)$ being the Pochhammer symbol, the functions $G^{(a)}_{i,j}$ and $X_{i,j}^{(a)}$
can be easily determined

\begin{align}
&G^{(-1)}_{0,0}(x)=\frac{(-1)^{s-1}[-2\beta \alpha_2]_{s}[-s-2\beta \alpha_3]_{s}x^{s} }{2\beta\Gamma(s)\Gamma(1+s)} \;  G^{(0)}_{1,0}(x),
 \label{G000}\\
&G^{(0)}_{1,0}(x)= (1-x)^{2\alpha_2\alpha_3}_2F_1(-2\beta \alpha_3,s-2\beta \alpha_2;1+s;x), \label{G011}\\
&G^{(0)}_{0,0}(x)=(1-x)^{2\alpha_2\alpha_3}\left[\sum_{l=0}^{s-1}\frac{[-2\beta \alpha_2]_l [-s-2\beta \alpha_3]_l}{l![1-s]_l} x^l+\frac{[-2\beta 
\alpha_2]_{s}[s-2\beta \alpha_3]_{s}x^{s} }{s![1-s]_{s-1}}H(x)\right] \label{G11} \\
&G^{(1)}_{1,0}(x)=-(1-x)^{2\alpha_2\alpha_3}\left[\sum_{l=1}^{\infty}\frac{[-2\beta \alpha_3]_{l} [1+s-2\beta \alpha_2]_l}{l![2+s]_l} x^l
\sum_{j=0}^{l-1}
\frac{1}{j+2+s}\right], \\
&H(x)=\left[\frac{1}{1+2\beta \alpha_3}-\sum_{l=0}^{s-2}\left(\frac{1}{l+1-s}-
\frac{1}{l-s-2\beta \alpha_3}\right)\right] _2F_1(-2\beta \alpha_3,s-2\beta \alpha_2;1+s;x)\nonumber \\
&+\sum_{l=1}^{\infty}\frac{[-2\beta \alpha_3]_{l} [1+2\beta \alpha_2]_l}{l![1+s]_l} x^l \sum_{k=0}^{l-1}
\left(\frac{1}{k+1}-\frac{1}{k-2\beta \alpha_3}\right),\label{HH}\\
&X^{(-1)}_{1,0}=-\frac{1}{2\beta}\left[\frac{\pi[-2\beta\alpha_2]_s}{\Gamma(1+s)}\right]^2,\\
&X^{(1)}_{0,0}=2\beta\left[\frac{\pi\Gamma(s)}{[-s-2\beta\alpha_3]_{s}}\right]^2.
\end{align}

\section{The special function $\Upsilon_{\beta}(x)$}
This appendix contains some properties of the  function $\Upsilon_{\beta}(x)$, for more details we remand to the comprehensive
review  \cite{Barnes}. The function $\Upsilon_{\beta}(x)$ is related to the Barnes
double Gamma function $\Gamma_2(x|\omega_1,\omega_2)$ defined for  $\text{Re}(\omega_1)>0$ and $\text{Re}(\omega_2)>0$
by 
\begin{equation}
\log\Gamma_2(x|\omega_1,\omega_2)\equiv\left.\frac{d}{ds}\right|_{s=0}\sum_{n,m=0}^{\infty}\frac{1}{(x+n\omega_1+m\omega_2)^s},\quad \text{Re}(s)\geq2. 
\end{equation}
The function $\Gamma_2(x|\omega_1,\omega_2)$ satisfies the shift relations
\begin{align}
\label{shift1}
&\Gamma_2(x+\omega_1|\omega_1,\omega_2)=\frac{\sqrt{2\pi}\omega_2^{1/2-x/\omega_2}}{\Gamma(x/\omega_2)}
\Gamma_2(x|\omega_1,\omega_2),\\
\label{shift2}
&\Gamma_2(x+\omega_2|\omega_1,\omega_2)=\frac{\sqrt{2\pi}\omega_1^{1/2-x/\omega_1}}{\Gamma\bigl(x/\omega_1)}
\Gamma_2(x|\omega_1,\omega_2),
\end{align}
which show that it has simple poles on the lattice points $x=-n\omega_1-m\omega_2$, for $n$ and $m$ non-negative.
The proof of (\ref{shift1}-\ref{shift2}) is sketched below. Observe that
\begin{equation}
\label{app1}
 \frac{1}{x^s}=-\frac{1}{2\pi i}\Gamma(1-s)\int_C dt (-t)^{s-1}e^{-xt},
\end{equation}
where the contour $C$ starts end at $\infty$ and encircles the origin counterclokwise. The branch cut  in \eqref{app1} is chosen
on the positive real axis and arguments in the complex $t$ plane are bounded in $[-\pi,\pi]$. Using \eqref{app1} and
performing explicitly the contour integral\footnote{One must regularize the divergent integrals when the contour has been split.}, the relation
\begin{equation}
\label{int_rep}
 \left.\frac{d}{ds}\right|_{s=0}\sum_{n=0}^{\infty}\frac{1}{(x+n\omega)^s}=\frac{1}{\omega}
 \int_{0}^{\infty}\frac{dt}{t}\left[-\frac{1}{t}-\frac{e^{-t}(\omega-2x)}{2}+\frac{\omega e^{-xt}}{1-e^{-\omega t}}\right]
\end{equation}
can be proved. Then recalling Binet formula for the logarithm of the Gamma function
\begin{equation}
\log\Gamma(x)=(x-1/2)\log x+\frac{1}{2}\log 2\pi-x+\int_{0}^{\infty}\frac{dt}{t}e^{-tx}\left(\frac{1}{2}-\frac{1}{t}+\frac{1}{e^t-1}\right), 
\end{equation}
we obtain
\begin{equation}
\left.\frac{d}{ds}\right|_{s=0}\sum_{n=0}^{\infty}\frac{1}{(x+n\omega)^s}=\frac{1}{2}\log 2\pi+\left(\frac{1}{2}-\frac{x}{\omega}\right)\log\omega-\log\Gamma(x/\omega) 
\end{equation}
and the shift relations are now easily verified. Analogously to  \eqref{int_rep} one derives the integral representation
for the Barnes double Gamma function, valid for $\text{Re}(x)>0$
\begin{align}
\label{doubleGamma}
\log\Gamma_2(x|\omega_1,\omega_2)&=\frac{1}{\omega_1\omega_2}\int_{0}^{\infty}\frac{dt}{t}\left[-\frac{1}{t^2}-\frac{Q-2x}
{2 t}-\frac{e^{-t}[(2x-Q)^2+2x(x-Q)+\omega_1\omega_2]}{12}+\right.\nonumber\\ 
&\hspace*{3cm}\left.+\frac{\omega_1\omega_2 e^{-xt}}{(1-e^{-\omega_1 t})(1-e^{-\omega_2t})}\right],
\end{align}
where we defined $Q\equiv\omega_1+\omega_2$. Following \cite{Zam_lectures}, we introduce the function
$\Upsilon(x|\omega_1,\omega_2)$ as
\begin{equation}
\Upsilon(x|\omega_1,\omega_2)=\frac{\Gamma_2^2(Q/2|\omega_1,\omega_2)}{\Gamma_2(x|\omega_1,\omega_2)\Gamma_2(Q-x|\omega_1,\omega_2)}.
\end{equation}
It is then straightforward to check that
for $\omega_1=\omega_2^{-1}=\beta$, $\Upsilon(x|\beta,\beta^{-1})\equiv\Upsilon_\beta(x)$ has the integral representation
\eqref{Zam_int} and that satisfies the shift relations \eqref{rec_Zam}.
Notice that the
function $\Upsilon_{\beta}(x)$ has zeros at $x=-n\beta-m\beta^{-1}$ or $x=(n+1)\beta+(m+1)\beta^{-1}$ with $n,m$ non-negative.
\end{appendices}


\begin{thebibliography}{10}

\bibitem{DiFra}
P.~Di Francesco, P.~Mathieu, and D.~S{\'e}n{\'e}chal.
\newblock {\em Conformal Field Theory}.
\newblock Springer New York, 1997.

\bibitem{ItSaZu}
C.~Itzykson, H.~Saleur, and J.B. Zuber.
\newblock {\em Conformal invariance and applications to statistical mechanics}.
\newblock World Scientific, 1988.

\bibitem{BPZ}
A.~Belavin, A.~Polyakov, and A.~B. Zamolodchikov.
\newblock Infinite conformal symmetry in two-dimensional quantum field theory.
\newblock {\em Nuclear Physics B}, 241(2):333--380, 1984.

\bibitem{Durev}
B.~Duplantier.
\newblock Conformal fractal geometry and boundary quantum gravity.
\newblock {\em \tt{ArXiv:math-ph/0303034}}, 2003.

\bibitem{FK1}
C.M. Fortuin and P.W. Kasteleyn.
\newblock On the {R}andom {C}luster {M}odel {I}: introduction and relation to
  other models.
\newblock {\em Physica}, 57(4):536--564, 1972.

\bibitem{FK2}
CM~Fortuin.
\newblock On the {R}andom {C}luster {M}odel {II}: the percolation model.
\newblock {\em Physica}, 58(3):393--418, 1972.

\bibitem{Ni}
B.~Nienhuis.
\newblock Coulomb {G}as formulation of two-dimensional phase transitions.
\newblock {\em Phase transitions and critical phenomena}, 11:1--53, 1987.

\bibitem{BaBe}
M.~Bauer and D.~Bernard.
\newblock {2D} growth processes: {SLE} and loewner chains.
\newblock {\em Physics reports}, 432(3):115--221, 2006.

\bibitem{GaKa}
M.~Gaberdiel and H.~G. Kausch.
\newblock A local logarithmic conformal field theory.
\newblock {\em Nuclear Physics B}, 538(3):631--658, 1999.

\bibitem{GaKao}
M.~Gaberdiel and H.~G. Kausch.
\newblock A rational logarithmic conformal field theory.
\newblock {\em Physics Letters B}, 386(1):131--137, 1996.

\bibitem{Ka}
H.~G. Kausch.
\newblock Symplectic fermions.
\newblock {\em Nuclear Physics B}, 583(3):513--541, 2000.

\bibitem{GaRuWo}
M.~Gaberdiel, I.~Runkel, and S.~Wood.
\newblock Fusion rules and boundary conditions in the c=0 triplet model.
\newblock {\em Journal of Physics A: Mathematical and Theoretical},
  42(32):325403, 2009.

\bibitem{Gu}
V.~Gurarie.
\newblock Logarithmic operators in conformal field theory.
\newblock {\em Nuclear Physics B}, 410(3):535--549, 1993.

\bibitem{Ca_rev_log}
J.~Cardy.
\newblock Logarithmic conformal field theories as limits of ordinary {CFTs} and
  some physical applications.
\newblock {\em \tt{ArXiv:1302.4279}}, 2013.

\bibitem{CrRi}
T.~Creutzig and D.~Ridout.
\newblock {L}ogarithmic {C}onformal {F}ield {T}heory: {B}eyond an
  {I}ntroduction.
\newblock {\em \tt{ArXiv:1303.0847}}, 2013.

\bibitem{Gu_rev}
V.~Gurarie.
\newblock Logarithmic operators and logarithmic conformal field theories.
\newblock {\em \tt{ArXiv:1303.1113}}, 2013.

\bibitem{GJRSV}
A.~Gainutdinov, J.~Jacobsen, N.~Read, H.~Saleur, and R.~Vasseur.
\newblock {L}ogarithmic {C}onformal {F}ield {T}heory: a {L}attice {A}pproach.
\newblock {\em \tt{ArXiv:1303.2082}}, 2013.

\bibitem{FuHwSeTi}
J.~Fuchs, S.~Hwang, A.~Semikhatov, and I.~Yu Tipunin.
\newblock Nonsemisimple fusion algebras and the {V}erlinde formula.
\newblock {\em Communications in mathematical physics}, 247(3):713--742, 2004.

\bibitem{FeGaSeTio}
B.~Feigin, A.~Gainutdinov, A.~Semikhatov, and I.~Yu Tipunin.
\newblock Modular group representations and fusion in logarithmic conformal
  field theories and in the quantum group center.
\newblock {\em Communications in mathematical physics}, 265(1):47--93, 2006.

\bibitem{FeGaSeTi}
B.~Feigin, A.~Gainutdinov, A.~Semikhatov, and I.~Yu Tipunin.
\newblock Logarithmic extensions of minimal models: characters and modular
  transformations.
\newblock {\em Nuclear Physics B}, 757(3):303--343, 2006.

\bibitem{PeRa}
P.~Pearce and J.~Rasmussen.
\newblock Solvable critical dense polymers.
\newblock {\em Journal of Statistical Mechanics: Theory and Experiment},
  2007(02):P02015, 2007.

\bibitem{PeRaZu}
P.~Pearce, J.~Rasmussen, and J.B. Zuber.
\newblock Logarithmic minimal models.
\newblock {\em Journal of Statistical Mechanics: Theory and Experiment},
  2006(11):P11017, 2006.

\bibitem{ReSa}
N.~Read and H.~Saleur.
\newblock Associative-algebraic approach to logarithmic conformal field
  theories.
\newblock {\em Nuclear Physics B}, 777(3):316--351, 2007.

\bibitem{GaJaVaSa}
A.~Gainutdinov, J.~Jacobsen, H.~Saleur, and R.~Vasseur.
\newblock A physical approach to the classification of indecomposable
  {V}irasoro representations from the blob algebra.
\newblock {\em Nuclear Physics B}, 2013.

\bibitem{VaGaJaSa}
R.~Vasseur, A.~Gainutdinov, J.~Jacobsen, and H.~Saleur.
\newblock Puzzle of bulk conformal field theories at central charge c= 0.
\newblock {\em Physical Review Letters}, 108(16):161602, 2012.

\bibitem{Ri}
D.~Ridout.
\newblock Non-chiral logarithmic couplings for the {V}irasoro algebra.
\newblock {\em Journal of Physics A: Mathematical and Theoretical},
  45(25):255203, 2012.

\bibitem{FuScSt}
J.~Fuchs, C.~Schweigert, and C.~Stigner.
\newblock From non-semisimple hopf algebras to correlation functions for
  logarithmic {CFT}.
\newblock {\em \tt{ArXiv:1302.4683}}, 2013.

\bibitem{DeVi}
G.~Delfino and J.~Viti.
\newblock On the three-point connectivity in two-dimensional percolation.
\newblock {\em Journal of physics. A, Mathematical and theoretical}, 44(3),
  2011.

\bibitem{Za}
Al. Zamolodchikov.
\newblock Three-point function in the minimal {L}iouville gravity.
\newblock {\em Theoretical and mathematical physics}, 142(2):183--196, 2005.

\bibitem{KoPe}
I.~Kostov and V.~Petkova.
\newblock Bulk correlation functions in {2D} quantum gravity.
\newblock {\em Theoretical and mathematical physics}, 146(1):108--118, 2006.

\bibitem{Ziff}
R.~Ziff, J.~Simmons, and P.~Kleban.
\newblock Factorization of correlations in two-dimensional percolation on the
  plane and torus.
\newblock {\em Journal of Physics A: Mathematical and Theoretical},
  44(6):065002, 2011.

\bibitem{DPSV}
M.~Picco, R.~Santachiara, J.~Viti, and G.~Delfino.
\newblock Connectivities of {P}otts {F}ortuin-{K}asteleyn clusters and
  time-like {L}iouville correlator.
\newblock {\em Nuclear Physics B 875 719}, 2013.

\bibitem{HaMaWi}
D.~Harlow, J.~Maltz, and E.~Witten.
\newblock Analytic continuation of {L}iouville theory.
\newblock {\em Journal of High Energy Physics}, 2011(12):1--105, 2011.

\bibitem{TeLiouville}
J.~Teschner.
\newblock Liouville theory revisited.
\newblock {\em Classical and Quantum Gravity}, 18(23):R153, 2001.

\bibitem{Zam_lectures}
A.~B. Zamolodchikov and Al. Zamolodchikov.
\newblock Lectures on {L}iouville {T}heory and {M}atrix {M}odels.
\newblock {\em \tt{http://qft.itp.ac.ru/ZZ.pdf}}, 2007.

\bibitem{DoFa}
Vl. Dotsenko and V.~Fateev.
\newblock Conformal algebra and multipoint correlation functions in {2D}
  statistical models.
\newblock {\em Nuclear Physics B}, 240(3):312--348, 1984.

\bibitem{DoFa2}
Vl. Dotsenko and V.~Fateev.
\newblock Four-point correlation functions and the operator algebra in {2D}
  conformal invariant theories with central charge $0<c<1$.
\newblock {\em Nuclear Physics B}, 251:691--734, 1985.

\bibitem{DoFa3}
Vl. Dotsenko and V.~Fateev.
\newblock Operator algebra of two-dimensional conformal theories with central
  charge $0<c<1$.
\newblock {\em Physics Letters B}, 154(4):291--295, 1985.

\bibitem{DPSV2}
G.~Delfino, M.~Picco, R.~Santachiara, and J.~Viti.
\newblock Spin clusters and conformal field theory.
\newblock {\em \tt{ArXiv:1307.6123}}.
\newblock Accepted for publication in JSTAT.

\bibitem{Zamhem}
Al. Zamolodchikov.
\newblock Higher equations of motion in {L}iouville field theory.
\newblock {\em International Journal of Modern Physics A}, 19(supp02):510--523,
  2004.

\bibitem{DoOt}
H.~Dorn and H.~Otto.
\newblock Two-and three-point functions in {L}iouville theory.
\newblock {\em Nuclear Physics B}, 429(2):375--388, 1994.

\bibitem{ZaZa}
A.~B. Zamolodchikov and Al. Zamolodchikov.
\newblock Conformal bootstrap in {L}iouville field theory.
\newblock {\em Nuclear Physics B}, 477(2):577--605, 1996.

\bibitem{Salog}
H.~Saleur.
\newblock Polymers and percolation in two dimensions and twisted {N=2}
  supersymmetry.
\newblock {\em Nuclear Physics B}, 382(3):486--531, 1992.

\bibitem{HaYa}
H.~Hata and S.~Yamaguchi.
\newblock Logarithmic behaviours in the {F}eigin-{F}uchs construction of the
  {c=-2} conformal field theory.
\newblock {\em Physics Letters B}, 482(1):283--286, 2000.

\bibitem{Ya}
S.~Yamaguchi.
\newblock Logarithmic correlation functions in {L}iouville field theory.
\newblock {\em Physics Letters B}, 546(3):300--304, 2002.

\bibitem{JDPSW}
J.~Jacobsen, P.~Le Doussal, M.~Picco, R.~Santachiara, and K.~Wiese.
\newblock Critical interfaces in the random-bond {P}otts model.
\newblock {\em Physical review letters}, 102(7):070601, 2009.

\bibitem{Zam_rec1}
Al. Zamolodchikov.
\newblock Conformal symmetry in two dimensions: {A}n explicit recurrence
  formula for the conformal partial wave amplitude.
\newblock {\em Communications in Mathematical Physics}, 96(3):419--422, 1984.

\bibitem{Zam_rec2}
Al. Zamolodchikov.
\newblock Conformal symmetry in two-dimensional space: recursion representation
  of conformal block.
\newblock {\em Theoretical and Mathematical Physics}, 73(1):1088--1093, 1987.

\bibitem{Yan}
S.~Yanagida.
\newblock Norm of logarithmic primary of virasoro algebra.
\newblock {\em Letters in Mathematical Physics}, 98(2):133--156, 2011.

\bibitem{Doco}
Vl. Dotsenko.
\newblock S{\'e}rie de cours sur la th{\'e}orie conforme.
\newblock {\em \tt{http://cel.archives-ouvertes.fr/cel-00092929/en/}}, 2004.

\bibitem{Lebedev}
http://en.wikipedia.org/wiki/Hypergeometric\_function.

\bibitem{JDPSW_un}
J.~Jacobsen, P.~Le Doussal, M.~Picco, R.~Santachiara, and K.~Wiese.
\newblock unpublished.

\bibitem{CaSi}
J.~Simmons and J.~Cardy.
\newblock Twist operator correlation functions in {O(n)} loop models.
\newblock {\em Journal of Physics A: Mathematical and Theoretical},
  42(23):235001, 2009.

\bibitem{AGT}
L.~Alday, D.~Gaiotto, and Y.~Tachikawa.
\newblock Liouville correlation functions from four-dimensional gauge theories.
\newblock {\em Letters in Mathematical Physics}, 91(2):167--197, 2010.

\bibitem{SaTa10}
R.~Santachiara and A.~Tanzini.
\newblock Moore-read fractional quantum hall wave functions and {SU(2)} quiver
  gauge theories.
\newblock {\em Physical Review D}, 82(12):126006, 2010.

\bibitem{Ca_rev}
J.~Cardy.
\newblock {SLE} for theoretical physicists.
\newblock {\em Annals of Physics}, 318(1):81--118, 2005.

\bibitem{VaJaSa}
R.~Vasseur, J.~Jacobsen, and H.~Saleur.
\newblock Indecomposability parameters in chiral logarithmic conformal field
  theory.
\newblock {\em Nuclear Physics B}, 851(2):314--345, 2011.

\bibitem{DuLERW}
B.~Duplantier.
\newblock Loop-erased self-avoiding walks in two dimensions: exact critical
  exponents and winding numbers.
\newblock {\em Physica A: Statistical Mechanics and its Applications},
  191(1):516--522, 1992.

\bibitem{DuSa3}
B.~Duplantier and H.~Saleur.
\newblock Exact surface and wedge exponents for polymers in two dimensions.
\newblock {\em Physical review letters}, 57(25):3179, 1986.

\bibitem{Maj}
S.~Majumdar.
\newblock Exact fractal dimension of the loop-erased self-avoiding walk in two
  dimensions.
\newblock {\em Physical review letters}, 68(15):2329, 1992.

\bibitem{GuLu1}
V.~Gurarie and A.~Ludwig.
\newblock Conformal algebras of two-dimensional disordered systems.
\newblock {\em Journal of Physics A: Mathematical and General}, 35(27):L377,
  2002.

\bibitem{GuLu2}
V.~Gurarie and A.~Ludwig.
\newblock Conformal field theory at central charge c= 0 and two-dimensional
  critical systems with quenched disorder.
\newblock {\em \tt{ArXiv: hep-th/0409105}}, 2004.

\bibitem{Barnes}
E.~Barnes.
\newblock The theory of the double gamma function.
\newblock {\em Philosophical Transactions of the Royal Society of London.
  Series A, Containing Papers of a Mathematical or Physical Character},
  196:265--387, 1901.

\end{thebibliography}
\end{document}